\documentclass[12 pt,a4paper, oneside, openany, notitlepage]{article}
\input epsf
\usepackage{subcaption}
\usepackage{graphicx}
\usepackage[usenames,dvipsnames]{xcolor}
\usepackage{soul}

\usepackage{amsmath, amssymb, graphics}
\usepackage{mathtools}
\newcommand{\mathsym}[1]{{}}
\newcommand{\unicode}[1]{{}}

\usepackage{amsthm}
\usepackage{amsfonts}
\usepackage{multicol}
\usepackage{lipsum}
\usepackage{ulem}
\usepackage{multirow}
\usepackage{makecell}
\usepackage{float}
\usepackage{array, makecell} 
\newcommand{\PreserveBackslash}[1]{\let\temp=\\#1\let\\=\temp}
\newcolumntype{C}[1]{>{\PreserveBackslash\centering}p{#1}}
\usepackage{longtable}
\usepackage{tabu}
\usepackage{threeparttable}
\usepackage{threeparttablex}
\usepackage{booktabs}
\usepackage{amsmath}
\usepackage{graphicx}   

\usepackage{tikz}
\def\checkmark{\tikz\fill[scale=0.4](0,.35) -- (.25,0) -- (1,.7) -- (.25,.15) -- cycle;} 

\usepackage[inner=0.7in,outer=0.7in,bottom=0.9in,top=0.9in]{geometry}
\usepackage{setspace}
\usepackage{lmodern}

\usepackage{lscape}

 \newtheoremstyle{mystyle1}
  {\topsep}
  {\topsep}
  {\normalfont}
  {}
  {\bfseries \color{black}}
  {.}
  {.5em}
  {}

\theoremstyle{mystyle1}

\newtheorem{theorem}{Theorem}[section]
  {\popQED\end{theorem}}

\newtheorem{proposition}{Proposition}
\newenvironment{prop}{\begin{proposition}%
 \pushQED{\qed}}%
  {\popQED\end{proposition}}

  \newtheorem{lemma}{Lemma}
\newenvironment{lm}{\begin{lemma}%
  \pushQED{\qed}}%
  {\popQED\end{lemma}}

    \newtheorem{definition}{Definition}
\newenvironment{def2}{\begin{definition}%
  \pushQED{\qed}}%
  {\popQED\end{definition}}
  
    \newtheorem{corollary}{Corollary}
  {\popQED\end{corollary}}

 \newtheoremstyle{mystyle}
  {\topsep}
  {\topsep}
  {\normalfont}
  {}
  {\bfseries \color{black}}
  {.}
  {.5em}
  {}

\theoremstyle{mystyle}
  
  \newtheorem{remarkex}{Remark}
\newenvironment{remark}
  {\pushQED{\qed}\remarkex}
  {\popQED\endremarkex}

\newtheorem{assumptionex}{Assumption}
\newenvironment{assumption}
  {\pushQED{\qed}\assumptionex}
 {\popQED\endassumptionex}

\usepackage{diagbox}

\usepackage[nouppercase]{frontespizio}
\usepackage{epsfig}
\usepackage{geometry}
\usepackage{morefloats}
\usepackage{bbm}

\usepackage{changepage}

\usepackage{enumerate}
\usepackage[shortlabels]{enumitem}
\newlist{Part}{enumerate}{2}
\setlist[Part]{label={ Part \arabic*} ,itemindent=*}

\newlist{Challenge}{enumerate}{2}
\setlist[Challenge]{label={Challenge \arabic*} ,itemindent=*}

\usepackage[colorlinks = true,
            allcolors  = blue, 
            linkcolor=blue,
            urlcolor=black
            ]{hyperref}

\usepackage{xparse}

\NewDocumentCommand{\MYref}{O{black}mo}{%
  \begingroup
  \hypersetup{linkcolor=#1}%
  \ref{#2}%
  \IfValueT{#3}{%
    \color{#1}{#3}%
  }%
  \endgroup
}

\usepackage{bigints}
\usepackage[font={footnotesize}]{caption}
\usepackage{chngcntr}
\usepackage{comment}

\usepackage[multiple]{footmisc}

\usepackage{xr}

\usepackage{mathrsfs}

\usepackage[gen]{eurosym}

\usepackage[T1]{fontenc}

\usepackage{pdflscape}

\renewenvironment{abstract}
 {\small
  \begin{center}
  \bfseries \abstractname\vspace{-.5em}\vspace{0pt}
  \end{center}
  \list{}{
    \setlength{\leftmargin}{.1cm}%
    \setlength{\rightmargin}{\leftmargin}%
  }%
  \item\relax}
 {\endlist}

\begin{document}
\title{{\scshape Partial identification in matching models for the marriage market\thanks{\scriptsize First version: February 2019. We would like to thank  Pierre-Andr\'e Chiappori, Thierry Magnac, and the seminar and conference participants at Toulouse School of Economics, the University of Warwick, IAAE 2019, EEA-ESEM 2019, the Bristol-TSE Econometrics Workshop, the 2019 Cowles Foundation conference on Matching: Optimal Transport and Beyond,  the 2019 Network Econometrics Juniors’ Conference at Northwestern University, and the University of Toronto.    Rossi Abi-Rafeh and Camila Comunello provided excellent research assistance. We acknowledge funding from the French National Research Agency (ANR) under the Investments for the Future (Investissements d'Avenir) program, grant ANR-17-EURE-0010. An earlier version of the paper was previously circulated under the title ``Partial Identification and Inference in
One-to-One Matching Models with
Transfers''.}}}
\author{Cristina Gualdani\thanks{\scriptsize Email: \href{mailto:c.gualdani@qmul.ac.uk}{c.gualdani@qmul.ac.uk}, Queen Mary University of London, London, UK.}\text{ } \text{ } Shruti Sinha\thanks{\scriptsize Email: \href{mailto:shruti.sinha@tse-fr.eu}{shruti.sinha@tse-fr.eu}, Toulouse School of Economics, University of Toulouse Capitole, Toulouse, France.}}
\date{July 2022}
\maketitle
\begin{abstract}
We study partial identification of the preference parameters in the one-to-one matching model with perfectly transferable utilities. We do so without imposing parametric distributional assumptions on the unobserved heterogeneity and with data on one large market. We provide a tractable characterisation of the identified set under various classes of nonparametric distributional assumptions on the unobserved heterogeneity. Using our methodology, we re-examine some of the relevant questions in the empirical literature on the marriage market, which have been previously studied under the Logit assumption. {Our results reveal that many findings in the aforementioned literature  are  primarily driven by such parametric restrictions.}
\vspace{1cm}

\end{abstract}
\newpage

\section{Introduction}
\label{introduction}
Matching markets are two-sided markets, where agents on each side have preferences
over matching with agents on the other side. For example, social interactions lead individuals to find marital partners, production tasks are assigned to workers, and auctions sort buyers with sellers.
While the economic theory of matching models has been around for more than five decades, the literature on empirical matching models is relatively recent  (\hyperlink{Chiappori}{Chiappori and Salani\'e, 2016}).

An important strand of this literature focuses on the one-to-one matching model, in which every agent forms at most one match. Each possible match generates a surplus. In the framework where utilities are perfectly transferable, agents can share the match surplus with their partners without frictions.
Since \hyperlink{becker}{Becker (1973)}, the one-to-one matching model with perfectly transferable utilities (hereafter, 1to1TU) has been extensively used in household economics to represent the marriage market (\hyperlink{Chiappori3}{Chiappori, 2017}). In particular, researchers have exploited the 1to1TU model to estimate the systematic part of the match surplus. Recovering the systematic match surplus is useful, for example, to investigate sorting patterns and how they change over time, to learn about the complementarities and substitutabilities of partner characteristics, to assess the efficiency and welfare implications of the status-quo assignment, and to measure the impact of pre-marital decisions on the sharing of the match surplus between spouses.\footnote{The 1to1TU model has also been used to study matching of CEOs to firms (\hyperlink{Chen}{Chen, 2017}), matching of academics to offices (\hyperlink{Baccara}{Baccara, et al., 2012}), merging of banks (\hyperlink{Akkus}{Akkus, Cookson, and Horta\c{c}su, 2016}), formation of research alliances (\hyperlink{Mindruda2}{Mindruda, Moeen, and Agarwal, 2016}), and collaboration between academics and firms (\hyperlink{Mindruda}{Mindruda, 2013}; \hyperlink{Banal}{Banal-Esta\~{n}ol, Macho-Stadler, and P\'{e}rez-Castrillo, 2018}).}

Most of the papers using the 1to1TU model to estimate the systematic match surplus proceed under strong parametric distributional restrictions on the agents' unobserved heterogeneity. These restrictions amount to imposing i.i.d. standard Extreme Value Type I  taste shocks, independently distributed from covariates. Along with data on {one large} market, these restrictions make the 1to1TU model {just} identified, leading to {point} identification of the systematic match surplus via standard Logit  formulas (\hyperlink{Choo_Siow}{Choo and Siow, 2006}). 
The motivation for using the Logit  1to1TU model is computational simplicity. However, this framework may lead to paradoxical conclusions that run against economic sense. For example, it is well known that the one-sided Logit  model is inherently linked to the independence of irrelevant alternatives  (IIA) axiom and severely restricts cross-elasticities. The same holds in two-sided markets and causes unintuitive 
 comparative static predictions, as explained in \hyperlink{Graham}{Graham (2013a)} and \hyperlink{Galichon_Salanie2}{Galichon and Salani\'e (2019)}.

The fact that widespread empirical practices rest on the Logit  1to1TU model raises several questions.  When we have data on one large market, does the 1to1TU model retain any identifying power on the systematic match surplus without restrictions on the taste shock distribution? If not, is it still possible to recover some information  on the systematic match surplus under nonparametric distributional assumptions  on the unobserved heterogeneity? How are the answers to relevant policy questions driven by the Logit assumption? The contribution of our paper is to address these issues.  By doing so, we also offer methodological guidance for researchers who wish to consider more robust alternatives to (or do sensitivity checks of) the Logit  1to1TU model. 

We start our analysis by observing that, if the taste shock distribution is not assumed to be {fully} known by the researcher, then the 1to1TU model is {under}-identified with data on one large market (\hyperlink{Galichon_Salanie}{Galichon and Salanie, 2021}; hereafter, GS). 
In  the absence of any restrictions on the taste shock distribution, we show that  the under-identification issue is severe, as the 1to1TU model is completely uninformative about the systematic match surplus. Formally, this means that, for every possible value of the systematic match surplus, there exists a taste shock distribution that rationalises the data when combined with that value of the systematic match surplus.

We proceed by investigating whether the 1to1TU model retains  some information  on the systematic match surplus under various classes of nonparametric distributional assumptions  on the unobserved heterogeneity (for instance, independence of taste shocks from covariates, quantile restrictions, symmetry restrictions, and identically distributed marginals). 
Answering this question poses the challenge of  tractably characterising the identified set of  the systematic match surplus. We do that by extending the linear programming computational approach of \hyperlink{Torgo}{Torgovitsky (2019)} to our framework. For a given value of the systematic match surplus, this method transforms a search over the space of infinite-dimensional cumulative distribution functions into a search over a space of cumulative distribution functions evaluated at a finite number of points. The latter search can be written as a simple linear program. Further, note that the analyst would need to solve the linear program for every admissible value of the systematic match surplus. Usually, this is carried out in the partial identification literature by generating a grid of points to approximate the parameter space and then repeating the exercise of interest for each grid point. However, the difficulty of doing so increases with the size of the grid, which, in the 1to1TU model, increases exponentially with the cardinality of the covariates' support,   leading quickly to a computational bottleneck. We alleviate this issue by showing that the parameter space
can be ex-ante partitioned into a  finite number of subsets such that, for each subset, every value belonging to that subset gives rise to the same linear program. Therefore, the analyst has to solve the linear program only once for each subset. These results are new and represent the methodological contribution of the paper.

We use our methodology to re-examine some of the relevant questions  in the empirical literature on the marriage market that have been previously answered by relying on the Logit  1to1TU model.  A key question that stands out in this literature is whether educational sorting  (i.e., the tendency of agents to marry  someone with similar or very different education levels) has changed over time. 
Answering this question is important because educational sorting may have a crucial impact on inequality by determining family formation and intergenerational transmission of human capital    (\hyperlink{Kremer}{Kremer, 1997};  \hyperlink{Fernandez2}{Fern\'andez and Rogerson, 2001}; \hyperlink{Fernandez1}{Fern\'andez, Guner, and Knowles, 2005}; \hyperlink{Heckman}{Heckman and Mosso, 2014}; \hyperlink{Dupuy_Weber}{Dupuy and Weber, 2019}; \hyperlink{Eika}{Eika, Mogstad, and Zafar, 2019}; \hyperlink{Chiappori_Meghir2}{Chiappori, et al., 2020}; \hyperlink{Ciscato_Weber}{Ciscato and Weber, 2020}). 
The literature proposes two approaches to measure changes in educational sorting. The first amounts to using indices of sorting  based on comparing the empirical match probabilities to a counterfactual world where matching happens randomly (\hyperlink{Fernandez2}{Fern\'andez and Rogerson, 2001}; \hyperlink{Greenwood1}{Greenwood, Guner, and Kocharkov, 2003}; \hyperlink{Liu}{Liu and Lu, 2006}; \hyperlink{Greenwood2}{Greenwood, et al., 2014}; \hyperlink{Abbott}{Abbott, et al., 2019}; \hyperlink{Eika}{Eika, Mogstad, and Zafar, 2019}; \hyperlink{Shen}{Shen, 2019}). The second consists of using a structural model of the marriage market  in order to estimate individual preferences and analyse how they evolve over time. The second approach has been implemented by \hyperlink{Siow}{Siow (2015)} and \hyperlink{CSW}{Chiappori, Salani\'e, and Weiss (2017)} (hereafter, CSW), based on the Logit  1to1TU model.

Both approaches suggest that, on average, positive educational sorting has increased in the U.S. in the past decades. However, there is some debate around this trend when we look closer at each education category. 
For instance,  \hyperlink{Eika}{Eika, Mogstad, and Zafar (2019)} find that positive educational sorting  has declined among the highly educated and increased among the less educated. Instead, \hyperlink{CSW}{CSW}  find that positive educational sorting has increased  particularly among the highly educated.\footnote{{See also \hyperlink{Chiappori_Meghir}{Chiappori, Costa-Dias, and Meghir (2020)} and   \hyperlink{Chiappori_Meghir3}{Chiappori, Costa-Dias, and Meghir (2021)} for similar conclusions.}}
Using   data from the American Community Survey between the years 1940 and 1967 as in \hyperlink{CSW}{CSW}, we  exploit our methodology to assess whether the conclusions  achieved via the structural approach are robust to the dropping of the Logit assumption.   {Under various classes of nonparametric distributional assumptions, we find that the 1to1TU
model is uninformative about the presence and trend of positive educational sorting among the highly educated. We   find  the presence of positive educational sorting among the less educated, although the model remains ambiguous about its evolution across cohorts. Overall, our results suggest that the previous findings on the increase in  positive educational sorting   based on the Logit  1to1TU model are,   in fact, driven by the Logit assumption.}


Lastly, we use our methodology to study the evolution of  marital returns to education. As discussed by \hyperlink{Chiappori5}{Chiappori, Iyigun, and Weiss (2009)} and \hyperlink{CSW}{CSW}, the increase in educational sorting makes a higher stock of human capital more valuable in the marriage market. Consequently, they predict an increase in the expected maximum payoff an agent can receive in the marriage market due to achieving a college degree (“marital college premium”), especially among women. Their empirical findings corroborate  such a prediction for the U.S., based on the Logit  1to1TU model.     {Without imposing parametric distributional assumptions, we find that the 1to1TU model is inconclusive about the evolution of  marital college premium over time. Further, it is   particularly uninformative about the women's side, indicating that any evidence on the increase in marital college premium  from  the Logit  1to1TU model is  a consequence of the arbitrary parametric restrictions.}

In what follows, Section \MYref{lit} summarises the related literature, Section \MYref{model} presents the model, Section \MYref{identification} discusses identification, Section \MYref{empirical} illustrates the empirical applications, and Section \MYref{conclusions} concludes.

\section{Literature review}
\label{lit}

The Logit  1to1TU model was introduced by \hyperlink{Choo_Siow}{Choo and Siow (2006)} and since then has become popular in  empirical research on the marriage market. Several papers use it to learn whether matching preferences are positive assortative by age, education, geographical location, etc. (\hyperlink{Choo_Siow}{Choo and Siow, 2006}; \hyperlink{Botticini}{Botticini and Siow, 2011}; \hyperlink{Bruze}{Bruze, Svarer, and Weiss, 2015}; \hyperlink{Choo}{Choo, 2015};\footnote{\hyperlink{Bruze}{Bruze, Svarer, and Weiss (2015)} and \hyperlink{Choo}{Choo (2015)} incorporate dynamic aspects into the framework of \hyperlink{Choo_Siow}{Choo and Siow (2006)}.} \hyperlink{Siow}{Siow, 2015}; \hyperlink{Galichon_Kominers_Weber}{Galichon, Kominers, and Weber, 2019}\footnote{\hyperlink{Galichon_Kominers_Weber}{Galichon, Kominers, and Weber (2019)} extend the framework of \hyperlink{Choo_Siow}{Choo and Siow (2006)} to imperfectly transferable utilities.}). Other papers use it to assess which of the partner characteristics are complements/substitutes in the production of the systematic match surplus and their relative strengths (\hyperlink{Dupuy_Galichon}{Dupuy and Galichon, 2014};\footnote{\hyperlink{Dupuy_Galichon}{Dupuy and Galichon (2014)} extend the framework of \hyperlink{Choo_Siow}{Choo and Siow (2006)} to continuous covariates.} \hyperlink{Ciscato}{Ciscato, Galichon, Gouss\'e, 2020}).  
The Logit  1to1TU model has been frequently used to investigate the evolution  of the link between education levels and marriage market outcomes over time. In particular, the   literature has studied questions like how educational sorting and the marital college premium have changed over time (\hyperlink{Chiappori5}{Chiappori, Iyigun, and Weiss, 2009}; \hyperlink{Siow}{Siow, 2015}; \hyperlink{CSW}{CSW}; \hyperlink{Chiappori_Meghir}{Chiappori, Costa-Dias, and Meghir, 2020}; \hyperlink{Chiappori_Meghir2}{Chiappori, et al., 2020}).
Other papers adopt the Logit  1to1TU model to measure the effect on marital choices of exogenous events that change the distribution of individual characteristics on each side of the market, such as  the famine in China  between 1958 and 1961 (\hyperlink{Brandt}{Brandt, Siow, and Carl, 2016}).

 The Logit  1to1TU model is often incorporated into bigger structural models. Examples of these include collective household models with marriage and labour supply (\hyperlink{Choo2}{Choo and Seitz, 2013}); life cycle models of education, marriage, labour supply, and consumption (\hyperlink{Chiappori6}{Chiappori, Costa-Dias, and Meghir, 2018}); collective household models with marriage, labour supply, home production choices, and  joint taxation (\hyperlink{Gayle}{Gayle and Shephard, 2019}\footnote{\hyperlink{Gayle}{Gayle and Shephard (2019)} allow for imperfectly transferable utilities.});  collective household models with marriage, fertility decisions, and child socialisation choices  (\hyperlink{Bisin}{Bisin and Tura, 2020}). \hyperlink{Mourifie}{Mourifi\'e and Siow (2021)} extend the Logit  1to1TU model to allow for peer effects and cohabitation.

 \hyperlink{Galichon_Salanie}{GS}   investigates the identification of the 1to1TU model when one dispenses with the Logit  assumption.  Under the assumption that the taste shock distribution is {fully} known by the analyst, they show that the 1to1TU model is {just} identified (and, thus, the systematic match surplus is {point} identified)  with data on {one large} market. They also provide closed-form expressions of the systematic match surplus for some parametric distributional families. 

The  literature has explored two ways to introduce unknown parameters in the taste shock distribution while maintaining point identification. The first approach exploits variations in matching patterns across {many i.i.d. markets}  (\hyperlink{Fox_2010}{Fox, 2010}; \hyperlink{Fox_2018}{Fox, 2018}; \hyperlink{Fox_Yang_Hsu}{Fox, Yang, and Hsu, 2018}; \hyperlink{Sinha}{Sinha, 2018}), as in the empirical IO tradition. With data on many i.i.d. markets, one can proceed without  parametric restrictions on the taste shock distribution. However,  in most datasets, it is unclear as to what truly defines i.i.d. markets.  
For instance, the majority of the  empirical applications of this approach assume that consecutive years represent i.i.d. markets, which can be often hard to justify (\hyperlink{Baccara}{Baccara, et al., 2012}; \hyperlink{Mindruda}{Mindruda, 2013}; \hyperlink{Akkus}{Akkus, Cookson, and Horta\c{c}su, 2016}; \hyperlink{Mindruda2}{Mindruda, Moeen, and Agarwal, 2016};  \hyperlink{Chen}{Chen, 2017};  \hyperlink{Banal}{Banal-Esta\~{n}ol, Macho-Stadler, and P\'{e}rez-Castrillo, 2018}). The second approach exploits variations of matching patterns across  {a few large cohorts} which feature different distributions of  covariates, independent matching processes, identical systematic match surplus up to some drifts or linear/quadratic trends, and identical taste shock distributions. This approach is implemented by  \hyperlink{CSW}{CSW} to introduce   gender heteroskedasticity in the Extreme Value Type I distribution.

Recent advances in the partial identification literature have pointed out an alternative route to avoid parametric assumptions on the taste shock distribution,  without  adding any further structure on the systematic match surplus, and while remaining within a {one large} market framework. In particular, \hyperlink{Graham_2011}{Graham (2011}; \hyperlink{Graham_errata}{2013b)} shows that if the taste shocks are i.i.d., then the signs of some complementarities between the spouses' observed characteristics  are identified.  \hyperlink{Fox_2018}{Fox (2018)} bounds the systematic match surplus under the assumption that the taste shocks are exchangeable across the observed characteristics of the potential partners. Our paper contributes to this strand of the literature by constructing the identified set of  the systematic match surplus without requiring the taste shocks to be i.i.d. or exchangeable, which can both be strong assumptions.   {Further, our paper showcases the usefulness of partial identification approaches for formally understanding empirical results that might otherwise be accepted less critically.}


\section{The model}
\label{model}
This section describes the 1to1TU model that has been previously studied in  \hyperlink{Choo_Siow}{Choo and Siow (2006)} and \hyperlink{Galichon_Salanie}{GS}. The model relies on four main assumptions that are standard in the current  empirical practice.  In what follows, we refer to agents on one side of the market as men and to agents on the other side of the market as women.
\begin{assumption}{\normalfont({\itshape  One large market})}
\label{large}
There is a two-sided market. One side  is populated by an (uncountably) infinite set of men, $\mathcal{I}$, with measure $d\tilde{\mu}_{\mathcal{I}}$. The other side is populated by an (uncountably) infinite set of women, $\mathcal{J}$, with measure $d\tilde{\mu}_{\mathcal{J}}$. The two sides   are stochastically independent.
\end{assumption}

\begin{assumption}{\normalfont({\itshape  Finite number of observed types})}
\label{types}
Each man $i\in \mathcal{I}$ is characterised by a type, $X_i$, with finite support, $\mathcal{X}$. The mass of men of type $x\in \mathcal{X}$ is denoted by $m_x$. Each woman $j\in \mathcal{J}$ is characterised by a type, $Y_j$, with finite support, $\mathcal{Y}$. The mass of women of type $y\in \mathcal{Y}$ is denoted by $w_y$. Without loss of generality, we normalise the total mass of agents to $1$, i.e., $\sum_{x\in \mathcal{X}}m_x+\sum_{y\in \mathcal{Y}}w_y=1$. The realisations of $X_i$ and $Y_j$ are observed by the researcher and all agents. 
We define the sets of partner types that are available to men and women by $\mathcal{Y}_0\equiv \mathcal{Y}\cup \{0\}$ and $\mathcal{X}_0\equiv \mathcal{X}\cup \{0\}$, respectively, where ``$0$'' represents the option not to match.
\end{assumption}

\begin{assumption}{\normalfont({\itshape  Taste shocks})}
\label{shocks}
Each man $i\in \mathcal{I}$ is endowed with a $|\mathcal{Y}_0|\times 1$ vector of  taste shocks, $\epsilon_i\equiv (\epsilon_{iy}: y\in \mathcal{Y}_0)$, where
$|\mathcal{Y}_0|$ denotes the cardinality of $\mathcal{Y}_0$ and $\epsilon_{iy}$ is the idiosyncratic preference of man $i$ for marrying a woman of type $y\in \mathcal{Y}_0$. Conditional on $X_i=x$ and for each $x\in \mathcal{X}$, $\epsilon_i$ has cumulative distribution function (hereafter, CDF) $F_x$. $F_x$ is absolutely continuous with respect to the Lebesgue measure and has support  in $\mathbb{R}^{|\mathcal{Y}_0|}$. Each woman $j\in \mathcal{J}$ is endowed with a $|\mathcal{X}_0|\times 1$ vector of  taste shocks, $\eta_j\equiv (\eta_{xj} : x\in \mathcal{X}_0)$, where
$\eta_{xj}$ is the idiosyncratic preference of woman $j$ for marrying a man of type $x\in \mathcal{X}_0$. Conditional on $Y_j=y$ and for each $y\in \mathcal{Y}$, $\eta_j$ has CDF $G_y$.  $G_y$ is absolutely continuous with respect to the Lebesgue measure and has support  in $\mathbb{R}^{|\mathcal{X}_0|}$. The realisations of $\epsilon_i$ and $\eta_j$ are observed by all agents but are not observed by the researcher. 
\end{assumption}

\begin{assumption}{\normalfont({\itshape  Separability})}
\label{sep}
A match between man $i\in \mathcal{I}$ of type $x\in \mathcal{X}$ and woman $j\in \mathcal{J}$ of type $y\in \mathcal{Y}$ generates a match surplus defined as 
$$
 \tilde{\Phi}_{ij}\equiv \Phi_{xy}+\epsilon_{iy} +\eta_{xj},
$$
where $\Phi\equiv (\Phi_{xy} : (x,y)\in \mathcal{X}\times \mathcal{Y})$ is the systematic match surplus. The payoff of man $i\in \mathcal{I}$  from remaining unmatched is 
$$
\tilde{\Phi}_{i0}\equiv \epsilon_{i0}.
$$
The payoff of woman $j\in \mathcal{J}$  from remaining unmatched is 
$$
\tilde{\Phi}_{0j}\equiv \eta_{0j}.
$$
\end{assumption}
 Assumption \MYref{large} outlines the {one} {large} {market} framework. 
The restriction on the stochastic independence of the two sides of the market is not crucial for our results and can be relaxed. 
Assumption \MYref{types} requires each agent to belong to one type. There is a {finite} number of types, which is defined by the Cartesian product of  the individual characteristics  observed by the researcher. 
  Assumption \MYref{shocks} requires each agent to have idiosyncratic marital preferences over the types of the potential partners and not over their identities. It implies that women (men) of the same type are perfect substitutes for a man (woman).
Assumption \MYref{sep} imposes that the match surplus is the sum of two components. One is the systematic match surplus, that is determined by the types of potential partners. The other is the sum of the taste shocks of the potential partners.  In particular, the latent heterogeneity entering the match surplus equation does not consist of an $ij$-indexed term. Instead, it is modelled through the sum of two terms, $\epsilon_{iy}+\eta_{xj}$, each of which only depends on the type of the potential partner. Assumption \MYref{sep} is typically referred to as ``separability'' (\hyperlink{Chiappori3}{Chiappori, 2017}). Lastly, observe that the systematic match surplus from remaining single is normalised to zero. This location normalisation is standard in the literature and   imposed, for instance, also by \hyperlink{Galichon_Salanie}{GS} and \hyperlink{CSW}{CSW}.

A matching consists of \begin{enumerate}[(i)]
\item A measure $d\tilde{\mu}$ on the set $\mathcal{I}\times \mathcal{J}$, such that the marginal of $d\tilde{\mu}$ over $\mathcal{I}$ ($\mathcal{J}$) is $d\tilde{\mu}_\mathcal{I}$ ($d\tilde{\mu}_\mathcal{J}$). 
\item A set of payoffs, $\{\tilde{U}_i\}_{i\in \mathcal{I}}$ and  $\{\tilde{V}_j\}_{j\in \mathcal{J}}$, such that
$$
\tilde{U}_i+\tilde{V}_j=\tilde{\Phi}_{ij} \text{ }\quad\forall (i,j)\in \text{supp}(d\tilde{\mu}).
$$
\end{enumerate}
(\hyperlink{Nesheim}{Chiappori, McCann, and Nesheim, 2010}; \hyperlink{Chiappori_McCann_Pass}{Chiappori, McCann, and Pass, 2020}). 
That is, a matching consists of a match assignment and a match surplus sharing rule. A match assignment is a description of who is matched with whom.  A match surplus sharing rule tells us how the match surplus is divided between spouses. This division of the match surplus relies on endogenously determined transfers, ensuring that every agent maximises their utility and the market clears. 

A matching, $d\tilde{\mu}, \{\tilde{U}_i\}_{i\in \mathcal{I}}, \{\tilde{V}_j\}_{j\in \mathcal{J}}$,  is stable when no agent has an incentive to change his partner, i.e.,
$$
\begin{aligned}
& \tilde{U}_i\geq \tilde{\Phi}_{i0} \text{ }\forall i \in \mathcal{I},\\
& \tilde{V}_j\geq \tilde{\Phi}_{0j} \text{ }\forall j \in \mathcal{J},\\
& \tilde{U}_i+\tilde{V}_j \geq \tilde{\Phi}_{ij}  \text{ }\forall (i,j)\in \mathcal{I}\times \mathcal{J}.
\end{aligned}
$$
The first two sets of inequalities imply that married agents would not prefer being single. The last set of inequalities states that no man or woman would get a strictly higher match surplus by matching together than what they get under the match assignment, $d\tilde{\mu}$. It can be shown that a stable matching exists under mild continuity assumptions (\hyperlink{Villani}{Villani, 2009}).  Moreover, in the limit of continuous and atomless populations, the stable matching is generically unique (\hyperlink{Gretsky}{Gretsky, Ostroy and Zame, 1992}). Importantly for the identification analysis, the resulting equilibrium mass of couples where the man is of type $x$ and the woman is of type $y$ is unique for every $(x,y)$. In what follows, we denote this  equilibrium mass of couples by
$$
\mu_{xy} \text{ for each } (x,y)\in \mathcal{Z}\equiv \mathcal{X}_0\times \mathcal{Y}_0\setminus\{0,0\}.
$$

\section{Identification}
\label{identification}

\subsection{Data and parameters of interest}
\label{policy}
For identification, we assume that the analyst knows $\{\mu_{xy}\}_{(x,y)\in \mathcal{Z}}$, $\{m_x\}_{x \in \mathcal{X}}$, and $\{w_y\}_{y\in \mathcal{Y}}$. For estimation, we replace $\{\mu_{xy}\}_{(x,y)\in \mathcal{Z}}$, $\{m_x\}_{x \in \mathcal{X}}$, and $\{w_y\}_{y\in \mathcal{Y}}$ with consistent sample analogues, resulting from sampling at random from the market at the individual
level or at the household level. 
Define $$p_{y|x}\equiv \frac{\mu_{xy}}{m_x}\quad\text{and}\quad p_{x|y}\equiv \frac{\mu_{xy}}{w_y},$$ as the equilibrium probability of marrying a woman of type $y\in \mathcal{Y}_0$ conditional on being a man of type $x\in \mathcal{X}$, and the equilibrium probability of marrying a man of type $x\in \mathcal{X}_0$ conditional on being a woman of type $y\in \mathcal{Y}$, respectively. Lastly, let $$p_{x}\equiv \frac{m_x}{\sum_{x\in \mathcal{X}}m_x}\quad\text{and}\quad p_{y}\equiv \frac{w_y}{\sum_{y\in \mathcal{Y}}w_y},$$ 
be the proportion of men of type $x\in \mathcal{X}$ and the proportion of women of type $y\in \mathcal{Y}$, respectively.

Our primary interest lies in recovering the systematic match surplus, $\Phi$. In fact, (partially) identifying $\Phi$ allows us to answer many important questions considered in the marriage market literature.
In particular, $\Phi$ can be used to learn whether agents tend to match with similar people, i.e., whether there is positive assortativeness. Investigating positive assortativeness and its evolution over time has been a focus of empirical research since \hyperlink{Becker}{Becker (1973)} because it is crucial to understand the sources of  inequality in intergenerational outcomes. For instance, if    parents' education level affects their children's school attainment and marriage is positive assortative by education, then inequality in the next generation may be higher. Formally, let $\mathcal{X}=\mathcal{Y}\equiv \{1,...,r\}$ collect $r$ education levels, ordered from lowest to highest.
For any $x,\tilde{x}\in \mathcal{X}$ with $x>  \tilde{x}$, consider the cross-difference operator,
\begin{equation}
\label{core_1}
D_{xx,\tilde{x}\tilde{x}}(\Phi)\equiv \Phi_{xx}+\Phi_{\tilde{x}\tilde{x}}-\Phi_{x\tilde{x}}-\Phi_{\tilde{x}x}. 
\end{equation}
$D_{xx,\tilde{x}\tilde{x}}(\Phi)$ measures how the incremental (dis)value of marrying a more educated man evolves as the education of the woman increases. Hence, one can evaluate changes in positive educational sorting across markets (for instance, across cohorts) by estimating the supermodular core of $\Phi$, 
$$
D
(\Phi)\equiv (D_{xx,\tilde{x}\tilde{x}}(\Phi) : x,\tilde{x}\in \mathcal{X}, x>\tilde{x}),
$$
within each market. As per Definition 1 in \hyperlink{CSW}{CSW}, if every component of the vector $D(\Phi)$ is positive within a given market, then that market exhibits positive educational sorting. Further, if every component of $D(\Phi)$ increases across markets, then  positive educational sorting increases across markets. The definition can be restricted to a subset of education categories. We can say that there is positive educational sorting among   more educated people if each component of the vector
$$
D^{\ell}
(\Phi)\equiv (D_{xx,\tilde{x}\tilde{x}}(\Phi) : x,\tilde{x}\in \{\ell,\ell+1,..., r\}, x>\tilde{x}),
$$
is positive, where $\ell$ is large enough.  Similarly,  there is positive educational sorting among less educated people if each component of the vector
$$
D^{\ell}
(\Phi)\equiv (D_{xx,\tilde{x}\tilde{x}}(\Phi) : x,\tilde{x}\in \{ 1,...,\ell\}, x>\tilde{x}),
$$
is positive, where $\ell$ is  low enough.  We can also say that positive educational sorting increases across markets among   more (less) educated people if every component of $D^{\ell}(\Phi)$ increases across markets.\footnote{In settings with multidimensional covariates, the cross-difference operator defined in (\MYref{core_1}) can   be used to learn which of the spouses' observed characteristics are complements or substitutes in the production of the systematic match surplus and, in turn,  discover the key drivers of the gains to matching. See, for instance, \hyperlink{Fox_2010}{Fox (2010)} and \hyperlink{Graham_2011}{Graham (2011}; \hyperlink{Graham_errata}{2013b)}.}

In addition to recovering $\Phi$, our methodology permits the analyst to (partially) identify how $\Phi$ is shared between spouses and hence assess the impact of pre-marital decisions on marriage market outcomes. In particular, let $U_{xy}$ be the part of $\Phi_{xy}$ that is gained by a man of type $x\in \mathcal{X}$ when matching with a woman of type $y\in \mathcal{Y}$. Analogously, let $V_{xy}$ be the part of $\Phi_{xy}$ that is gained by a woman of type $y\in \mathcal{Y}$ when matching with a man of type $x\in \mathcal{X}$.  
Define $\bar{U}_x$ as  the expected payoff that a man  of type $x\in \mathcal{X}$ gets when marrying,
$$
\bar{U}_x\equiv E_{F_x}(\max_{y\in \mathcal{Y}_0} U_{xy}+\epsilon_{iy}|X_i=x).
$$
For any $\tilde{x}\in \mathcal{X}$ with $\tilde{x}<x$, the difference $\Delta_{x\tilde{x}}(U)\equiv \bar{U}_x-\bar{U}_{\tilde{x}}$ denotes the gain in expected utility   from reaching education level $x$ instead of $\tilde{x}$. Therefore, it represents the marital {\it education} premium. When $x=r,$ a college degree, and $\tilde{x}=r-1$, such a difference is called the marital {\it college} premium. These quantities have received particular attention because they capture the value of human capital on the marriage market (\hyperlink{Chiappori5}{Chiappori, Iyigun, and Weiss, 2009}; \hyperlink{CSW}{CSW}). As shown in  \hyperlink{Galichon_Salanie}{GS}, the marital education premium is equal to 
\begin{equation}
\label{marital_our}
\Delta_{x\tilde{x}}(U)=\sum_{y\in \mathcal{Y}_0} p_{y|x}U_{xy}-\sum_{y\in \mathcal{Y}_0} p_{y|\tilde{x}}U_{\tilde{x}y}+E_{F_x}(\epsilon_{iy^*_i}|X_i=x)-E_{F_{\tilde{x}}}(\epsilon_{i\tilde{y}^*_i}|X_i=\tilde{x}),
\end{equation}
where $y^*_i\in \mathcal{Y}_0$ is the optimal choice of man $i$ of type $x$ and $\tilde{y}^*_i\in \mathcal{Y}_0$ is the optimal choice of man $i$ of type $\tilde{x}$.
Note from (\MYref{marital_our}) that  computing (bounds for) $\Delta_{x\tilde{x}}(U)$ requires the specification of (finite bounds for) $E_{F_x}(\epsilon_{iy^*_i}|X_i=x)-E_{F_{\tilde{x}}}(\epsilon_{i\tilde{y}^*_i}|X_i=\tilde{x})$. The latter are typically not obtained within a  nonparametric framework like ours without further assumptions. Nevertheless, we   will provide bounds for the difference 
$$
C_{x\tilde{x}} (U)\equiv \sum_{y\in \mathcal{Y}_0} p_{y|x}U_{xy}-\sum_{y\in \mathcal{Y}_0} p_{y|\tilde{x}}U_{\tilde{x}y},
$$
which represents the change in the weighted average systematic payoff due to reaching education level $x$ instead of $\tilde{x}$.  Such bounds will help us make certain conclusions regarding $\Delta_{x\tilde{x}}(U)$. In fact, we will see that in all the empirical cases of interest, the estimates of 
$$
C(U)\equiv (C_{x\tilde{x}}(U) : x,\tilde{x}\in \mathcal{X}, x>\tilde{x}) \quad\text{and}\quad
C(V)\equiv (C_{y\tilde{y}}(V) : y,\tilde{y}\in \mathcal{Y}, y>\tilde{y}),
$$
are   unbounded above and below. By (\MYref{marital_our}), this implies that the marital education premia are unbounded as well, and therefore, their evolution over time cannot be inferred.\footnote{Note that imposing $E_{F_x}( \epsilon_{iy}|X_i=x)<+\infty$ and $E_{F_{\tilde{x}}}( \epsilon_{iy}|X_i=\tilde{x})<+\infty$ for each $y\in \mathcal{Y}_0$ is necessary to ensure finiteness of $E_{F_x}(\epsilon_{iy^*_i}|X_i=x)$ and $E_{F_{\tilde{x}}}(\epsilon_{i\tilde{y}^*_i}|X_i=\tilde{x})$ in (\MYref{marital_our}).  In our framework, we do not consider such a class of nonparametric assumptions on $\{F_x\}_{x\in \mathcal{X}}$. This is without loss of generality. In fact, as suggested by Example 2 in \hyperlink{Torgo}{Torgovitsky (2019)},  requiring $E_{F_x}( \epsilon_{iy}|X_i=x)$ to be equal to some finite number for each $(x,y)\in \mathcal{X}\times \mathcal{Y}_0$  does not place any restrictions on the set of underlying distributions that determines the identified set for $U$, thus leaving the bounds for $C(U)$ unchanged.  }

\subsection{Two multinomial choice models}
\label{multinomial}
Based on the  separability restriction (Assumption \MYref{sep}), Proposition 1 of \hyperlink{Galichon_Salanie}{GS} provides a key result for our identification analysis.\footnote{This result also appears in  previous working paper versions of \hyperlink{Galichon_Salanie}{GS}, in \hyperlink{Chiappori_UV}{Chiappori, et al. (2008)}, and in \hyperlink{CSW}{CSW}.} This result states that, given the primitives $\Phi$, $\{p_x\}_{x\in \mathcal{X}}$, $\{p_y\}_{y\in \mathcal{Y}}$, $\{F_x\}_{x\in \mathcal{X}}$, $\{G_y\}_{y\in \mathcal{Y}}$ generating the stable matching $d\tilde{\mu}, \{\tilde{U}_i\}_{i\in \mathcal{I}}, \{\tilde{V}_j\}_{j\in \mathcal{J}}$,  there exist vectors $$U\equiv (U_{xy} : (x,y)\in \mathcal{X}\times \mathcal{Y}_0)\in \mathbb{R}^{|\mathcal{X}\times \mathcal{Y}_0|}\text{ and }{V} \equiv (V_{xy} : (x,y)\in \mathcal{X}_0\times \mathcal{Y})\in \mathbb{R}^{|\mathcal{X}_0\times \mathcal{Y}|},$$ such that 
\begin{align}
&\tilde{U}_i=\max_{y\in \mathcal{Y}_0} (U_{xy}+\epsilon_{iy}) & \text{ $\forall i\in \mathcal{I}$ of type $x\in \mathcal{X}$, $\forall x \in \mathcal{X}$,}\label{multinomial1}\\
&\tilde{V}_j=\max_{x\in \mathcal{X}_0} (V_{xy}+\eta_{xj}) & \text{ $\forall j\in \mathcal{J}$ of type $y\in \mathcal{Y}$, $\forall y \in \mathcal{Y}$,}\label{multinomial2}\\
&U_{xy}+V_{xy}=\Phi_{xy}\text{, }U_{x0} = 0\text{, }V_{0y}= 0 & \text{ $\forall (x,y) \in \mathcal{X}\times \mathcal{Y}$.} \label{sum}
\end{align}

Proposition 1 of \hyperlink{Galichon_Salanie}{GS} allows us to rewrite the framework of Section \MYref{model} as two separate one-sided multinomial choice models linked by market-clearing transfers that are implicitly embedded into the vectors $U$ and $V$. This alternative representation of the problem is useful as it immediately suggests a way to investigate the identification of $\Phi$: the researcher can study separate identification of $U$ and $V$ from (\MYref{multinomial1}) and  (\MYref{multinomial2}) using various restrictions on the unobserved heterogeneity, and then obtain identification results for $\Phi$ through (\MYref{sum}).\footnote{The restrictions $U_{x0}=V_{0y}=0$ in (\MYref{sum}) come from the fact that  the systematic match surplus from remaining single is normalised to zero  in Assumption  \MYref{sep}.}

Based on Proposition 1, Proposition 2 of \hyperlink{Galichon_Salanie}{GS}   shows that if $\{F_x\}_{x\in \mathcal{X}}$ and $\{G_y\}_{y\in \mathcal{Y}}$ are {fully} known by the analyst, then the 1to1TU model is {just} identified and, thus, $\Phi$ is {point} identified. Note that fully knowing $\{F_x\}_{x\in \mathcal{X}}$ and $\{G_y\}_{y\in \mathcal{Y}}$ requires either such conditional CDFs to be parameter-free, or the analyst to fix the value of each parameter governing them. In particular, a widespread practice in the empirical literature amounts to assuming that the  taste shocks are  i.i.d. standard Extreme Value Type I, independently distributed from  types, so that  $\Phi$ can be recovered via standard Logit   arguments applied to each side of the market  (\hyperlink{Choo_Siow}{Choo and Siow, 2006}).

Unfortunately, the Logit  1to1TU model suffers from the same limitations of the  one-sided Logit    framework. It exhibits IIA which has counterintuitive predictions by implying proportional substitution across types. This is illustrated by \hyperlink{Galichon_Salanie2}{Galichon and Salani\'e (2019)} with an example that resembles the blue-bus/red-bus example of \hyperlink{McFadden}{McFadden (1974)}.\footnote{See also \hyperlink{Graham}{Graham (2013a)} for another example on violation of IIA.}
While most of the applied literature on one-sided markets has replaced the Logit  assumption with the  Generalised Extreme Value (GEV) framework, such a transition is yet to occur in the two-sided literature. This is because, due to the just identification result mentioned above, it would be  necessary to arbitrarily specify the value of {each} parameter governing the GEV distribution, at the risk of incurring serious misspecification. There are two possible ways to solve this impasse with data on one large market: the researcher could either place assumptions on $\Phi$ (for instance, by restricting complementarities/substitutabilities among observed characteristics) so as to get degrees of freedom for estimating the GEV parameters, or adopt a {partial} identification perspective. The first approach is rare in the marriage market literature because most of the empirical studies consider marital sorting on a single dimension (one attribute at a time). In this paper, we  explore the second approach  and  construct {bounds} for $\Phi$. In doing so, we do not restrict the distribution of the taste shocks to belong to any specific parametric family.

\subsection{The extent of under-identification}
\label{under}

As shown by  \hyperlink{Galichon_Salanie}{GS}, if  $\{F_x\}_{x\in \mathcal{X}}$ and $\{G_y\}_{y\in \mathcal{Y}}$ are not assumed to be fully known by the analyst, then  the 1to1TU model is under-identified. As a first step, this section investigates the extent
of under-identification by answering the following question: does the 1to1TU model retain some identifying power on $\Phi$ without imposing any restrictions on $\{F_x\}_{x\in \mathcal{X}}$ and $\{G_y\}_{y\in \mathcal{Y}}$? 
Lemma \MYref{result0} below gives a negative answer. 

Let  $\mathcal{U}$, $\mathcal{V}$, and $\Theta$ denote the parameter spaces of $U$, $V$, and $\Phi$, respectively, i.e., 
$$
\begin{aligned}
&\mathcal{U}\equiv \{ U\in \mathbb{R}^{|\mathcal{X}\times \mathcal{Y}_0|}: U_{x0}=0 \text{ }\forall x \in \mathcal{X}\},\\
&\mathcal{V}\equiv \{ V\in \mathbb{R}^{|\mathcal{X}_0\times \mathcal{Y}|}: V_{0y}=0 \text{ }\forall y \in \mathcal{Y}\},\\
&\Theta \equiv \mathbb{R}^{|\mathcal{X}\times \mathcal{Y}|}.\\
\end{aligned}
$$
Further, let $\mathcal{F}$ and $\mathcal{G}$ be the function spaces of all admissible  taste shock distributions, $\{F_x\}_{x\in \mathcal{X}}$ and $\{G_y\}_{y\in \mathcal{Y}}$, respectively.\footnote{Note that one element of $\mathcal{F}$ is a family of $|\mathcal{X}|$ conditional CDFs, $\{F_x\}_{x\in \mathcal{X}}$. Similarly, one element of $\mathcal{G}$ is a family of $|\mathcal{Y}|$ conditional CDFs, $\{G_y\}_{y\in \mathcal{Y}}$.} Lastly, for any $y\in \mathcal{Y}_0$, $U\in \mathcal{U}$, and  $\{F_x\}_{x\in \mathcal{X}}\in \mathcal{F}$, let $\kappa(U,F_x,y)$ be the model-implied probability of marrying a woman of type $y\in \mathcal{Y}_0$ conditional on being a man of type $x\in \mathcal{X}$, i.e., 
$$
\kappa(U,F_x,y)\equiv \lambda_{F_x}(\{(e_y : y\in \mathcal{Y}_0)\in \mathbb{R}^{|\mathcal{Y}_0|}: U_{xy}+e_y \geq U_{x\tilde{y}}+e_{\tilde{y}} \text{ }\forall \tilde{y}\in \mathcal{Y}_0\setminus\{y\}\}), 
$$
where $\lambda_{F_x}$ is the probability measure associated with $F_x$. 
 Similarly, for any  $x\in \mathcal{X}_0$, $V\in \mathcal{V}$, and  $\{G_y\}_{y\in \mathcal{Y}}\in \mathcal{G}$, let $\kappa(V,G_y,x)$ be the model-implied probability of marrying a man of type $x\in \mathcal{X}_0$ conditional on being a woman of type $y\in \mathcal{Y}$, i.e., 
$$
\kappa(V,G_y,x)\equiv \lambda_{G_y}(\{ (e_x : x\in \mathcal{X}_0)\in \mathbb{R}^{|\mathcal{X}_0|}: V_{xy}+e_x \geq V_{\tilde{x}y}+e_{\tilde{x}} \text{ }\forall \tilde{x}\in \mathcal{X}_0\setminus\{x\}\}), 
$$
where $\lambda_{G_y}$ is the probability measure associated with $G_y$.

\begin{lm}{\normalfont({\itshape Under-identification})}
\label{result0}
{\normalfont 
For every data, $\{p_{y|x}\}_{(x,y)\in \mathcal{X}\times \mathcal{Y}_0}$ and $\{p_{x|y}\}_{(x,y)\in \mathcal{X}_0\times \mathcal{Y}}$, and   systematic match surplus, $\Phi\in \Theta$, there exist  $(U,V)\in \mathcal{U}\times \mathcal{V}$, $\{F_x\}_{x\in \mathcal{X}}\in \mathcal{F}$, and $\{G_y\}_{y\in \mathcal{Y}}\in \mathcal{G}$ such that
\begin{align}
&p_{y|x}=\kappa(U,F_x,y) && \forall (x,y) \in \mathcal{X}\times \mathcal{Y}_0, \label{man_id}\\
& p_{x|y}=\kappa(V,G_y,x)&& \forall (x,y) \in \mathcal{X}_0\times \mathcal{Y},  \label{woman_id}\\
& U_{xy}+V_{xy}=\Phi_{xy} &&  \forall (x,y) \in \mathcal{X}\times \mathcal{Y}. \label{clearing}
\end{align}
}
\end{lm}
Lemma \MYref{result0} is a straightforward application of Theorem 1 of \hyperlink{HHK}{Haile, Horta\c{c}su, and Kosenok (2008)} to our two-sided setting. It shows that in the absence of restrictions on  $\{F_x\}_{x\in \mathcal{X}}$ and $\{G_y\}_{y\in \mathcal{Y}}$, the 1to1TU model    leads to uninformative bounds on $\Phi$. Therefore, one needs to impose at least some distributional assumptions on the unobserved heterogeneity to get  information on $\Phi$.

\subsection{Adding distributional assumptions on unobserved heterogeneity}
\label{general_steps}

In this section, we ask ourselves whether the 1to1TU model retains some identifying power on $\Phi$ under various classes of nonparametric distributional assumptions on the unobserved heterogeneity, so as to still be able to address relevant policy matters while maintaining a certain  degree of robustness. To answer this question, we adopt a computational approach. In particular, we start from observing that  if $\{F_x\}_{x\in \mathcal{X}}$ and $\{G_y\}_{y\in \mathcal{Y}}$ are not assumed to be fully known by the analyst, then the 1to1TU model is under-identified, leading to partial identification of $\Phi$. Hence, we  provide a methodology to   construct the identified set of  $\Phi$ under various classes of nonparametric distributional assumptions on the unobserved heterogeneity.

The identified set of  $\Phi$ (hereafter, $\Theta^*$) is the set of values of $\Phi$ such that there exists  $U$, $V$, $\{F_x\}_{x\in \mathcal{X}}$, and $\{G_y\}_{y\in \mathcal{Y}}$ that satisfy (\MYref{man_id})-(\MYref{clearing}). By Proposition 1 of \hyperlink{Galichon_Salanie}{GS}, we can construct $\Theta^*$ by separately focusing on each side of the market. First, we construct the identified set of  $U$ (hereafter,   $\mathcal{U}^*$), i.e.,  the set of values of $U$ such that there exists  $\{F_x\}_{x\in \mathcal{X}}$ that satisfies (\MYref{man_id}). Then, we construct the identified set of  $V$ (hereafter,  $\mathcal{V}^*$), i.e., the  set of values of $V$ such that there exists  $\{G_y\}_{y\in \mathcal{Y}}$ that satisfies (\MYref{woman_id}). Finally, we  obtain $\Theta^*$ from  (\MYref{clearing}). In what follows, we explain the construction of $\mathcal{U}^*$. The construction of $\mathcal{V}^*$ is analogous. 

Recall that  in multinomial choice models what matters is differences in utilities. Therefore, as a preliminary step, we rewrite the identification problem using the taste shock differences. Without loss of generality, we label the women's types as $\mathcal{Y}\equiv \{1,...,r\}$. Let $\Delta{\epsilon_i}$ be the vector of differences between every pair of taste shocks of man $i\in \mathcal{I}$,
\begin{equation}
\label{delta_epsilon}
\Delta{\epsilon_i}\equiv (\epsilon_{i1}-\epsilon_{i0},..., \epsilon_{ir}-\epsilon_{i0}, \epsilon_{i1}-\epsilon_{i2}, ..., \epsilon_{i1}-\epsilon_{ir}, \epsilon_{i2}-\epsilon_{i3}, ..., \epsilon_{i2}-\epsilon_{ir}, ..., \epsilon_{ir-1}-\epsilon_{ir}),
\end{equation}
 with length $d\equiv \binom{r+1}{2}$. 

Observe that each $\{F_x\}_{x\in \mathcal{X}}\in \mathcal{F}$ determines a corresponding family of $d$-dimensional conditional CDFs of $\Delta \epsilon_i$, which we denote by $\{\Delta F_x\}_{x\in \mathcal{X}}$.  Further, note that the first $r$ components of $\Delta{\epsilon_i}$ can be arbitrary, while the remaining $(d-r)$ components are linear combination of the first $r$ components. Define  the set
$$
\begin{aligned}
\mathcal{B}\equiv \{(b_1,b_2,..., b_d)\in \mathbb{R}^{d}: \text{ } & b_{r+1}=b_1-b_2, b_{r+2}=b_1-b_3, ...,b_{2r-1}=b_1-b_r, \\
&b_{2r}=b_2-b_3, ..., b_{3r-3}=b_2-b_r,...,\\
& b_d=b_{r-1}-b_r\}.
\end{aligned}
$$
By the above arguments, $\Delta F_x$ has support contained in $\mathcal{B}$, i.e., 
\begin{equation}
\label{support_restriction}
\lambda_{\Delta F_x}(\mathcal{B})=1,
\end{equation}
for every $ x \in \mathcal{X}$.
We denote by $ \Delta \mathcal{F}$ the function space of all admissible $\{\Delta F_x\}_{x\in \mathcal{X}}$, each with support contained in $\mathcal{B}$.
Moreover, one may want to impose various nonparametric restrictions on  $\{\Delta F_x\}_{x\in \mathcal{X}}$  in order to obtain informative bounds on $U$ (and ultimately, $\Phi$), as highlighted by Lemma \MYref{result0}. We denote by  $\Delta \mathcal{F}^{\dagger}\subset  \Delta \mathcal{F}$ the restricted collection of   conditional CDFs.  We describe later which classes of nonparametric restrictions on  $\{\Delta F_x\}_{x\in \mathcal{X}}$ are considered. 

In summary, our objective is to construct  the identified set of  $U$, defined as
$$
\begin{aligned}
\mathcal{U}^* \equiv \{U\in \mathcal{U}:  \exists \text{ } \{\Delta F_x\}_{x\in \mathcal{X}}\in & \Delta \mathcal{F}^{\dagger} \text{ s.t. } \\
&p_{y|x}=\kappa(U,\Delta F_x,y) \text{ } \forall (x,y) \in \mathcal{X}\times \mathcal{Y}_0 \},
\end{aligned}
$$
where, with slight abuse of notation, we have replaced the argument $F_x$ of $\kappa(U,F_x,y)$ with $\Delta F_x$.
We split the discussion into two steps. First, for any given $U\in \mathcal{U}$, Section \MYref{first_step} explains  how to verify whether $U\in \mathcal{U}^*$. Second, Section \MYref{second_step} provides a result which reduces the computational burden of repeating the first step for every $U\in \mathcal{U}$.
Lastly, we introduce some useful notation adopted in the forthcoming arguments. $\bar{\mathbb{R}}$ denotes the extended real line. $0_d$ is the $d\times 1$ vector of zeros. $\Delta \epsilon_{i,l}$ is the $l$-th component of $\Delta \epsilon_i$ and $\Delta F_{x,l}$ is the $l$-th marginal of $\Delta F_{x}$. $\Delta \epsilon^y_{i}$ is the $r\times 1$ subvector of $\Delta \epsilon_i$ collecting the taste shock differences that are relevant when choosing $y\in \mathcal{Y}_0$, with conditional CDF $\Delta F^y_{x}$.\footnote{For instance, consider $r=2$ (hence, $d=3)$. When choosing $0$, man $i$ evaluates $\epsilon_{i1}-\epsilon_{i0}$ and $ \epsilon_{i2}-\epsilon_{i0}$. Thus,  given the definition of $\Delta \epsilon_i$ in (\MYref{delta_epsilon}), $ \Delta \epsilon^0_{i}\equiv (\epsilon_{i1}-\epsilon_{i0},\epsilon_{i2}-\epsilon_{i0})$. Similarly, $\Delta \epsilon^1_{i}\equiv (\epsilon_{i1}-\epsilon_{i0}, \epsilon_{i1}-\epsilon_{i2})$ and $\Delta \epsilon^2_{i}\equiv (\epsilon_{i2}-\epsilon_{i0}, \epsilon_{i1}-\epsilon_{i2})$.} 

\subsubsection{A linear program}
\label{first_step}
As discussed above,  $U\in \mathcal{U}^*$ if and only if 
\begin{equation}
\label{step1_general}
\exists \text{ } \{\Delta F_x\}_{x\in \mathcal{X}}\in \Delta \mathcal{F}^{\dagger} \text{ s.t. } p_{y|x}=\kappa(U,\Delta F_x,y) \text{ } \forall (x,y) \in \mathcal{X}\times \mathcal{Y}_0.
\end{equation}
Without parametric restrictions on the   unobserved heterogeneity, (\MYref{step1_general}) is an infinite-dimensional existence problem. 
In what follows, we  use and extend  \hyperlink{Torgo}{Torgovitsky (2019)} to transform (\MYref{step1_general}) into a   linear program. We   illustrate the result in the easiest case where $r=2$ (hence, $d=3$). Although notationally more cumbersome, the result for a generic $r$ follows the same intuition and is illustrated in Appendix \MYref{deg_prop_general_section}.

\subsubsection*{The two type case ($\pmb{r=2}$)} 
For simplicity, assume that $\Delta \mathcal{F}^{\dagger}=\Delta \mathcal{F}$. Hence, (\MYref{step1_general}) can be more explicitly written as 
 \begin{equation}
 \small
\label{existence2}
\begin{aligned}
&\exists \text{ } \{\Delta F_x\}_{x\in \mathcal{X}}\in \Delta \mathcal{F} \text{ s.t. } \forall x \in \mathcal{X},\\
&p_{1|x}=1+\Delta F_x(-U_{x1}, +\infty, U_{x2}-U_{x1})-\Delta F_x(+\infty, +\infty, U_{x2}-U_{x1})-\Delta F_x(-U_{x1}, +\infty, +\infty),\\
&p_{2|x}=\Delta F_x(+\infty, +\infty, U_{x2}-U_{x1})-\Delta F_x(+\infty, -U_{x2}, U_{x2}-U_{x1}),\\
&p_{0|x}=\Delta F_x(-U_{x1}, -U_{x2}, +\infty).
\end{aligned}
\end{equation}%

 Note that, for each $x\in \mathcal{X}$, (\MYref{existence2}) depends on the values of $\Delta F_x$ at a {\it finite} number of $3$-tuples. We collect such $3$-tuples in the following three sets: 
\begin{equation}
\label{A_sets}
\begin{aligned}
& \mathcal{A}_{x,1,U}\equiv \{-U_{x1}, +\infty, -\infty\}, \mathcal{A}_{x,2,U}\equiv \{-U_{x2}, +\infty, -\infty\},\mathcal{A}_{x,3,U}\equiv \{U_{x2}-U_{x1}, +\infty, -\infty\},\\
\end{aligned}
\end{equation}
where  $\mathcal{A}_{x,1,U}$ collects the elements at which $\Delta F_x$  is evaluated  along the first dimension, $\mathcal{A}_{x,2,U}$ collects the elements at which $\Delta F_x$  is evaluated along the second dimension, and  $\mathcal{A}_{x,3,U}$ collects the elements at which $\Delta F_x$  is evaluated along the third dimension. We add $-\infty$ to each set because it will be key later to outline the defining properties of CDFs. Lastly, we define 
$
\mathcal{A}_{x,U}\equiv \mathcal{A}_{x,1,U}\times \mathcal{A}_{x,2,U}\times \mathcal{A}_{x,3,U}
$.

 Thus,  the infinite-dimensional existence problem (\MYref{existence2}) is equivalent to verifying whether there exists a   {\it finite-domain} function $ \Delta \bar{F}_x:  \mathcal{A}_{x,U} \rightarrow [0,1]$ that satisfies the equations in (\MYref{existence2}) and that can be {\it extended} to a proper CDF $\Delta F_x: \bar{\mathbb{R}}^3\rightarrow [0,1]$, for every $x\in \mathcal{X}$. That is, (\MYref{existence2}) is equivalent to 
\par\nobreak
\vspace{-0.7cm}
{\small \begin{align}
& \exists \text{ }   \Delta \bar{F}_x:  \mathcal{A}_{x,U} \rightarrow [0,1] \text{ s.t. }\forall x \in \mathcal{X}, \nonumber \\
&p_{1|x}=1+ \Delta \bar{F}_x(-U_{x1}, +\infty, U_{x2}-U_{x1})- \Delta \bar{F}_x(+\infty, +\infty, U_{x2}-U_{x1})- \Delta \bar{F}_x(-U_{x1}, +\infty, +\infty), \label{LP_1}\\
&p_{2|x}= \Delta \bar{F}_x(+\infty, +\infty, U_{x2}-U_{x1})- \Delta \bar{F}_x(+\infty, -U_{x2}, U_{x2}-U_{x1}), \label{LP_2}\\
&p_{0|x}= \Delta \bar{F}_x(-U_{x1}, -U_{x2}, +\infty), \label{LP_3}\\
&\text{and $\{ \Delta \bar{F}_x\}_{x\in \mathcal{X}}$ can be extended to a proper family of conditional CDFs in $\Delta \mathcal{F}$ }.\label{extension_naive}
\end{align}}%
Importantly, observe that (\MYref{LP_1})-(\MYref{LP_3}) are {\it linear} in $ \Delta \bar{F}_x$. Further, we  show below that (\MYref{extension_naive}) can be expressed as  a {\it finite} collection of equations and inequalities that are also {\it linear} in $ \Delta \bar{F}_x$. Therefore, we can transform (\MYref{existence2}) into a linear program.

 We now explain how to write (\MYref{extension_naive})  as  a finite collection of linear equations and inequalities. It is clear that $ \Delta \bar{F}_x$ can be extended to a proper  CDF $\Delta F_x$ {\it only if}   
\begin{equation}
\label{ext_statement}
\text{$ \Delta \bar{F}_x$ satisfies the defining properties of  a CDF.}
\end{equation} 
Consider first  the  case where the support restriction (\MYref{support_restriction}) is ignored in the definition of $\Delta \mathcal{F}$. Then, based on   Sklar's Theorem (\hyperlink{Sklar1}{Sklar, 1959}; \hyperlink{Sklar2}{1996}; \hyperlink{Nelsen}{Nelsen, 2006}), Lemma 2 of \hyperlink{Torgo}{Torgovitsky (2019)} proves that (\MYref{ext_statement}) is also {\it sufficient} for extendibility. In particular, the defining properties of  CDFs are:\\
 (i) $ \Delta \bar{F}_x(a_1,a_2,a_3)=0$ for every $(a_1,a_2,a_3)\in\mathcal{A}_{x,U}$ that has at least one component equal to $-\infty$. That is,
 \begin{equation}
 \label{subdistribution_ext_0}
\begin{aligned}
& \Delta \bar{F}_x(-\infty,a_2,a_3)=0, \Delta \bar{F}_x(a_1,-\infty,a_3) =0,  \Delta \bar{F}_x(a_1,a_2,-\infty)=0 \quad  \forall (a_1,a_2,a_3)\in \mathcal{A}_{x,U}.  \\
\end{aligned}
\end{equation}
(ii) $ \Delta \bar{F}_x(a_1,a_2,a_3)=1$  when $a_l=+\infty$ for every $l\in\{1,2,3\}$. That is,
\begin{equation}
 \label{subdistribution_ext_1}
 \Delta \bar{F}_x(+\infty, +\infty, +\infty)=1.
\end{equation}
(iii) $ \Delta \bar{F}_x$ is $3$-increasing. Formally, given a pair of   $3$-tuples, $ (a_1,a_2,a_3),(a_1',a_2',a_3')$ in $\mathcal{A}_{x,U}$ with $(a_1,a_2,a_3)\leq (a_1',a_2',a_3')$, let $\mbox{Vol}_{ \Delta \bar{F}_x}([a_1,a_1']\times[a_2,a_2']\times[a_3,a'_3])$ denote the volume of the $3$-box $[a_1,a_1']\times[a_2,a_2']\times[a_3,a'_3]$. $ \Delta \bar{F}_x$ is called $3$-increasing if 
 \begin{equation}
  \label{prop_incr}
 \begin{aligned}
& \mbox{Vol}_{ \Delta \bar{F}_x}([a_1,a_1']\times[a_2,a_2']\times[a_3,a'_3])\geq 0,\\
& \text{for every }  (a_1,a_2,a_3), (a_1',a_2',a_3')\in \mathcal{A}_{x,U},  \\
&  \text{s.t. } \hspace{1cm} (a_1,a_2,a_3)\leq (a_1',a_2',a_3').\footnotemark\\
 \end{aligned}
\end{equation}
\footnotetext{Take  $(a_1,a_2,a_3), (a_1',a_2',a_3')\in \mathcal{A}_{x,U}$ with $ (a_1,a_2,a_3)\leq (a_1',a_2',a_3')$. Then, $$\mbox{Vol}_{ \Delta \bar{F}_x}([a_1,a_1']\times[a_2,a_2']\times[a_3,a'_3]) \equiv \sum_{v\in \text{vert}((a_1,a_2,a_3), (a_1',a_2',a_3'))} \Delta \bar{F}_x(v)*\mbox{sgn}(v),$$ where  $\text{vert}((a_1,a_2,a_3), (a_1',a_2',a_3'))$  is the set of the box's vertices, $v\equiv (v_1,v_2,v_3)$ denotes a generic vertex, $\mbox{sign}(v)$ is equal to $1$ if $v_l=a_l$ for an even number of $l\in \{1,2,3\}$, and equal to $-1$ otherwise; 0 is considered even.}%
Note that (\MYref{subdistribution_ext_0})-(\MYref{prop_incr}) constitute a   finite collection of  equations and inequalities that are linear in $ \Delta \bar{F}_x$.
Therefore, if (\MYref{support_restriction}) was absent, we could rewrite (\MYref{existence2}) as a linear program by  direct application of Lemma 2 of \hyperlink{Torgo}{Torgovitsky (2019)}.

The presence of (\MYref{support_restriction}) slightly complicates our analysis as we need to  extend  the latter result to handle such a support restriction.\footnote{\hyperlink{Torgo}{Torgovitsky (2019)} discusses how to handle support restrictions in the case of one-dimensional CDFs. We provide similar findings for  multidimensional CDFs.}  To do so, we first rewrite  (\MYref{support_restriction}) in a more convenient way. Specifically, note that (\MYref{support_restriction}) is equivalent to $\lambda_{\Delta F_x}(\mathcal{B}^c)=0$ where $\mathcal{B}^c$ is the complement of $\mathcal{B}$ in $\bar{\mathbb{R}}^3$. Also observe that, when $r=2$,   $\mathcal{B}$ is simply the plane
$$
\mathcal{B}\equiv \{(b_1,b_2,b_3)\in \mathbb{R}^3: b_1=b_2+b_3\}.
$$
Hence, $\mathcal{B}^c$ can be written as the union of the infinite collection of  $3$-boxes that lie fully above or below such a plane. In particular, for each $(b_1,b_2,b_3)\in \mathcal{B}$, one can construct a $3$-box $\mathcal{B}_{b_1, b_2, b_3}$ such that $\mathcal{B}^c=\cup_{(b_1,b_2,b_3)\in \mathcal{B}} \mathcal{B}_{b_1, b_2, b_3}$. See Appendix \MYref{box} for the representation of $\mathcal{B}_{b_1, b_2, b_3}$.

In turn,  assuming $\lambda_{\Delta F_x}(\mathcal{B}^c)=0$ is equivalent to assuming 
\begin{equation}
\label{deg}
\begin{aligned}
&\mbox{Vol}_{\Delta F_x}(\mathcal{B}_{b_1,b_2,b_3})=0 \quad \text{$\forall (b_1,b_2,b_3)\in \mathcal{B}$.}
\end{aligned}
\end{equation}

Therefore, it is clear that  $ \Delta \bar{F}_x$ can be extended to a  proper CDF $\Delta F_x$  satisfying (\MYref{deg}) {\it only if} the increasingness condition  (\MYref{prop_incr}) holds as  {\it equality} for every  pair of $3$-tuples, $ (a_1,a_2,a_3),(a_1',a_2',a_3')$ in $\mathcal{A}_{x,U}$ with $(a_1,a_2,a_3)< (a_1',a_2',a_3')$, such that the $3$-box $[a_1,a_1']\times [a_2,a_2']\times [a_3,a'_3]$  is  contained in a box $\mathcal{B}_{b_1,b_2,b_3}$ for some $(b_1,b_2,b_3)\in \mathcal{B}$. Proposition \MYref{deg_prop} shows that such a condition is also {\it sufficient} for extendibility when combined with (\MYref{subdistribution_ext_0})-(\MYref{prop_incr}).

\begin{prop}{\normalfont({\itshape Extendibility})}
\label{deg_prop}
Given $U\in \mathcal{U}$ and $x\in \mathcal{X}$, let $\mathcal{A}_{x,U}\equiv  \mathcal{A}_{x,1,U}\times \mathcal{A}_{x,2,U}\times \mathcal{A}_{x,3,U} $, where $ \mathcal{A}_{x,l,U}$ is a finite subset of $\bar{\mathbb{R}}$ and contains $\{+\infty,-\infty\}$ for each $l\in \{1,2,3\}$. Let $ \Delta \bar{F}_x:  \mathcal{A}_{x,U} \rightarrow[0,1]$ be a function satisfying 
(\MYref{subdistribution_ext_0})-(\MYref{prop_incr}), and   
\begin{align}
\label{supp_constraint}
&&&\mbox{Vol}_{ \Delta \bar{F}_x}([a_1,a_1']\times[a_2,a_2']\times[a_3,a'_3])=0,\\
 &&& \text{for every } (a_1,a_2,a_3), (a_1',a_2',a_3')\in \mathcal{A}_{x,U}, \nonumber\\
&&& \text{s.t. } \hspace{1cm} (a_1,a_2,a_3)<(a_1',a_2',a_3'), \nonumber\\
&&& \hspace{1.8cm}\text{and } [a_1,a_1']\times [a_2,a_2']\times [a_3,a'_3] \subset \mathcal{B}_{b_1,b_2,b_3} \text{ for some $(b_1,b_2,b_3)\in \mathcal{B}$}. \nonumber 
\end{align}
Then, there exists a proper CDF $\Delta F_x: \bar{\mathbb{R}}^3\rightarrow [0,1]$  such that: (i)   $ \Delta \bar{F}_x$ can be extended to $\Delta F_x$, i.e.,  $\Delta F_x(a_1,a_2,a_3)=   \Delta \bar{F}_x(a_1,a_2,a_3)$ for each $(a_1,a_2,a_3)\in \mathcal{A}_{x,U}$; (ii) $ \lambda_{\Delta F_x}(\mathcal{B})=1$. 
\end{prop}
Note that (\MYref{supp_constraint})  constitutes a   finite collection of  equations and inequalities that are linear in $ \Delta \bar{F}_x$.
Therefore,  by combining (\MYref{LP_1})-(\MYref{LP_3}), (\MYref{subdistribution_ext_0})-(\MYref{prop_incr}), and (\MYref{supp_constraint}), we can transform (\MYref{existence2}) into a linear program. By verifying the linear program for each $U\in \mathcal{U}$, we can obtain the {\it sharp} identified set.

In addition, we give a simple condition to verify if a  box $[a_1,a_1']\times [a_2,a_2']\times [a_3,a'_3]$  is  contained in a box $\mathcal{B}_{b_1,b_2,b_3}$ for some $(b_1,b_2,b_3)\in \mathcal{B}$, as required by (\MYref{supp_constraint}).
\begin{lm}{\normalfont({\itshape Zero-volume boxes})}
\label{deg_lemma}
Given $U\in \mathcal{U}$ and $x\in \mathcal{X}$, let $\mathcal{A}_{x,U}\equiv  \mathcal{A}_{x,1,U}\times \mathcal{A}_{x,2,U}\times \mathcal{A}_{x,3,U} $, where $ \mathcal{A}_{x,l,U}$ is a finite subset of $\bar{\mathbb{R}}$ and contains $\{+\infty,-\infty\}$ for each $l\in \{1,2,3\}$. Take $ (a_1,a_2,a_3),(a_1',a_2',a_3')$ in $\mathcal{A}_{x,U}$ with $(a_1,a_2,a_3)< (a_1',a_2',a_3')$. Define the $3$-box $\mathcal{H}\equiv [a_1,a_1']\times [a_2,a_2']\times [a_3,a'_3]$. Then, $\mathcal{H} \subset \mathcal{B}_{b_1,b_2,b_3}$ for some $(b_1,b_2,b_3)\in \mathcal{B}$ if and only if 
\begin{equation}
\label{deg_lemma_condition}
a_1>a_2'+a_3'\quad \text{or}\quad a'_1<a_2+a_3.
\end{equation}
\end{lm}

As shown by Theorem 1 of  \hyperlink{Torgo}{Torgovitsky (2019)}, the above methodology remains valid under various classes of nonparametric restrictions on $\{\Delta F_x\}_{x\in \mathcal{X}}$, which can be simply imposed on  $\{ \Delta \bar{F}_x\}_{x\in \mathcal{X}}$ as linear constraints. In particular, in the  simulations of Appendix \MYref{simulations} and the empirical application, we explore the identifying power of the following restrictions (not necessarily all maintained simultaneously): 
\begin{assumption}{\normalfont({\itshape Nonparametric assumptions on $\{\Delta F_x\}_{x\in \mathcal{X}}$})}
\label{nonpar_assumption}
\begin{enumerate}
\item[5.1.] $\Delta \epsilon_i$ is  independent of $X_i$.
\item[5.2.] Conditional on $X_i$ and for each $l\in \{1,...,d\}$, $\Delta \epsilon_{i,l}$ has a distribution symmetric at $0$.
\item[5.3.] Conditional on $X_i$,  $\{\Delta \epsilon_{i,l}\}_{l\in\{1,...,d\}}$ are identically distributed.
\item[5.4.] Conditional on $X_i$,  $\{\Delta \epsilon^y_{i}\}_{y\in \mathcal{Y}_0}$ are identically distributed.
\end{enumerate}
\end{assumption}
In Appendix \MYref{Torgo_formal}, we provide a formal statement of Theorem 1 of  \hyperlink{Torgo}{Torgovitsky (2019)} and  a list of nonparametric distributional assumptions on the taste shock differences that can be generally accommodated. Finally,   Appendix \MYref{example_LP} contains an example of a linear program to solve.


\subsubsection{Simplifying grid search}
\label{second_step}

To construct   $\mathcal{U}^{*}$, the analyst has to solve the linear program  of Section \MYref{first_step} for every $U\in \mathcal{U}$. Typically, this is done in the partial identification literature by constructing a grid of points to approximate $\mathcal{U}$ and then repeating the exercise of interest for each grid point. However, the difficulty of implementing this approach increases with the size of the grid, which in turn, increases exponentially with $r$,  quickly leading  to a computational bottleneck. In what follows, we give a characterisation of $\mathcal{U}$ so that  the issue of solving the linear program for every $U\in\mathcal{U}$ is reduced to solving the linear program for a handful of $U\in\mathcal{U}$. This mitigates the   burden of doing grid search. 
We first provide an intuition of the result and then a more formal statement.

For simplicity, we continue the example of Section \MYref{first_step} with $r=2$ (hence, $d=3$).
For any given $x\in \mathcal{X}$, the only pieces of the linear program  that might induce different sets of solutions for different values of $U$ are the $3$-increasingness constraint, (\MYref{prop_incr}), and the support constraint, (\MYref{supp_constraint}). In fact, note that (\MYref{prop_incr}) is activated only for the pairs of $3$-tuples in $ \mathcal{A}_{x,U}$ that are componentwise comparable. Similarly, (\MYref{supp_constraint}) is activated only for the pairs of $3$-tuples in $ \mathcal{A}_{x,U}$ that are componentwise comparable and satisfy (\MYref{deg_lemma_condition}). We refer to such pairs as the {\it critical pairs}.  Let $h\equiv \Pi_{l=1}^3 h_l$ be the cardinality of $\mathcal{A}_{x,U}$. Fix an order of the $3$-tuples in $ \mathcal{A}_{x,U}$  and list them in an   $h\times 3$ matrix, $\alpha_{x,U}$. 
If the  positions (i.e., row-indices) of the critical pairs  in $\alpha_{x,U}$ are  different across two  values of $U$, then these two values of $U$  will  induce potentially different sets of solutions to the linear program.  Conversely, if  the   positions  are  the same, then the two values  will  induce the same  (possibly, empty)  set of solutions.
This idea can be used to ex-ante partition the parameter space, $\mathcal{U}$,  into equivalence classes so that the researcher can solve  the linear program only once for each   class.\footnote{\hyperlink{Torgo}{Torgovitsky (2019)} suggests  to ex-ante partition the parameter space in order to simplify grid search, even though no algorithm is provided.}

 In what follows,   we provide some sufficient conditions to establish whether  the  positions  of the critical pairs  in $\alpha_{x,U}$ are  equal across two  values of $U$. 
For every $l\in \{1,2,3\}$,  fix an order of the elements of $\mathcal{A}_{x,l,U}$ and list them in an   $h_l \times 1$ vector, $\alpha_{x,l,U}$.   Similarly, construct an $(h_1+h_2h_3)\times 1$ vector, $\beta_{x,U}$, listing $ \alpha_{x,1,U}$ and the sum of every possible element of $\alpha_{x,2,U}$ with every possible element of $\alpha_{x,3,U}$.
Let $\Pi_{1}$ be the set of all possible permutations without repetition of $\{1,...,h_l\}$ and let $\Pi_{2}  \equiv \{<,=\}^{h_l-1}$. 
 Define the functions
$$
\pi_1: \bar{\mathbb{R}}^{h_l}\to \Pi_{1}  \quad\text{and}\quad \pi_2: \bar{\mathbb{R}}^{h_l}\to \Pi_{2},
$$
where $\pi_{1}(\omega)$ sorts the $h_l$ elements of $\omega$ from smallest to largest and reports their positions in the original vector; $\pi_{2}(\omega) $ reports the relational operators, $<$ or $=$, among the sorted elements of $\omega$. When $\omega$ contains multiple elements with the same value or indeterminate forms (like $+\infty-\infty$), then we can adopt any convention on which element to sort first.   Lastly, let $
 \pi(\omega)\equiv (\pi_{1}(\omega), \pi_{2}(\omega))\text{ } \forall \omega\in \bar{\mathbb{R}}^{h}$. 
 For instance, suppose $\omega=(100,99,+\infty)$. Then, $\pi(\omega)=\{(2,1,3), (<,<)\}$. Suppose $\omega=(5,5, -\infty)$. Then, $\pi(\omega)=\{(3,2,1), (<,=)\}$.

 \begin{def2}{\normalfont({\itshape $\pi$-ordering})}
\label{part2}
 Take any $U, \tilde{U}\in \mathcal{U}$. $U$ and $\tilde{U}$ have the same $\pi$-ordering if 
 $$
\begin{aligned}
& \pi(\alpha_{x,l,U})=\pi(\alpha_{x,l,\tilde{U}}) && \forall l\in \{1,...,d\}, x\in \mathcal{X},\\
&   \pi(\beta_{x,U})=\pi(\beta_{x,\tilde{U}})&&\forall   x\in \mathcal{X}. 
\end{aligned}
$$
\end{def2}
 
Proposition \MYref{part2} shows that if $U, \tilde{U}\in \mathcal{U}$ have the same $\pi$-ordering, then the  positions of the critical pairs  in $\alpha_{x,U}$ are equal to the positions of the  critical pairs  in $\alpha_{x,\tilde{U}}$. Therefore,  either both, $U$ and $\tilde{U}$, lie inside or outside the identified set, $\mathcal{U}^*$.

\begin{prop}{\normalfont({\itshape Simplify grid search over $\mathcal{U}$})}
\label{part2}
 Take any $U, \tilde{U}$ in $\mathcal{U}$ with the same   $\pi$-ordering.
Then, $U\in \mathcal{U}^*$ if and only if $\tilde{U}\in \mathcal{U}^*$.
\end{prop}
\noindent Remark \MYref{practice}  in Appendix \MYref{proofs_identification} explains how Proposition \MYref{part2} is used in practice   to ex-ante partition the parameter space.

\section{Empirical application}
\label{empirical}
In this section, we use our methodology to re-examine some of the questions considered in the empirical literature on the marriage market that have been previously answered by relying on the  Logit  1to1TU model. 

An important question is whether educational sorting   has changed over time, as this can be key to understanding the sources of inequality in intergenerational outcomes (see references in Section \MYref{introduction}).
Detecting changes in educational sorting   is challenging because it requires disentangling the effect of changes in the marginal probability distribution of education categories from potential structural changes in the match surplus. In fact, men and especially women have become more educated over time. This implies that individuals with higher education levels are mechanically more likely to marry. We are thus interested in capturing the changes in educational sorting after having accounted for the variations naturally arising due to    distributional shifts in education. 

The literature offers two approaches to do this. The first consists of using indices of sorting, based on comparing empirical matches to a random matching counterfactual (e.g., \hyperlink{Eika}{Eika, Mogstad, and Zafar, 2019}). The second consists of using a structural model of the marriage market to estimate individual preferences. For instance, one can take the 1to1TU model with $\mathcal{X}=\mathcal{Y}\equiv \{1,...,r\}$ listing education categories and study the evolution of  $D(\Phi)$, as discussed in Section \MYref{policy}. The second approach has been implemented  by \hyperlink{CSW}{CSW}  based on the Logit  assumption.
Both approaches conclude that positive educational sorting has overall increased in the U.S. in the past decades, although there is some debate about this trend when we distinguish among education categories. For example,  \hyperlink{Eika}{Eika, Mogstad, and Zafar (2019)} find that positive educational sorting  has declined among the highly educated and increased among the less educated. Instead, \hyperlink{CSW}{CSW} find that positive educational sorting has  increased particularly at the top of the education distribution. 
We use our methodology to investigate the robustness of  the  conclusions achieved via the structural Logit approach to the dropping of the Logit assumption.

In addition to studying the evolution of educational sorting, this section touches upon another important question in the empirical literature on the marriage market. In particular, as discussed by  \hyperlink{CSW}{CSW}, the increase in educational sorting makes a higher stock of human capital more valuable on the marriage market. Therefore, one should also expect  an increase in the marital education premium, especially at the highest levels of education and for women. Based on the Logit  1to1TU model, \hyperlink{CSW}{CSW} empirically confirm such a prediction for the U.S. We  apply our methodology to verify whether the same findings can be achieved without  the Logit assumption.

The remainder of the section is organised as follows: in Section \MYref{data}, we describe the data; in Section \MYref{results}, we present and interpret our results.

\subsection{Data}
\label{data}
We focus on the U.S. marriage market and take our data from the American Community Survey, which is a representative extract of the census. To construct the final dataset, we follow the steps outlined in Section I.A and Appendix B of  \hyperlink{CSW}{CSW}. In particular, from the $21,583,529$ households in the 2008 to 2014 waves, we take all white adults  out of school. We record the education level of each adult by distinguishing four categories: high school dropouts (HSD, or ``1''); high school graduates (HSG, or ``2''); some college (SC, or ``3'');  four-year college graduates and graduate degrees (CG, or ``4'').\footnote{For the white population, \hyperlink{CSW}{CSW} further distinguish between four-year college graduates (CG, or ``4'') and graduate degrees (CG+, or ``5''). In the Logit case, the main conclusions remain unchanged even without this distinction, as  shown in \hyperlink{Chiappori_Meghir}{Chiappori, Costa-Dias, and Meghir (2020)}.  In our analysis too, the conclusions do not change when distinguishing between CG and CG+, as we remark in Section \MYref{results}.} We treat individuals as married if they define themselves as such, without including cohabitation. We focus on first marriages and never-married singles. The final sample consists of $1,502,157$ couples and $136,052$ singles.  

We define cohorts by using year of birth and take women to be one year younger. For instance, cohort 1940 includes all men born in year 1940 and all women born in year 1941. In turn, the sample analogue of $p^{1940}_{y|x}$ is computed by taking the ratio between the number of men of type $x\in \mathcal{X}$ who are born in year 1940 and marry a woman of type $y\in \mathcal{Y}_0$ born in any year, and the number of men of type $x\in \mathcal{X}$ who are born in year 1940. Similarly,  the sample analogue of $p^{1940}_{x|y}$ is computed by taking the ratio between the number of women of type $y\in \mathcal{Y}$ who are born in year 1941 and marry a man of type $x\in \mathcal{X}_0$ born in any year, and the number of women of type $y\in \mathcal{Y}$ who are born in year 1941.\footnote{We ignore the issue of cohort mixing  to exactly mimic the data construction process of  \hyperlink{CSW}{CSW} and  make our conclusions as comparable as possible. In particular, given that the modal age difference within couples is one year in the data, \hyperlink{CSW}{CSW} concentrate their analysis on couples in which the age difference takes one year.} In what follows, we consider $28$ cohorts, from 1940 to $1967$, as in \hyperlink{CSW}{CSW}. 

\begin{figure}[htbp]
 \centering
 \captionsetup{justification=centering}
 \includegraphics[scale=0.6]{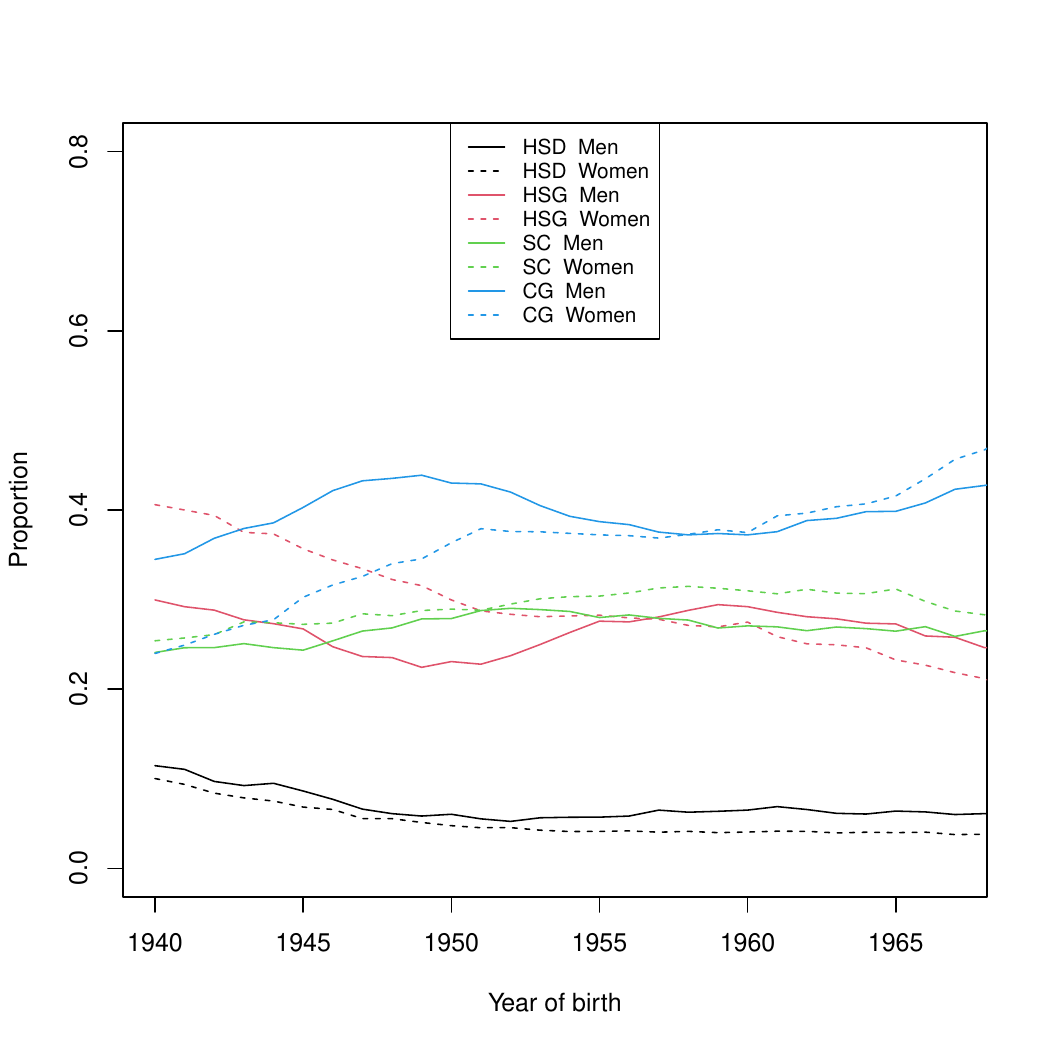}
 \caption{Education of men and women.}
 \label{fig1}
 \end{figure}

\begin{figure}
\begin{adjustwidth}{-1.5cm}{}
    \begin{subfigure}[t]{0.57\textwidth}          
            \includegraphics[width=\textwidth]{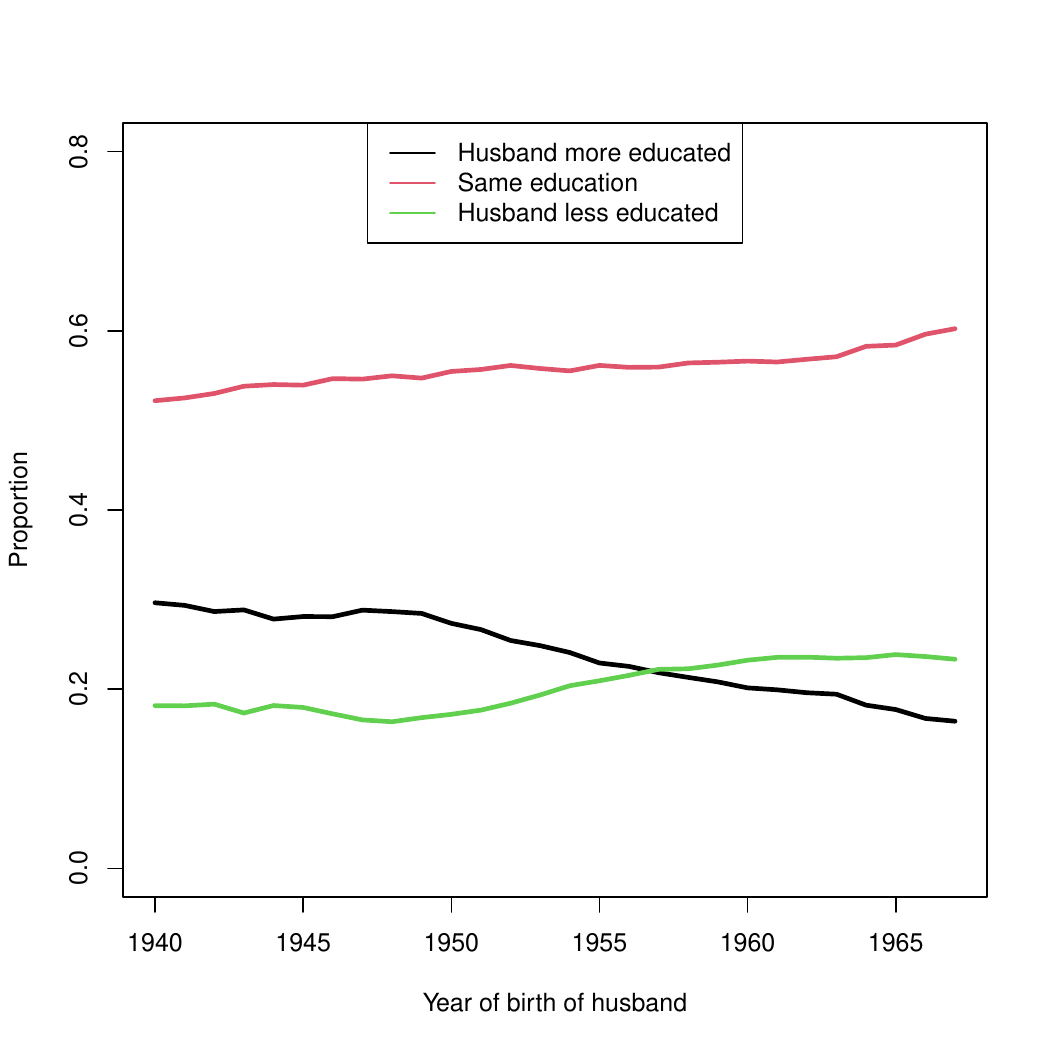}
            \caption{}
    \end{subfigure}%
    \begin{subfigure}[t]{0.57\textwidth}          
            \includegraphics[width=\textwidth]{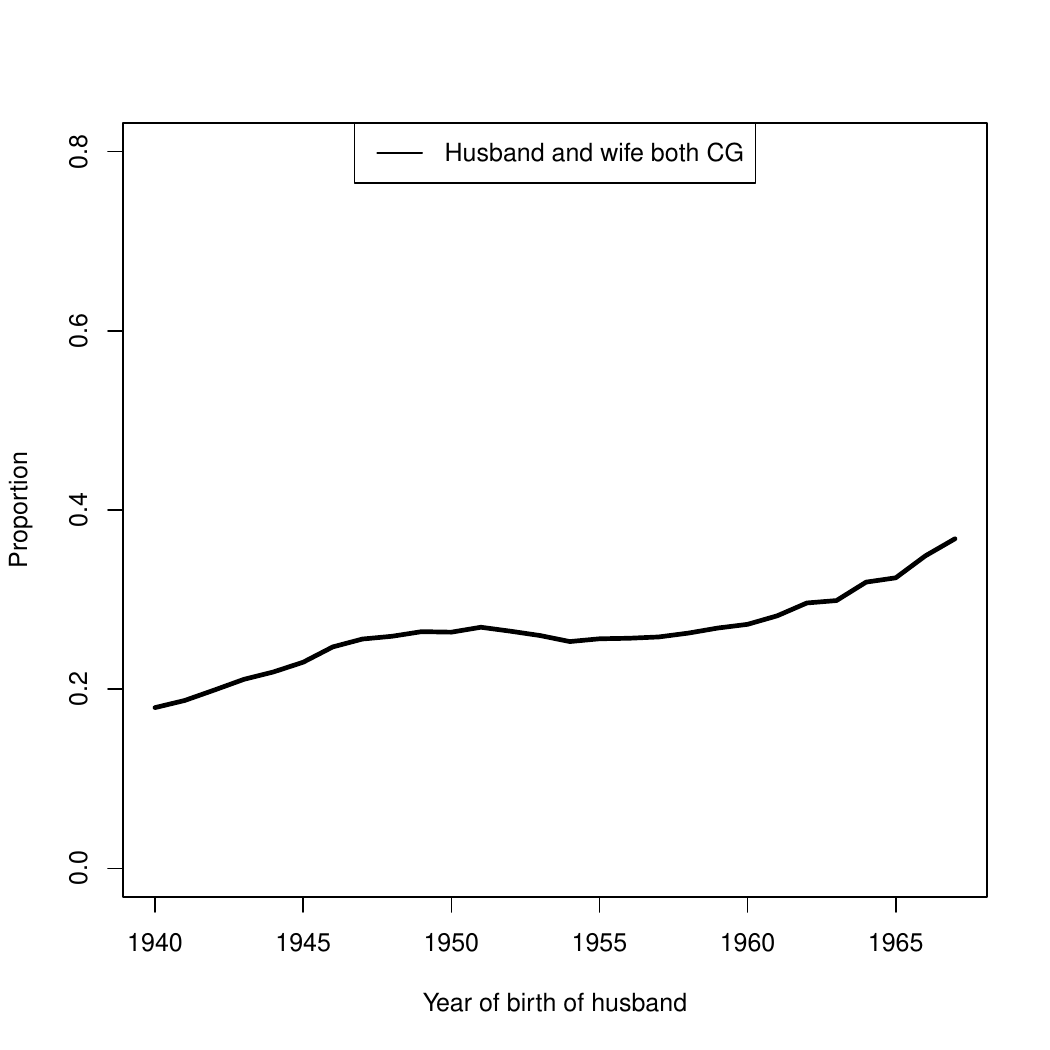}
            \caption{}
    \end{subfigure}%
\end{adjustwidth}
\caption{Comparing spouses.}
\label{fig2}
\end{figure}

Figures  \MYref{fig1}  and   \MYref{fig2}  below are similar to Figures 1 and 2 of \hyperlink{CSW}{CSW} and provide some key descriptive facts.
 Figure \MYref{fig1} reveals that the proportion of college educated men increases until 1950, then drops, and finally reverses into an increase around 1960. Instead, the proportion of college educated women always increases. Moreover, the proportion of  college educated women is lower than that of men in 1940, while the opposite is true by 1967. These changes imply that the evolution of educational sorting cannot be inferred by simply comparing matching patterns across cohorts. 
Figure \MYref{fig2} (a) shows an increase in the proportion of marriages of {\it like} with {\it like}. A substantial surge is also registered when focusing on the proportion of couples where both spouses have a college degree, as shown in Figure \MYref{fig2} (b).  However,   these figures are not proof of an increase in positive educational sorting because they may be mechanically driven by changes in the proportions of individuals  in each education category. 

\subsection{Results}
\label{results}

For each of the 28 cohorts, we estimate the identified sets of  $\Phi$, $D(\Phi)$, $C(U)$, and $C(V)$ under two classes of nonparametric distributional assumptions on the taste shocks,  which we refer to as specifications [A] and [B].  Specification [A] imposes Assumption 5.4. Specification [B] imposes Assumptions 5.2, 5.3, and 5.4. According to our simulations in  Appendix \MYref{simulations}, such specifications tend to deliver the tightest bounds among the various combinations of assumptions explored.

\begin{figure}[!htbp]
\centering
\captionsetup[subfigure]{justification=centering}
\begin{adjustwidth}{-1.7cm}{}
\vspace{-1.1cm}
    \begin{subfigure}[t]{0.43\textwidth}          
            \includegraphics[width=\textwidth, height=5.3cm]{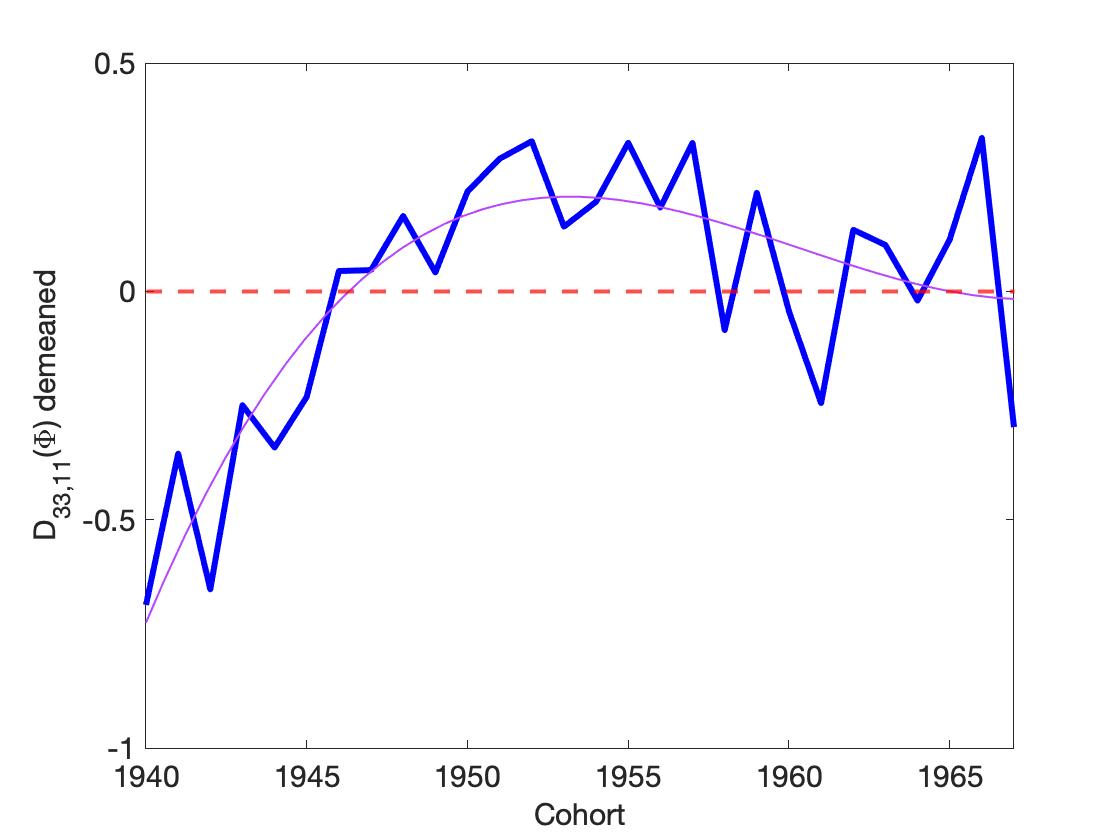}
            \subcaption{\hspace*{-1em}}
    \end{subfigure}%
    \begin{subfigure}[t]{0.43\textwidth}
        \hspace{-0.7cm}
            \includegraphics[width=\textwidth, height=5.3cm]{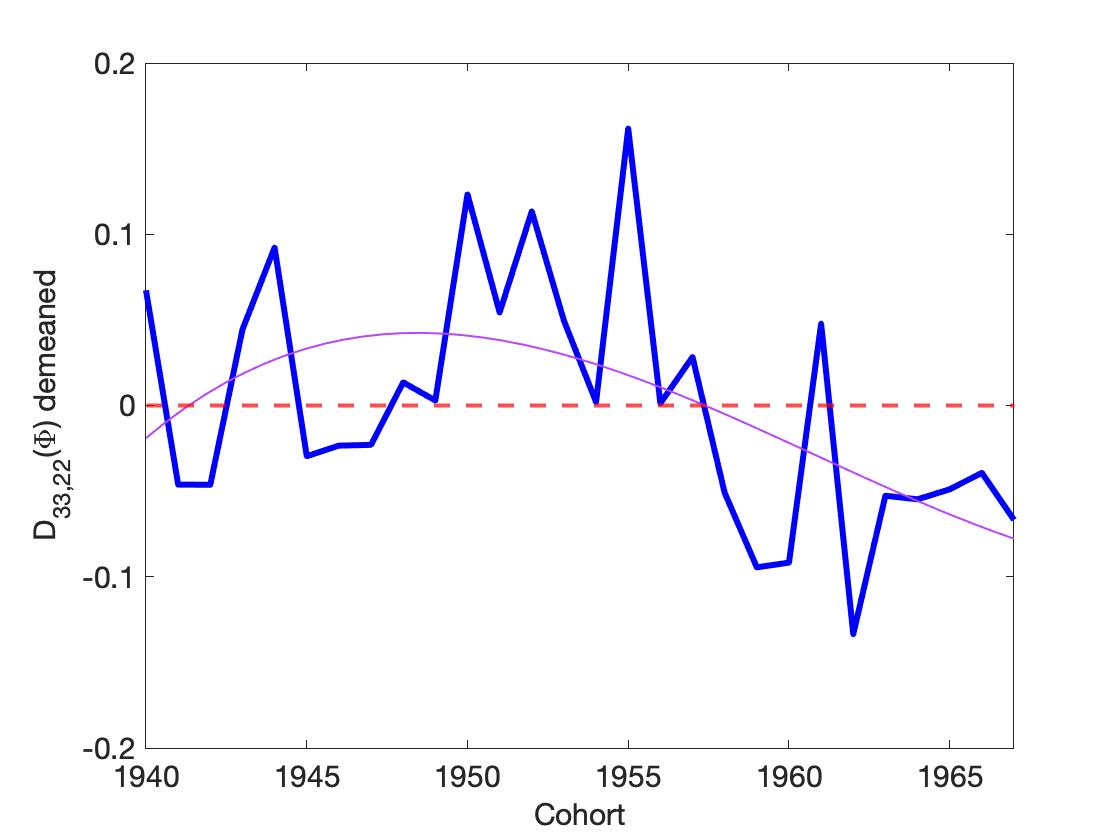}
             \subcaption{\hspace*{1.1em}}
    \end{subfigure}%
   \begin{subfigure}[t]{0.43\textwidth}
           \hspace{-1.3cm}
            \includegraphics[width=\textwidth, height=5.3cm]{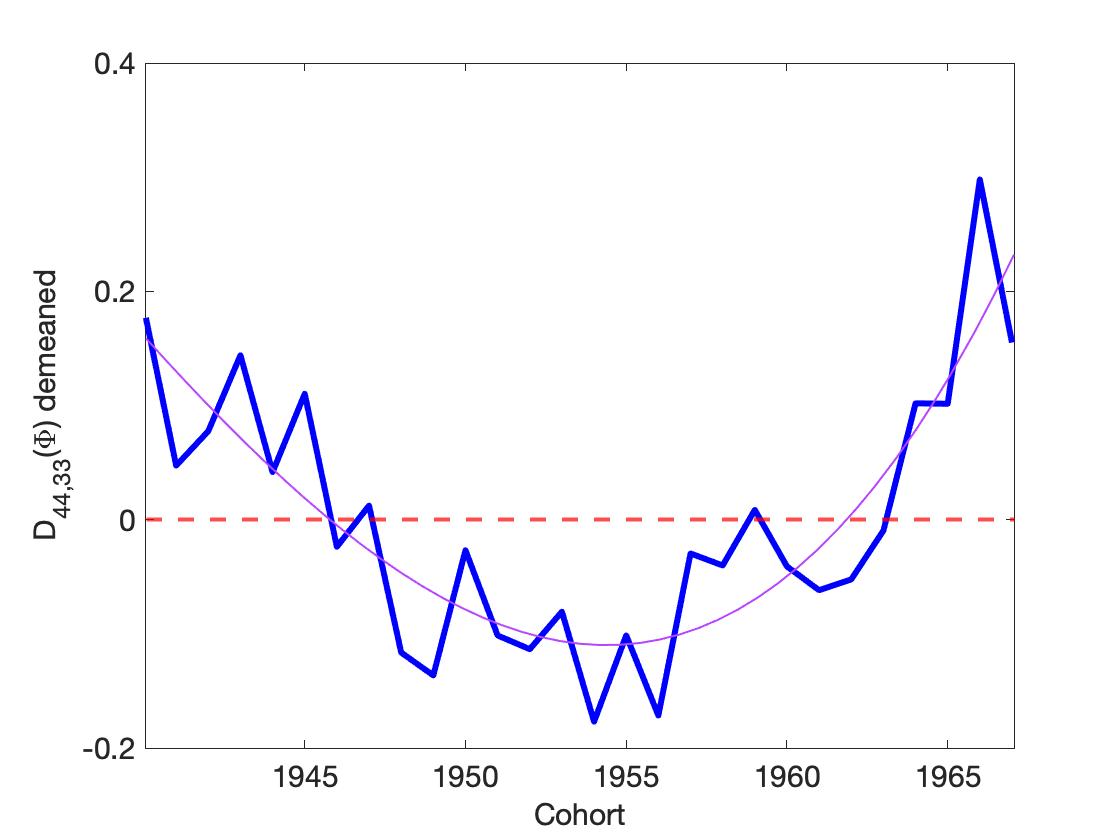}
        \subcaption{\hspace*{4.2em}}
    \end{subfigure}

    \begin{subfigure}[t]{0.43\textwidth}          
            \includegraphics[width=\textwidth, height=5.3cm]{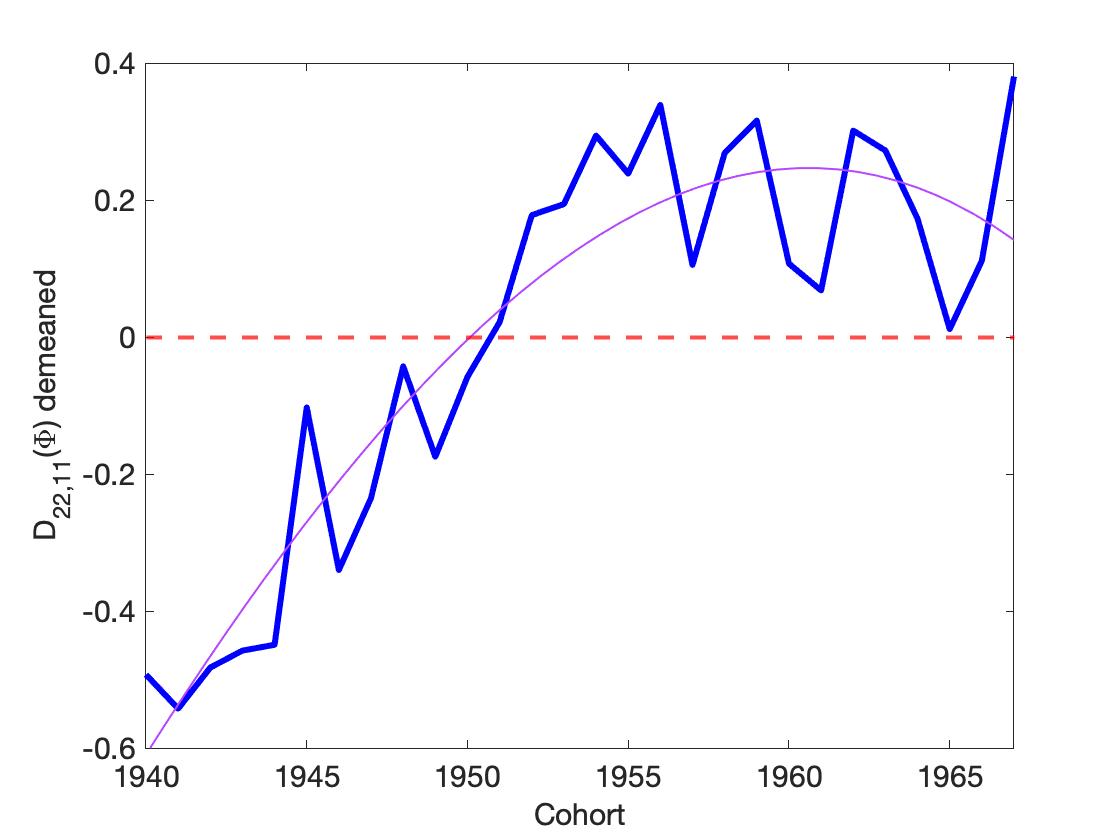}
            \subcaption{\hspace*{-1em}}
    \end{subfigure}%
    \begin{subfigure}[t]{0.43\textwidth}
            \hspace{-0.7cm}
            \includegraphics[width=\textwidth, height=5.3cm]{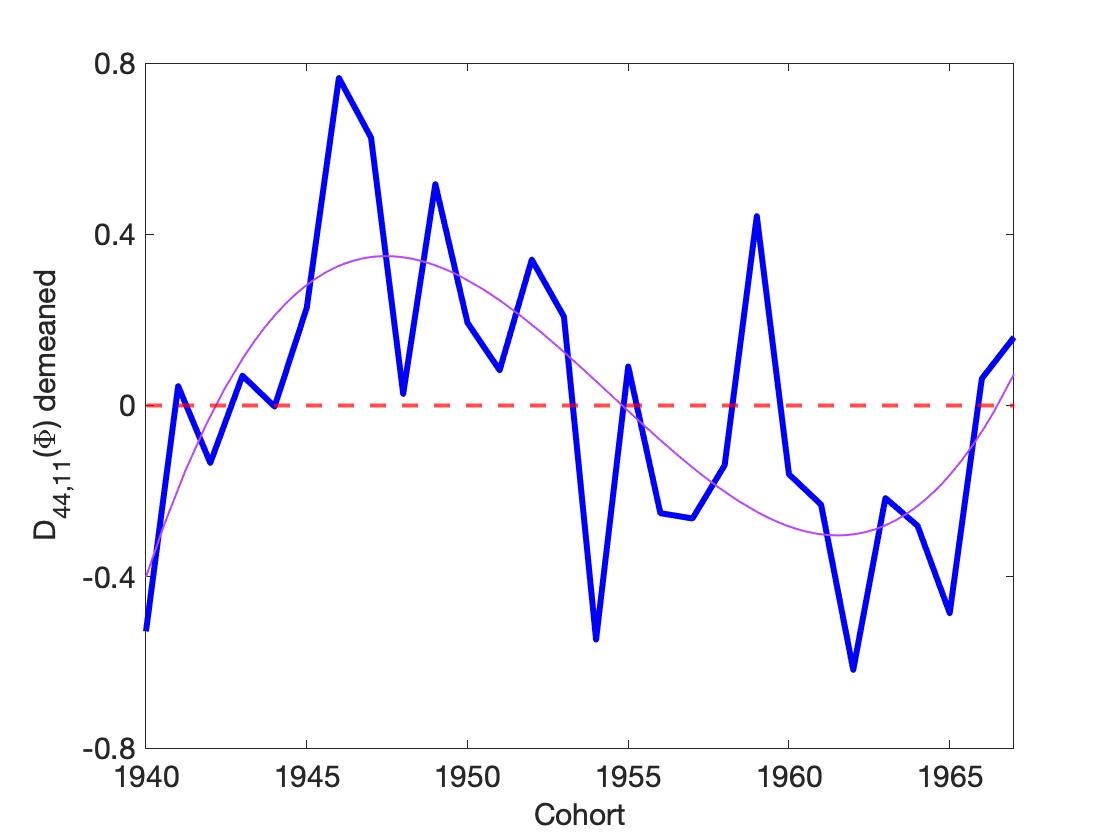}
           \subcaption{\hspace*{1.1em}}
    \end{subfigure}%
   \begin{subfigure}[t]{0.43\textwidth}
           \hspace{-1.3cm}
            \includegraphics[width=\textwidth, height=5.3cm]{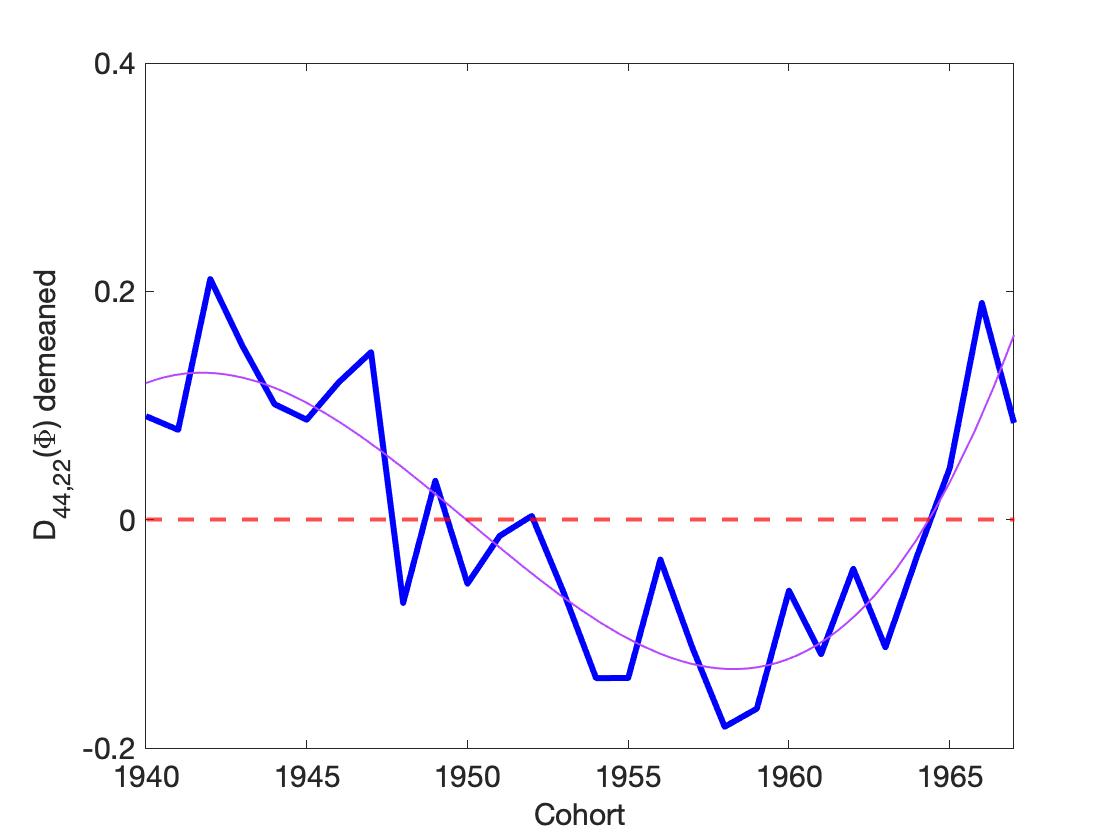}
           \subcaption{\hspace*{4.2em}}
    \end{subfigure}
    \end{adjustwidth}
\caption{Estimates of $D(\Phi)$, demeaned over cohorts, under the Logit  assumption.}
\label{core_original}
\end{figure}

We start by discussing the results on educational sorting.  The Logit  estimates of $D(\Phi)$ are  positive,   suggesting the presence of positive educational sorting in each education category and cohort. 
In particular, Figure \MYref{core_original}  plots  the Logit   estimates of $D(\Phi)$ demeaned over cohorts (blue curves).  If  educational  sorting has not changed over time, then the blue curves (and the smooth violet  curves representing trends) should be  identical to the horizontal $0$ line. The property is   violated for the highly educated, as the trend for  $D_{44,33}(\Phi)$  is  increasing in the most recent decades.\footnote{As discussed in Section \MYref{policy},  there is positive  educational sorting among more   educated people if  $D^\ell(\Phi)\equiv D_{44,33}(\Phi)>0$ and among less   educated people if  $D^\ell(\Phi)\equiv D_{22,11}(\Phi)>0$. Further, positive educational sorting increases across cohorts among more  educated people if   $D_{44,33}(\Phi)$  increases across cohorts. Similarly, positive educational sorting increases across cohorts among less  educated people    if   $D_{22,11}(\Phi)$  increases across cohorts.} We can thus  conclude that there has been an increase in positive assortativeness, at least among the highly educated, under the Logit assumption. More formally, based on the test described in Section IV.A of \hyperlink{CSW}{CSW}, the null hypothesis that educational  sorting has not changed over time is rejected: the Chi-squared test statistic has value $1,047.725$ with 243 degrees of freedom and the p-value is below $10^{-99}$.\footnote{When distinguishing between  CG and CG+, the conclusions on educational sorting based on the Logit assumption are similar, as shown by Figure 15 of \hyperlink{CSW}{CSW} and subsequent discussion. In particular, the null hypothesis that educational  sorting has not changed over time is rejected also with 5 types (the Chi-squared test statistic has value $1,573.717$ with $432$ degrees of freedom and the p-value is below $10^{-100}$).}

\begin{figure}[!htbp]
\centering
\begin{adjustwidth}{-1.8cm}{}
\vspace{-1cm}
    \begin{subfigure}[t]{0.43\textwidth}          
            \includegraphics[width=\textwidth, height=5.3cm]{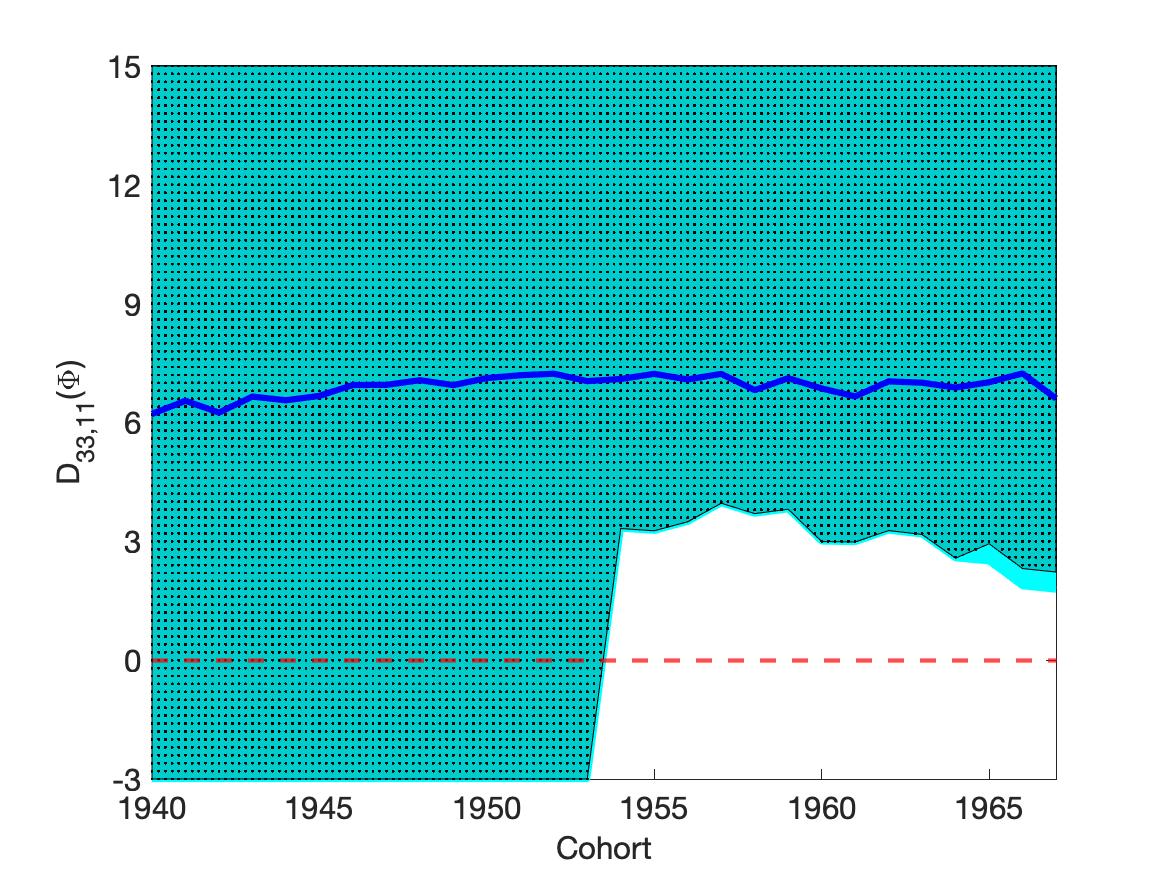}
            \subcaption{\hspace*{-1em}}
    \end{subfigure}%
    \begin{subfigure}[t]{0.43\textwidth}
        \hspace{-0.7cm}
            \includegraphics[width=\textwidth, height=5.3cm]{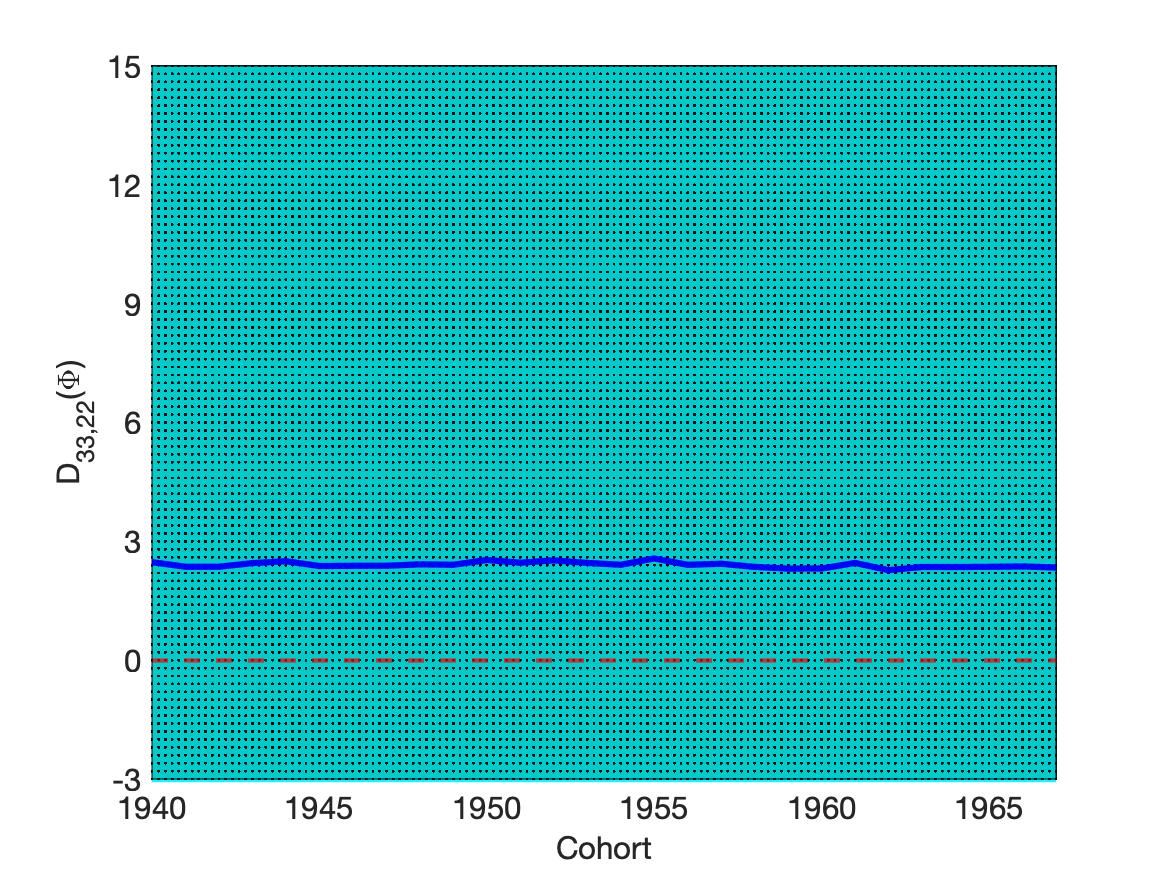}
           \subcaption{\hspace*{1.1em}}
    \end{subfigure}%
   \begin{subfigure}[t]{0.43\textwidth}
           \hspace{-1.3cm}
            \includegraphics[width=\textwidth, height=5.3cm]{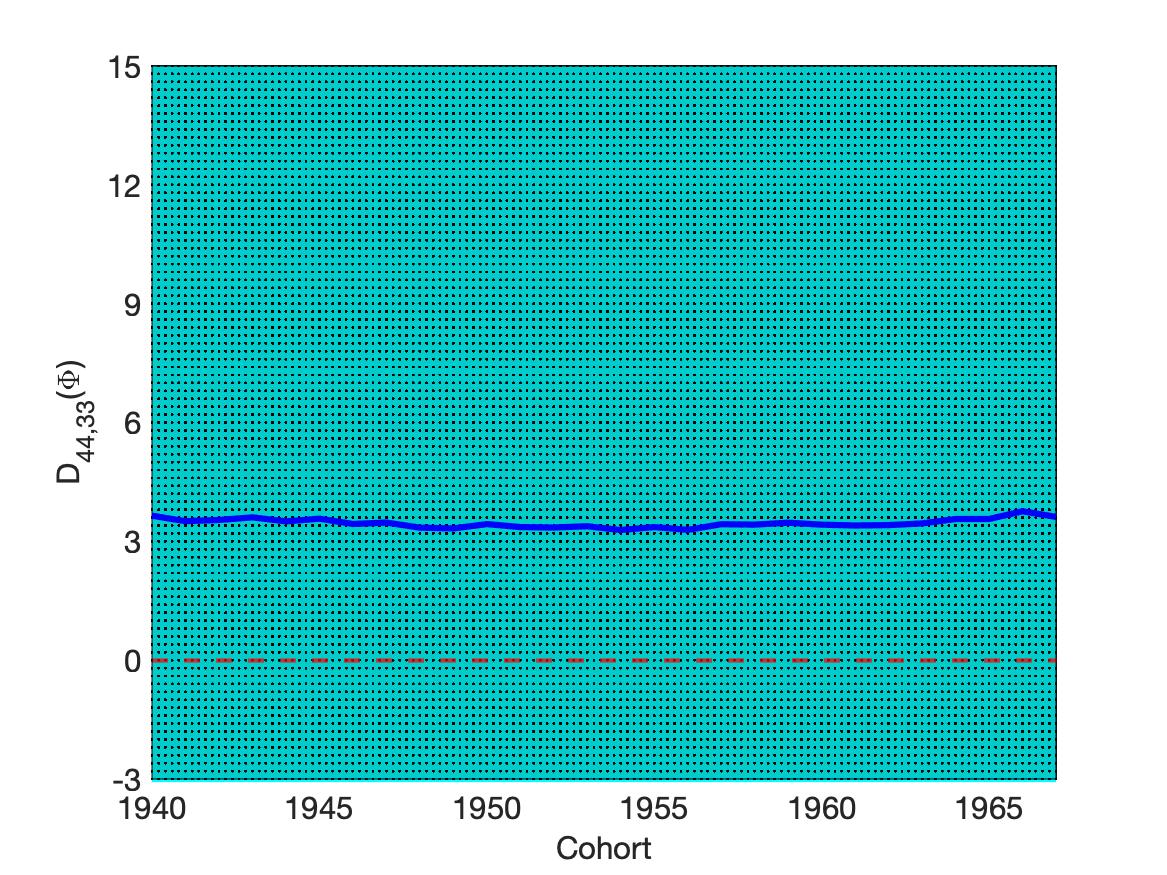}
             \subcaption{\hspace*{4.2em}}
    \end{subfigure}

    \begin{subfigure}[t]{0.43\textwidth}          
            \includegraphics[width=\textwidth, height=5.3cm]{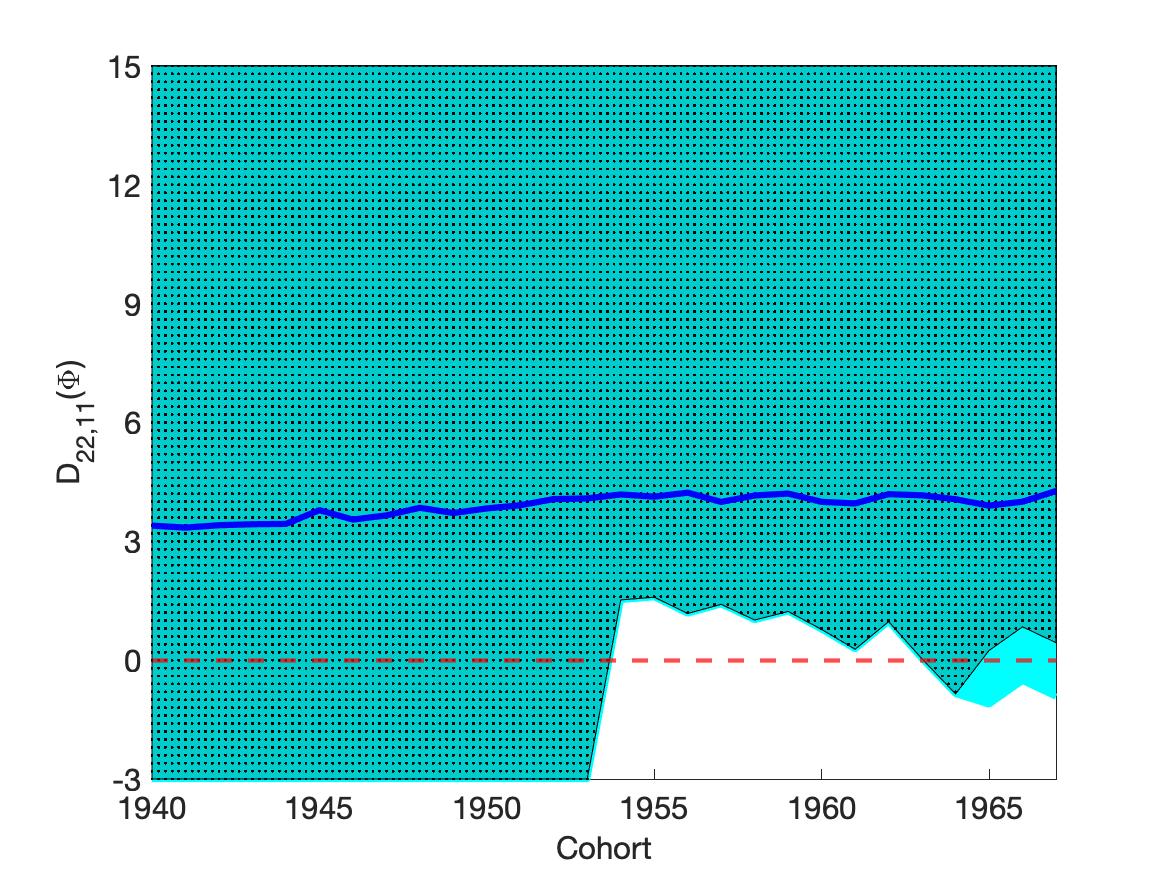}
             \subcaption{\hspace*{-1em}}
    \end{subfigure}%
    \begin{subfigure}[t]{0.43\textwidth}
            \hspace{-0.7cm}
            \includegraphics[width=\textwidth, height=5.3cm]{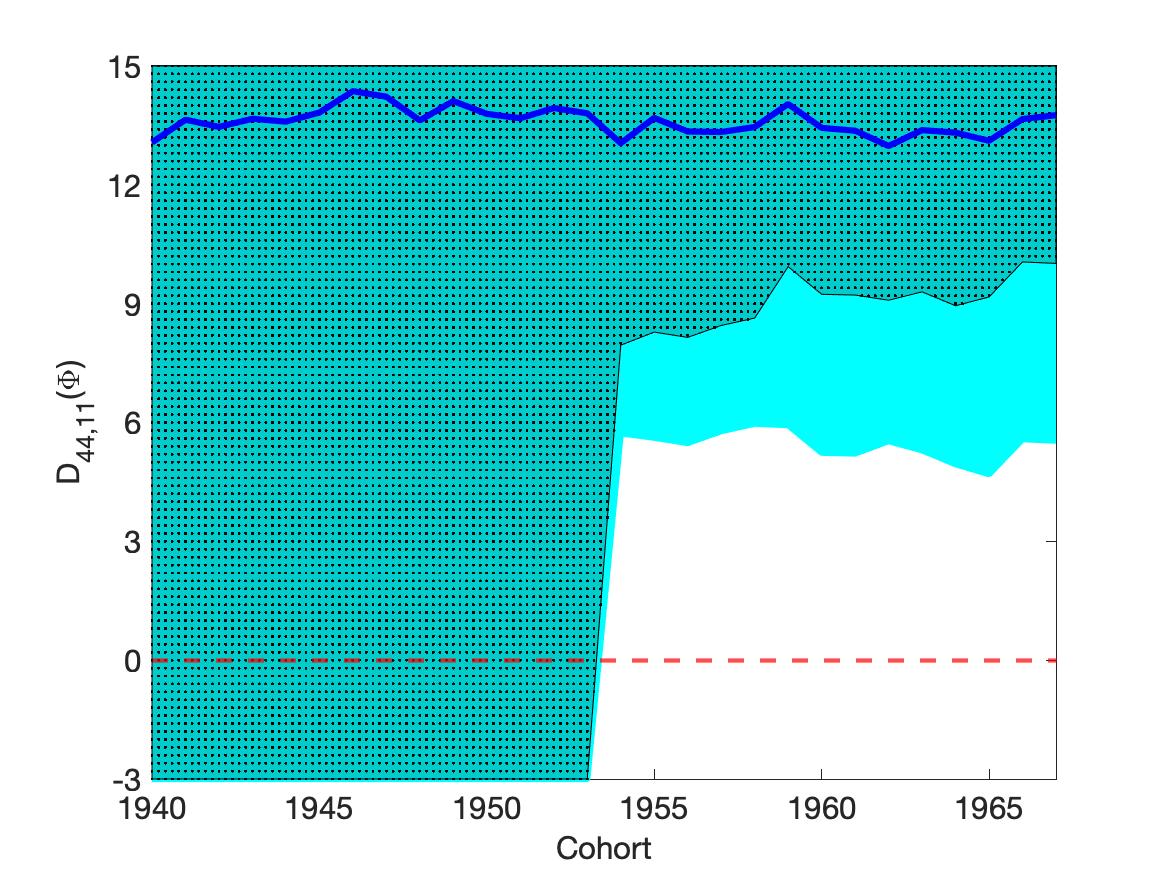}
              \subcaption{\hspace*{1.1em}}
    \end{subfigure}%
   \begin{subfigure}[t]{0.43\textwidth}
           \hspace{-1.3cm}
            \includegraphics[width=\textwidth, height=5.3cm]{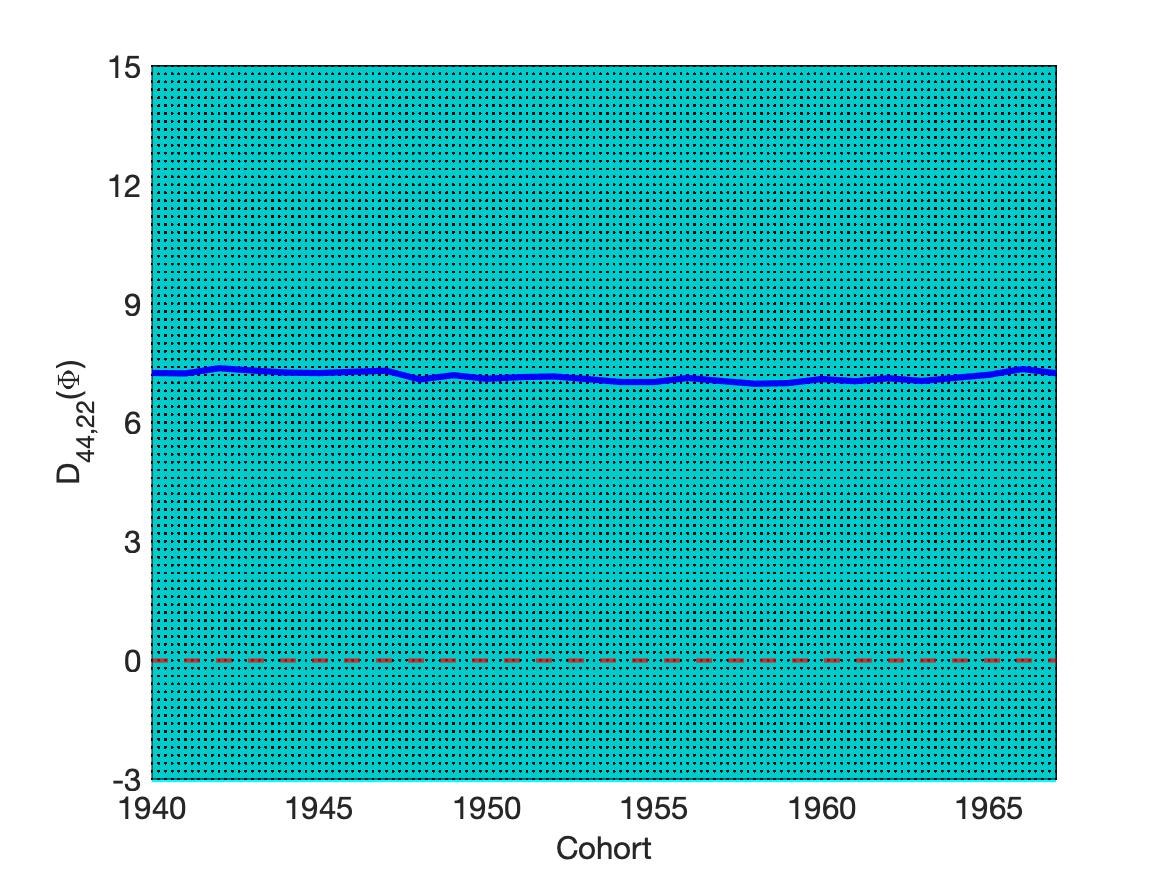}
            \subcaption{\hspace*{4.2em}}
    \end{subfigure}
    \end{adjustwidth}
\caption{The  blue and dotted regions are the estimated identified sets of $D(\Phi)$ under specifications [\text{A}] and [B], respectively.   The dark blue line represents the estimates of $D(\Phi)$ under the Logit  assumption.}
\label{core_ours}
\end{figure}

Figure \MYref{core_ours} reports our estimates of the identified set of $D(\Phi)$,  under specifications [A] (blue region) and [B] (dotted region).\footnote{We do not demean the estimates  in Figure \MYref{core_ours} in order to study their signs.} By construction, the dotted region is contained in (or is equal to) the blue region. Further, the Logit   estimates of $D(\Phi)$ (dark blue line) are  contained in the blue and dotted regions because  Assumptions 5.2, 5.3, and 5.4 are satisfied when imposing the Logit assumption.
As in  \hyperlink{CSW}{CSW}, we obtain our estimates by assuming that the cohorts feature  independent matching
processes.  However, our analysis is more robust in many ways. Importantly, we allow the taste shocks to have any distribution within   specifications [\text{A}] and [B]. For instance, the taste shocks can be correlated among each other, their distribution may freely vary across education categories, and there could be heteroskedasticity. 


{Figure \MYref{core_ours} reveals that, under the classes of nonparametric distributional assumptions considered, the 1to1TU model is uninformative about the presence and trend of  positive educational sorting among the highly educated, as the estimates of $D_{44,33}(\Phi)$   are unbounded  above and below.\footnote{The estimates are unbounded when the blue or dotted region hits the vertical axis limit.}  We find the presence of positive educational sorting among the less educated, as indicated by the  jump to positive values of the lower bound of $D_{22,11}(\Phi)$ around 1954.  However, once the  lower bound reaches the positive values, it does not exhibit any clear  trend, thereby remaining inconclusive about the evolution of positive educational sorting   among the less educated. These results suggest that the previous findings on  educational sorting  based on the Logit  1to1TU model are  driven by the Logit assumption.\footnote{When distinguishing between  CG and CG+, our estimates of $D_{55,44}(\Phi)$, $D_{55,33}(\Phi)$, and $D_{44,33}(\Phi)$ are unbounded above and below.    {Therefore, the 1to1TU model still does not allow us to conclude anything about the presence and trend of  educational sorting among the highly educated, as in Figure \MYref{core_ours}.}}\footnote{Figures  \MYref{core_ours_M}  and  \MYref{core_ours_W} in Appendix \MYref{empirical_appendix} further disentangle the men and women's contribution to $D(\Phi)$. They highlight that the unboundedness of $D_{22,11}(\Phi)$ (above)   and $D_{44,33}(\Phi)$  (above and below)  is mostly driven by the limited empirical content of the 1to1TU model on the women's side.}}

Table \MYref{Tab6} confirms the above conclusions. The first section of the table reports the projections of the estimated identified sets of $\Phi$, averaged over cohorts 1940, 1941, and 1942 (``early cohorts''), under specifications [A] and [B]. The second section of the table reports the  projections of the estimated identified sets of  $\Phi$, averaged over cohorts 1965, 1966, and 1967 (``late cohorts''), under specifications [A] and [B]. The last section of the table reports the changes in estimates between early  and late cohorts. The average estimates of $\Phi$ under the Logit  assumption (``Logit'') are also included. When using the Logit  assumption, the decline in surplus is always smaller (or inverted) for more educated couples, which is in line with the increase in positive educational sorting at the top of the distribution seen in Figure \MYref{core_original}.  This conclusion cannot be confirmed once the Logit  assumption is relaxed, as highlighted by the many unbounded intervals.\footnote{The Logit estimates in Table \MYref{Tab6} for early and late cohorts are numerically different from the Logit estimates in Table 6 of  \hyperlink{CSW}{CSW} due to  two reasons. First,  \hyperlink{CSW}{CSW} construct those estimates by using the assumptions that the evolution of the systematic match surplus is driven by education-specific drifts, which is not assumed here. Second,  \hyperlink{CSW}{CSW} distinguish between  CG and CG+. Nevertheless, the changes in the Logit estimates between early and late cohorts that we obtain (last section of the table) are very close to  \hyperlink{CSW}{CSW}'s findings and, importantly, suggest the same conclusions.}

\begin{figure}[!htbp]
\centering
\begin{adjustwidth}{-1.7cm}{}
\vspace{-1.1cm}
    \begin{subfigure}[t]{0.43\textwidth}          
            \includegraphics[width=\textwidth, height=5cm]{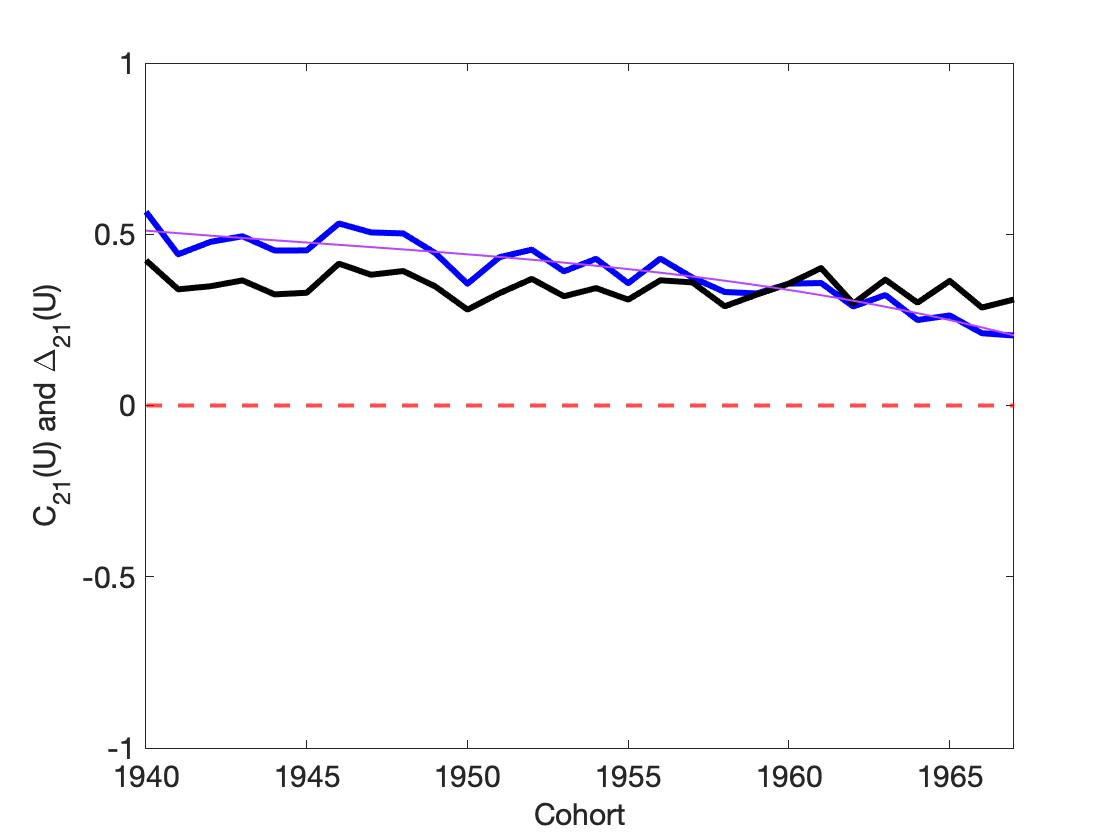}
            \subcaption{\hspace*{-1em}}
    \end{subfigure}%
    \begin{subfigure}[t]{0.43\textwidth}
        \hspace{-0.7cm}
            \includegraphics[width=\textwidth, height=5cm]{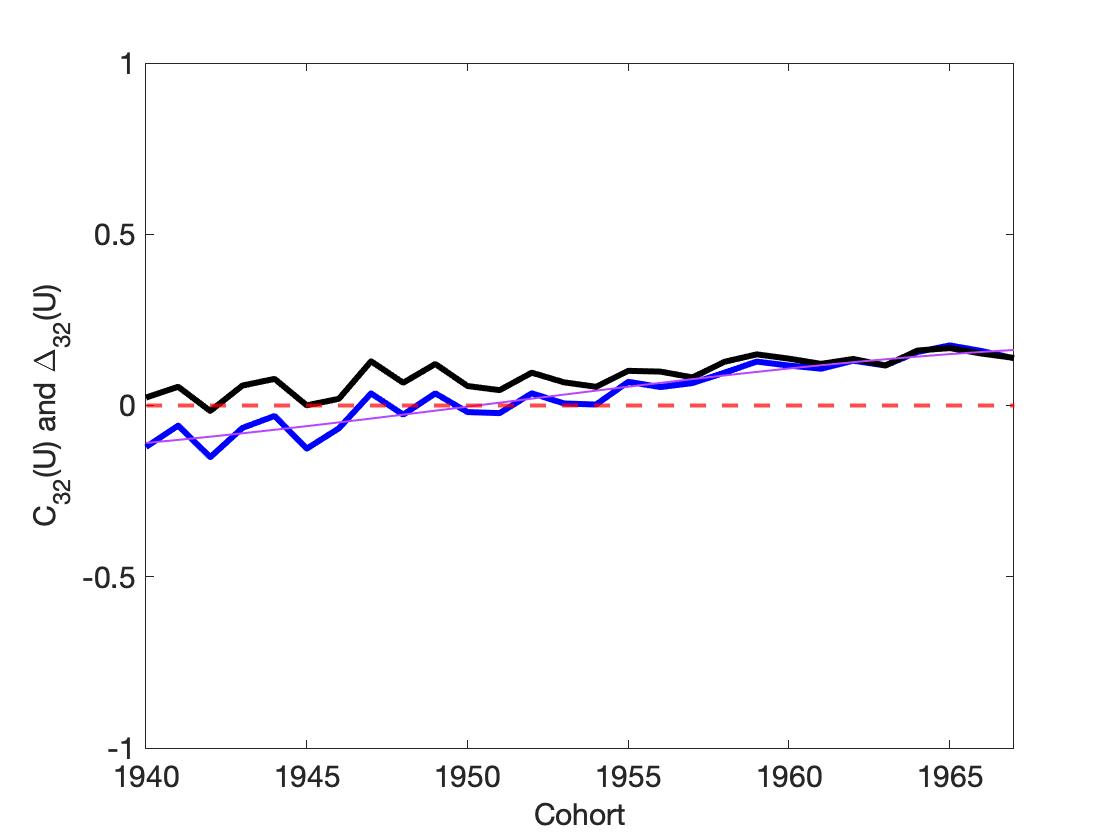}
            \subcaption{\hspace*{1.1em}}
    \end{subfigure}%
   \begin{subfigure}[t]{0.43\textwidth}
           \hspace{-1.3cm}
            \includegraphics[width=\textwidth, height=5cm]{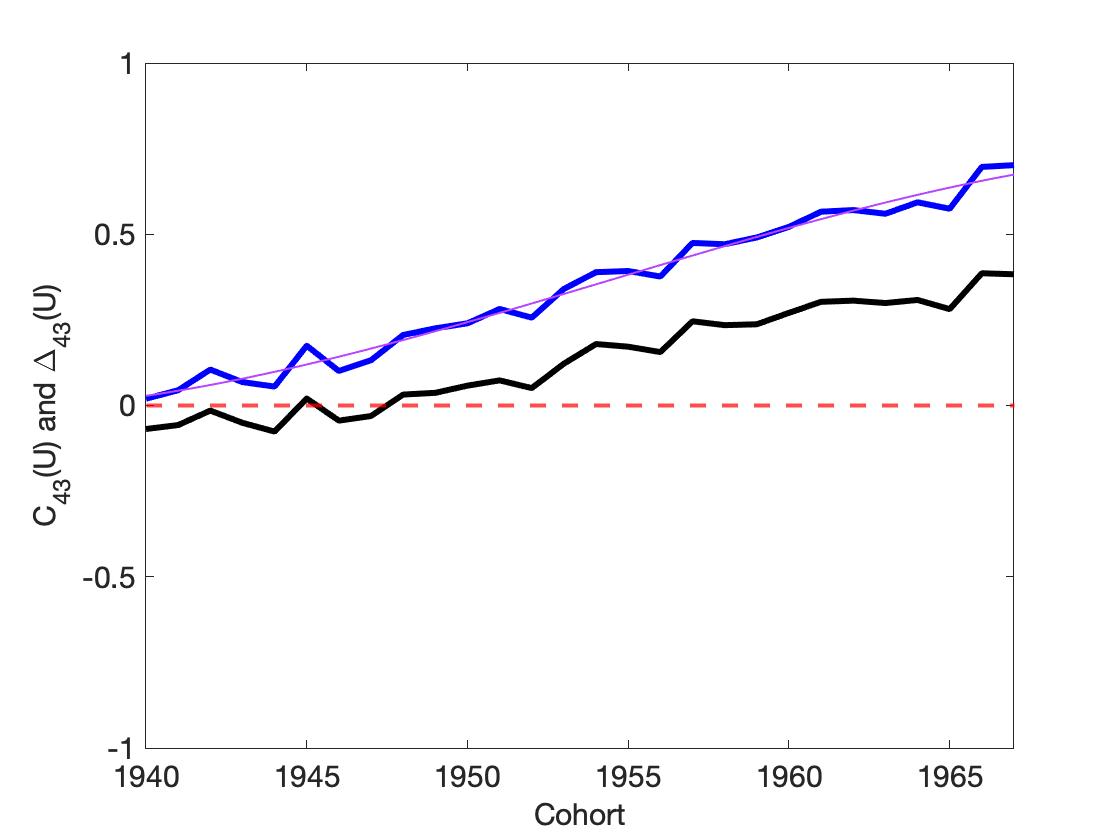}
            \subcaption{\hspace*{4.2em}}
    \end{subfigure}

    \begin{subfigure}[t]{0.43\textwidth}          
            \includegraphics[width=\textwidth, height=5cm]{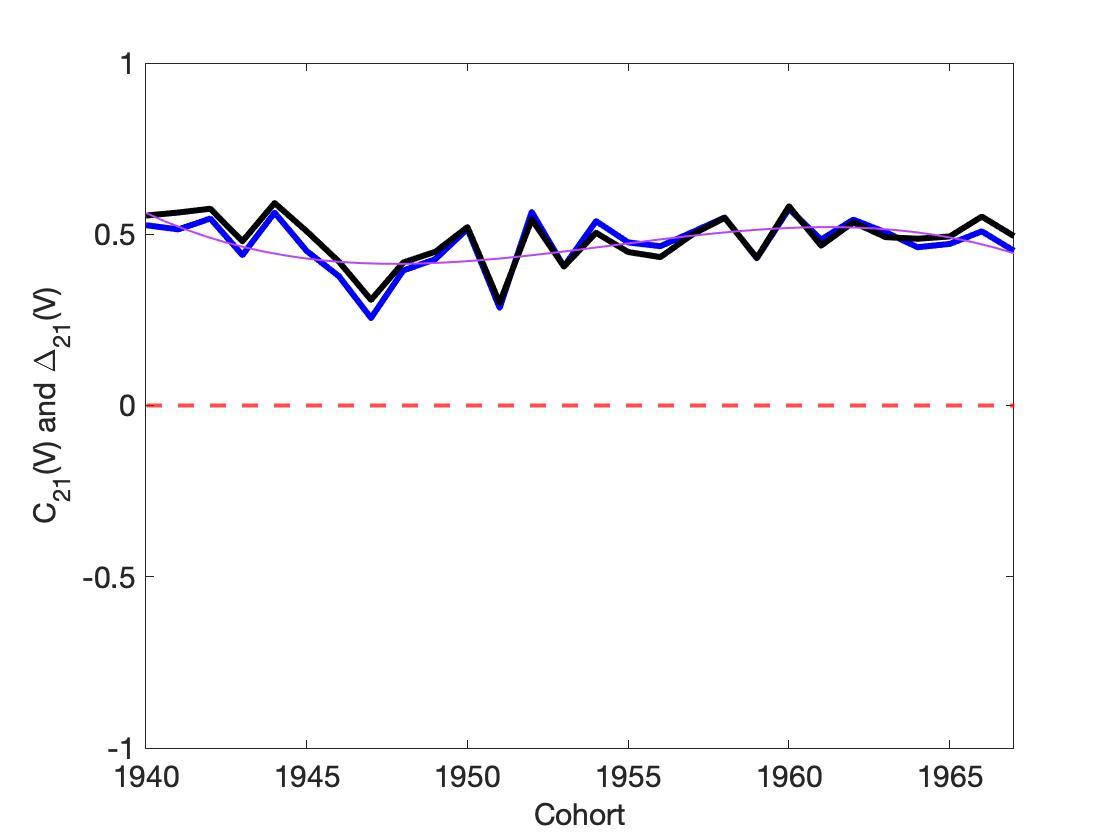}
           \subcaption{\hspace*{-1em}}
    \end{subfigure}%
    \begin{subfigure}[t]{0.43\textwidth}
            \hspace{-0.7cm}
            \includegraphics[width=\textwidth, height=5cm]{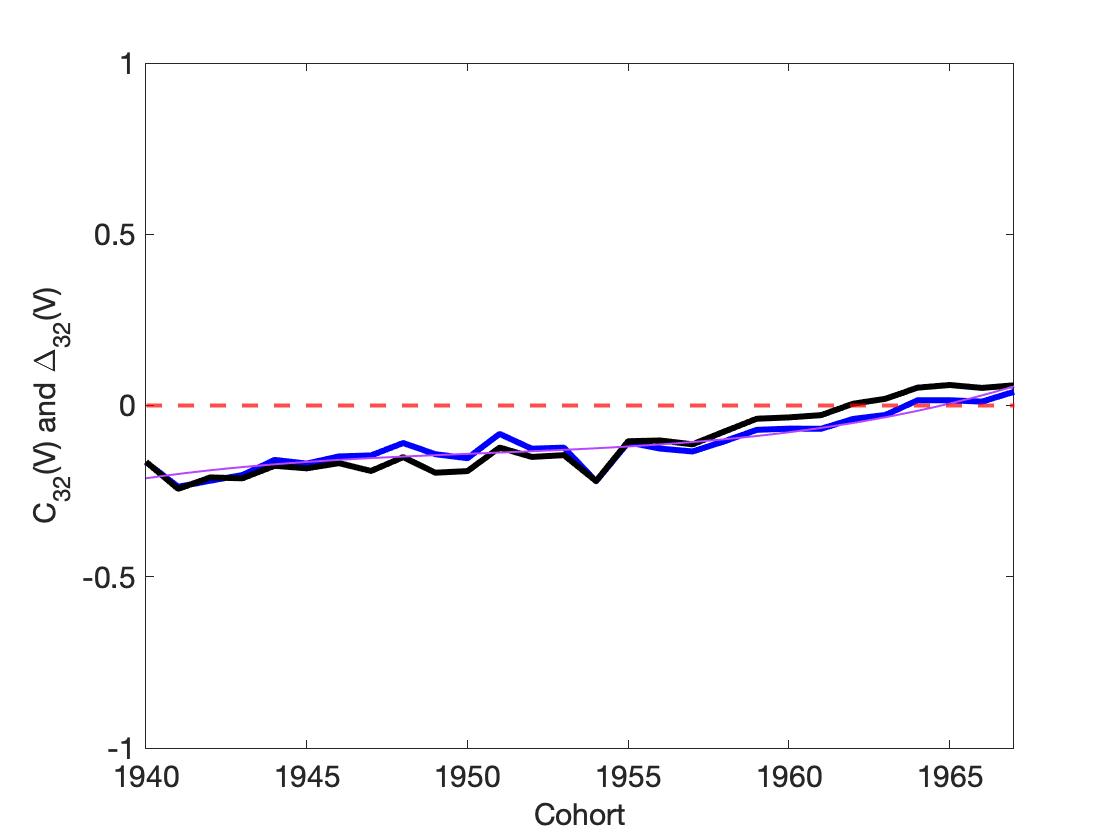}
          \subcaption{\hspace*{1.1em}}
    \end{subfigure}%
   \begin{subfigure}[t]{0.43\textwidth}
           \hspace{-1.3cm}
            \includegraphics[width=\textwidth, height=5cm]{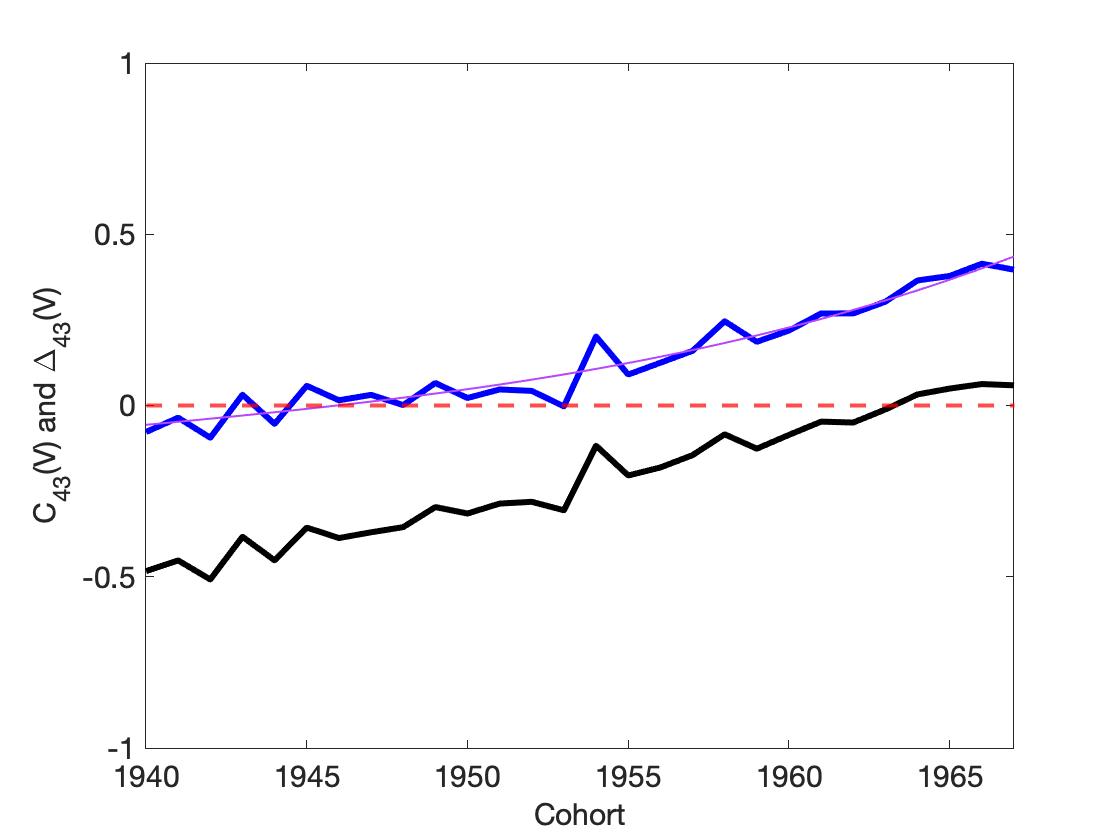}
            \subcaption{\hspace*{4.2em}}
    \end{subfigure}
    \end{adjustwidth}
\caption{The blue line represents the estimates of $C(U)$ and $C(V)$ under the Logit  assumption. The black line represents the estimates of the marital education premia under the Logit  assumption.}
\label{C_original}
\end{figure}

We now move to discuss the results on the marital education premia. The black curves in Figure \MYref{C_original} are the estimates of the marital education premia  under the Logit  assumption for men and women. Panels (c) and (f)  suggest that  the marital college premium  has increased for both men ($\Delta_{43}(U)$) and women ($\Delta_{43}(V)$). The increase is particularly pronounced for women. Further, while women of older cohorts had a negative marital college premium, this has become positive for recent cohorts.\footnote{When distinguishing between  CG and CG+, the conclusions on the marital education premia based on the Logit assumption are similar, as shown by Figures 20 and 21 of \hyperlink{CSW}{CSW} and subsequent discussion.} The blue curves in Figure \MYref{C_original} are the estimates of $C(U)$ and $C(V)$  under the Logit  assumption. The blue curves mimic  the trends of the black curves closely, although they are quite shifted from the black curves in panels (c) and (f).

\begin{figure}[!htbp]
\centering
\begin{adjustwidth}{-1.7cm}{}
\vspace{-1cm}
    \begin{subfigure}[t]{0.43\textwidth}          
            \includegraphics[width=\textwidth, height=5cm]{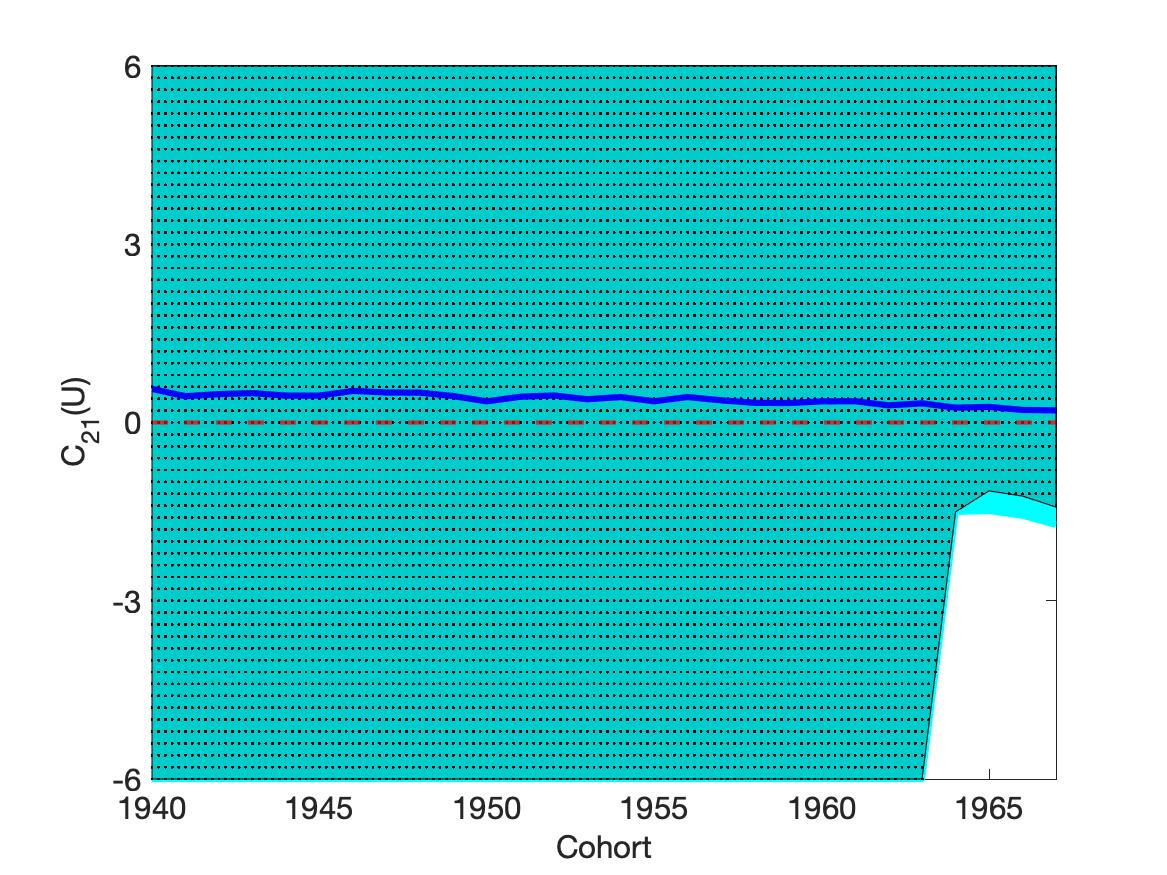}
            \subcaption{\hspace*{-1em}}
    \end{subfigure}%
    \begin{subfigure}[t]{0.43\textwidth}
        \hspace{-0.7cm}
            \includegraphics[width=\textwidth, height=5cm]{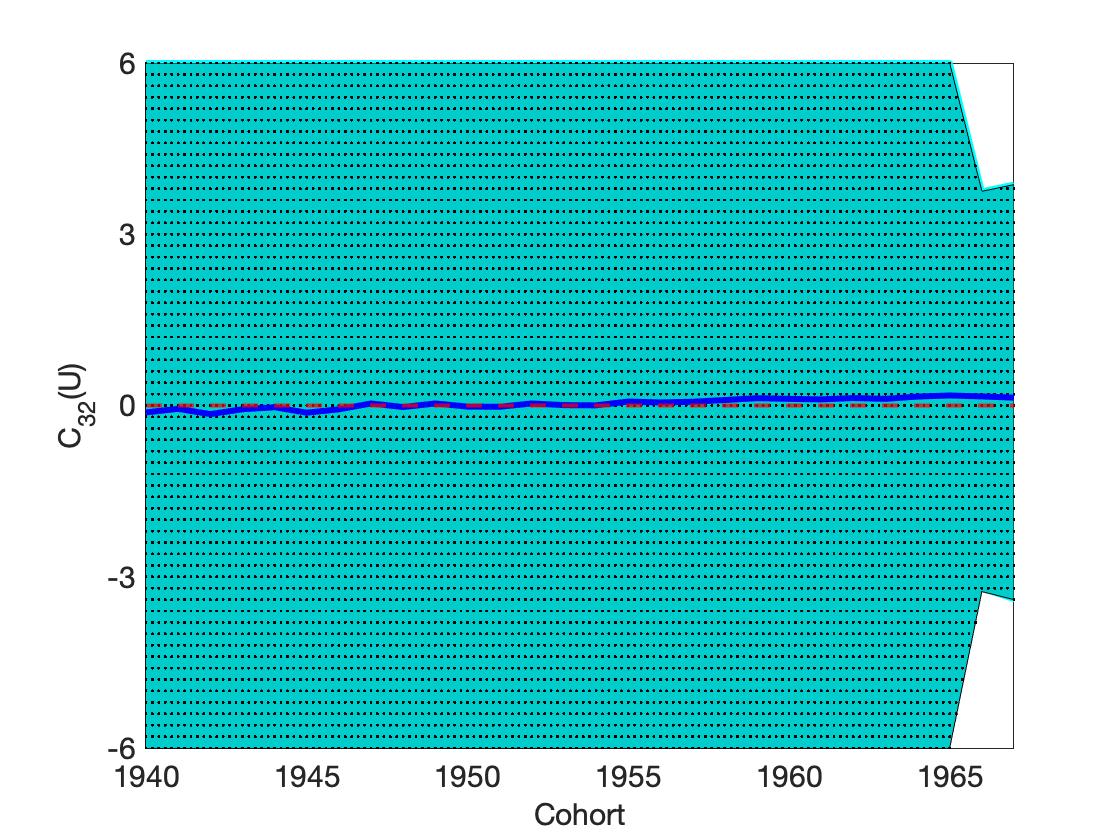}
             \subcaption{\hspace*{1.1em}}
    \end{subfigure}%
   \begin{subfigure}[t]{0.43\textwidth}
           \hspace{-1.3cm}
            \includegraphics[width=\textwidth, height=5cm]{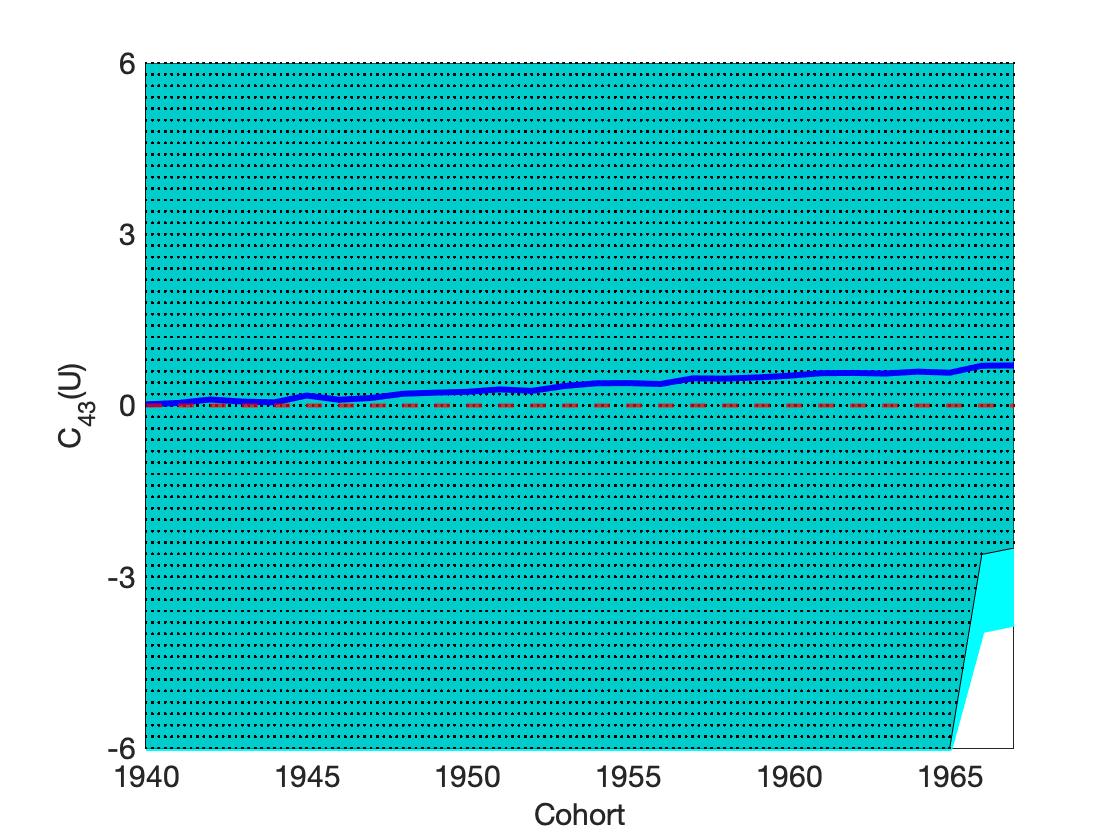}
            \subcaption{\hspace*{4.2em}}
    \end{subfigure}

    \begin{subfigure}[t]{0.43\textwidth}          
            \includegraphics[width=\textwidth, height=5cm]{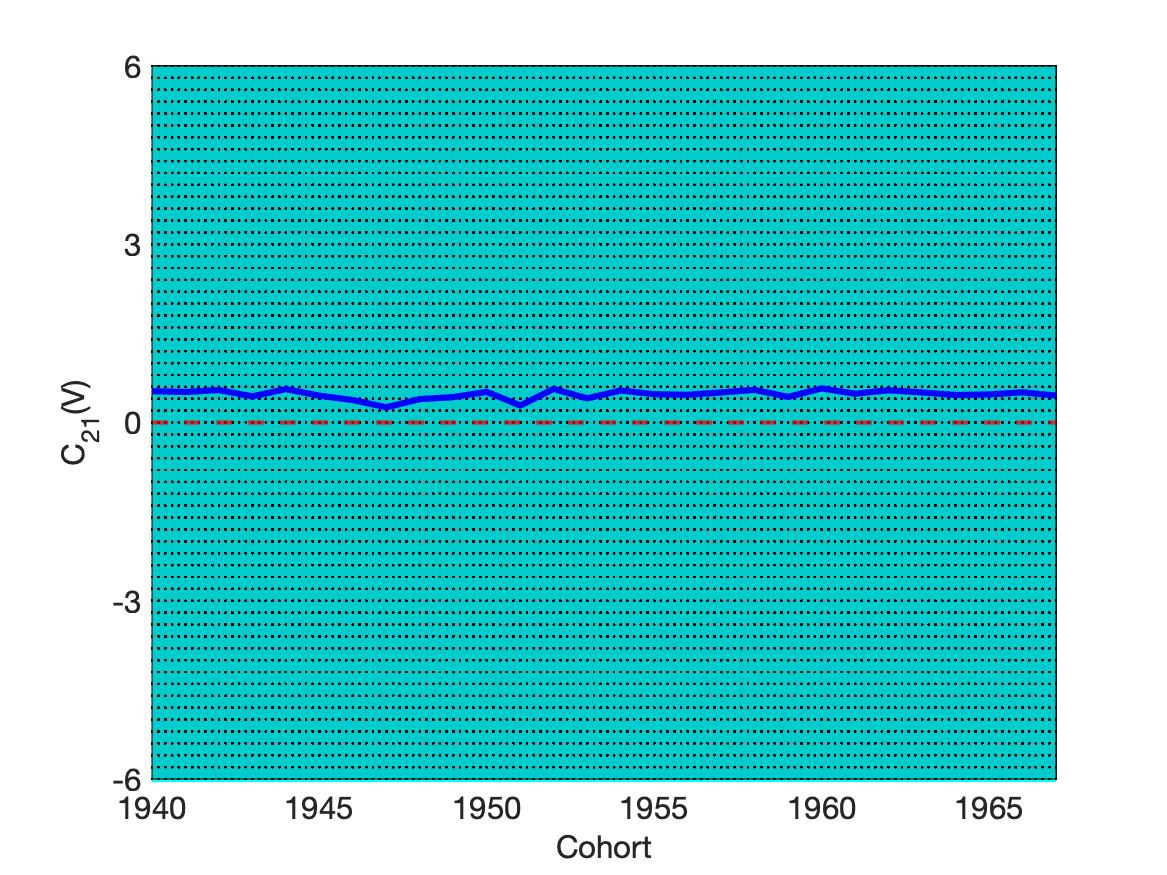}
            \subcaption{\hspace*{-1em}}
    \end{subfigure}%
    \begin{subfigure}[t]{0.43\textwidth}
            \hspace{-0.7cm}
            \includegraphics[width=\textwidth, height=5cm]{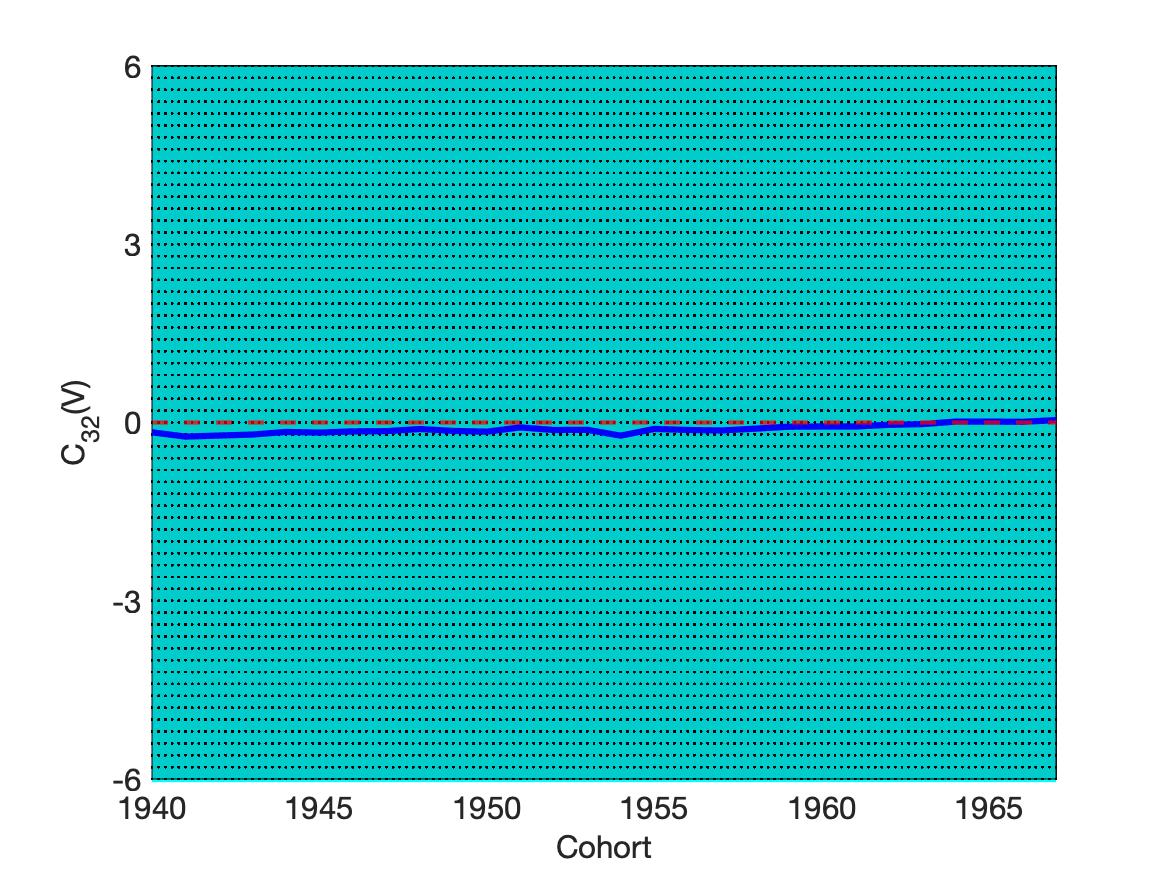}
             \subcaption{\hspace*{1.1em}}
    \end{subfigure}%
   \begin{subfigure}[t]{0.43\textwidth}
           \hspace{-1.3cm}
            \includegraphics[width=\textwidth, height=5cm]{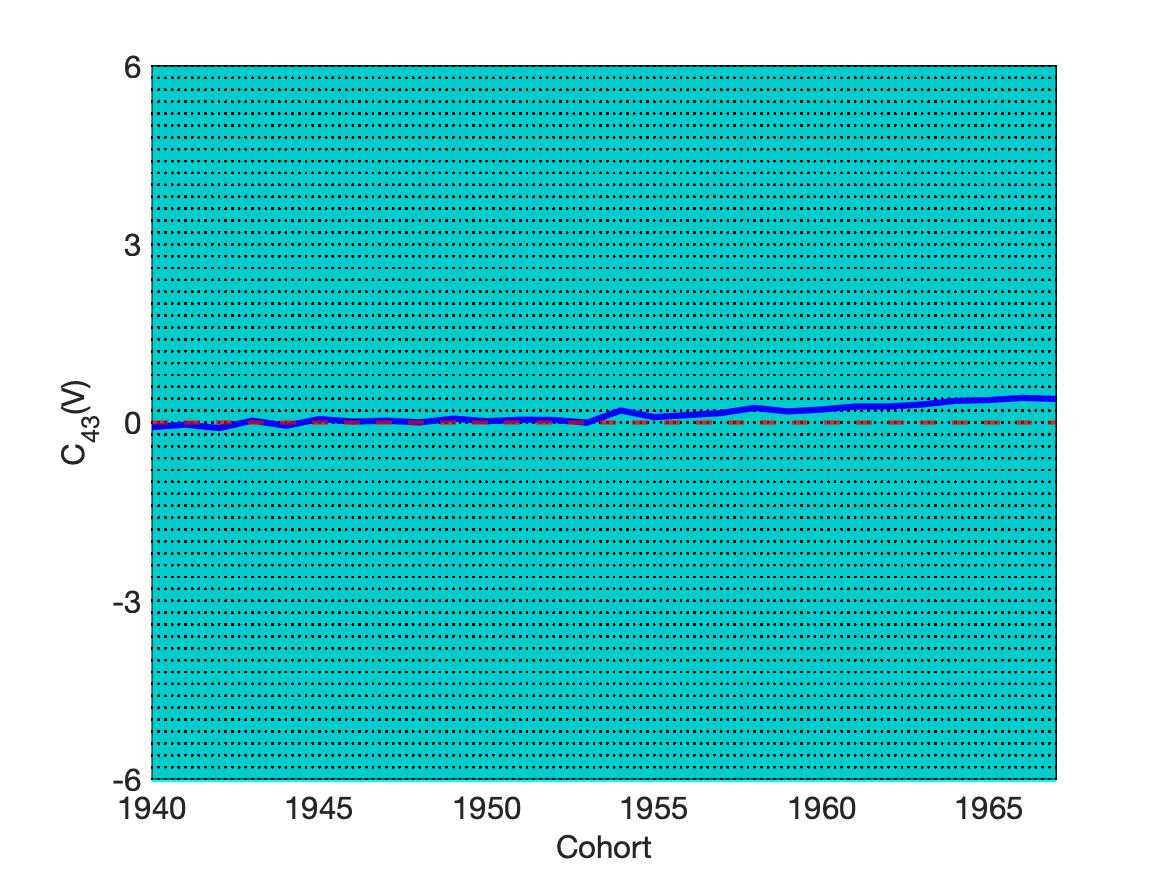}
             \subcaption{\hspace*{4.2em}}
    \end{subfigure}
    \end{adjustwidth}
\caption{The  blue region is the estimated identified set of $C(U)$ and $C(V)$ under specifications [\text{A}].  The dotted region is the estimated identified set of $C(U)$ and $C(V)$ under specifications [\text{B}]. By construction, the dotted region is contained in (or is equal to) the blue region. The dark blue line represents the estimates of $C(U)$ and $C(V)$ under the Logit  assumption. }
\label{C_ours}
\end{figure}

{Figure \MYref{C_ours} reports our estimates of the identified sets of $C(U)$ and $C(V)$  under specifications [A] (blue region) and [B] (dotted region).  Observe that we obtain unbounded intervals in almost every cohort. This is particularly true for the women's side, where the estimates of $C_{21}(V)$, $C_{32}(V)$, and $C_{43}(V)$ remain constantly unbounded above and below.\footnote{We obtain the same results when distinguishing between  CG and CG+. In particular, the estimates of $C_{21}(V)$, $C_{32}(V)$,  $C_{43}(V)$, and $C_{54}(V)$ remain constantly unbounded above and below.} In turn, the estimates of the marital education premia will be unbounded as well, and nothing can be said about their evolution over time.  As earlier, this indicates that the previous evidence on increasing marital college premium  based on the Logit  1to1TU model is a consequence of the Logit assumption.}  No evidence of an increase in
the marital education premia  has also been recently found by \hyperlink{CC}{Christensen and Connault (2022)} using a different methodology.


\section{Conclusions}
\label{conclusions}
This paper investigates the identifying power of the 1to1TU model for the systematic match surplus and related policy-relevant quantities when no parametric distributional assumptions are imposed on the unobserved heterogeneity. We conclude our analysis by highlighting three main findings.
First, we formally show that  the 1to1TU model  contains no   information about the  systematic match surplus without restricting the distribution of the unobserved heterogeneity.  Second, we propose a computational approach for constructing the identified set of the systematic match surplus that is based on principles of linear programming and works  under various classes of nonparametric distributional assumptions on the unobserved heterogeneity. Third, we use our methodology to re-examine some relevant questions in the empirical literature on the marriage market, which have been  previously studied  under the Logit assumption.  {Our estimates show that, without parametric distributional assumptions, the 1to1TU model is inconclusive about the evolution of  educational   sorting and marital education premia across cohorts. Therefore, most of the previous evidence on increasing positive educational sorting   and marital college premium is likely to be driven by the Logit assumption.
Our paper illustrates the usefulness of partial identification approaches in testing the robustness of empirical results based on strong parametric assumptions.}

\newpage

\newpage

 \begin{table}[!htbp]
\centering
\begin{adjustwidth}{2.5cm}{}
\scalebox{.7}{\begin{tabular}{cccccc}
\toprule
Assumptions                                        &&&& & \\ 
on unobservables  &Wife $\rightarrow$ & $1$ &$2$  &$3$ &$4$\\ 
\midrule
                & Husband $\downarrow$  & \multicolumn{4}{c}{Early cohorts}\\
                \cmidrule(l{3pt}r{3pt}){3-6}
Logit &  \multirow{3}{*}{$1$}   & $-2.06$&$-3.07$&$-5.22$&$-8.55$\\
$[\text{A}]$ &   & $-2.06$&$[-3.07, +\infty)$&$(-\infty ,+\infty)$&$(-\infty, +\infty)$\\
$[\text{B}] $& & $-2.06$&$[-3.07, +\infty)$&$(-\infty , +\infty)$&$(-\infty, +\infty)$\\
\\
Logit &  \multirow{3}{*}{$2$}& $-3.73$&$-1.35$&$-3.4$&$-5.76$\\
$[\text{A}] $&   &$[-5.3,-3.48]$&$[-3.79, +\infty)$&$(-\infty , +\infty)$&$(-\infty, +\infty)$\\
$[\text{B}]$ & &$[-5.3,-3.48]$&$[-3.63, +\infty)$&$(-\infty , +\infty)$&$(-\infty, +\infty)$\\
\\    
Logit &  \multirow{3}{*}{$3$} &$-5.29$&$-2.47$&$-2.12$&$-4.32$\\
$[\text{A}]$ &   &$(-\infty,-4.69]$&$(-\infty , +\infty)$&$(-\infty , +\infty)$&$(-\infty, +\infty)$\\
$[\text{B}]$ & &$(-\infty,-4.69]$&$(-\infty , +\infty)$&$(-\infty , +\infty)$&$(-\infty, +\infty)$\\
\\
Logit &  \multirow{3}{*}{$4$}& -8.01&-4&-2.46&-1.11\\
$[\text{A}]$ &   &$(-\infty ,-6.59]$&$(-\infty , +\infty)$&$(-\infty , +\infty)$&$[-5.74, +\infty)$\\
$[\text{B}]$ & &$(-\infty ,-6.59]$&$(-\infty , +\infty)$&$(-\infty , +\infty)$&$[-4.4, +\infty)$\\
\\\\
                & Husband $\downarrow$  & \multicolumn{4}{c}{Late cohorts}\\
                 \cmidrule(l{3pt}r{3pt}){3-6}
Logit &  \multirow{3}{*}{$1$}&$-3.34$&$-4.41$&$-5.61$&$-8.75$\\
$[\text{A}]$ &   &$-3.34 $&$[-4.56, -2.29]$&$(-\infty ,-4.6]$&$(-\infty, -6.79]$\\
$[\text{B}]$ & &$-3.34 $&$[-4.56, -3.65]$&$(-\infty ,-5.05]$&$(-\infty, -7.26]$\\
\\   
Logit &  \multirow{3}{*}{$2$}&$-5.1$&$-2.14$&$-3.18$&$-4.95$\\
$[\text{A}] $&   &$(-\infty,-5.1]$&$[-5.02, +\infty)$&$[-11.58,1.38]$&$[-13.46,2.67]$\\
$[\text{B}]$ & &$(-\infty,-5.1]$&$[-5.02, +\infty)$&$[-11.58,1.38]$&$[-13.46,2.54]$\\
\\
Logit &  \multirow{3}{*}{$3$}&$-6.98$&$-3.61$&$-2.3$&$-3.52$\\
$[\text{A}]$ &   &$(-\infty, -6.27]$&$(-\infty, +\infty)$&$[-5.48 , +\infty)$&$(-\infty, +\infty)$\\
$[\text{B}] $& &$(-\infty, -6.27]$&$(-\infty, +\infty)$&$[-5.48 , +\infty)$&$(-\infty, +\infty)$\\
\\
Logit &  \multirow{3}{*}{$4$}&$-9.13$&$-5.52$&$-3.49$&$-1.07$\\
$[\text{A}]$ &   &$(-\infty, -7.92]$&$(-\infty, +\infty)$&$(-\infty, +\infty)$&$[-6.16 , +\infty)$\\
$[\text{B}]$ & &$(-\infty, -7.92]$&$(-\infty, +\infty)$&$(-\infty, +\infty)$&$[-2.16 , +\infty)$\\
\\\\
                & Husband $\downarrow$  & \multicolumn{4}{c}{Change}\\
                 \cmidrule(l{3pt}r{3pt}){3-6}
Logit &  \multirow{3}{*}{$1$}& $-1.28$&$-1.34$&$-0.39$&$-0.2$\\
$[\text{A}]$ &   &$-1.28$&$(-\infty, -0.66]$&$(-\infty , +\infty)$&$(-\infty, +\infty)$\\
$[\text{B}] $& &$-1.28$&$(-\infty, -0.7]$&$(-\infty, +\infty)$&$(-\infty, +\infty)$\\
\\ 
Logit &  \multirow{3}{*}{$2$} &$-1.37$&$-0.78$&$0.22$&$0.81$\\
$[\text{A}]$ &   &$(-\infty, 0.1]$&$(-\infty , +\infty)$&$(-\infty, +\infty)$&$(-\infty, +\infty)$\\
$[\text{B}]$ & &$(-\infty, 0.1]$&$(-\infty, +\infty)$&$(-\infty , +\infty)$&$(-\infty , +\infty)$\\
\\     
Logit &  \multirow{3}{*}{$3$}&$-1.69$&$-1.14$&$-0.18$&$0.8$\\
$[\text{A}]$ &   &$(-\infty , +\infty)$&$(-\infty , +\infty)$&$(-\infty, +\infty)$&$(-\infty, +\infty)$\\
$[\text{B}] $& &$(-\infty, +\infty)$&$(-\infty , +\infty)$&$(-\infty, +\infty)$&$(-\infty , +\infty)$\\
\\
Logit &  \multirow{4}{*}{$4$}&$-1.12$&$-1.52$&$-1.03$&$0.04$\\
$[\text{A}]$ &   &$(-\infty, +\infty)$&$(-\infty, +\infty)$&$(-\infty, +\infty)$&$(-\infty , +\infty)$\\
$[\text{B}]$ & &$(-\infty , +\infty)$&$(-\infty , +\infty)$&$(-\infty, +\infty)$&$(-\infty, +\infty)$\\
\bottomrule
\end{tabular}}
 \end{adjustwidth}
 \caption{{\footnotesize The first section of the table reports the projections of the estimated identified sets of $\Phi$, averaged early cohorts, under specifications [A] and [B]. The second section of the table reports the  projections of the estimated identified sets of  $\Phi$, averaged over late cohorts, under specifications [A] and [B]. The last section of the table reports the change in estimates between early  and late cohorts. Some intervals are singleton because of the scale normalisations imposed (see Appendix \MYref{simulations}). The average estimates of $\Phi$ under the Logit  assumption are also included. }}
\label{Tab6}
\end{table}

\newpage
\begin{center}
\section*{References}
\end{center}

\begin{description}

\item \hypertarget{Abbott}{}  Abbott, B., G. Gallipoli, C. Meghir, and G. L. Violante (2019): ``Education Policy and Intergenerational Transfers in Equilibrium,'' {\it Journal of Political Economy}, 127(6), 2569--2624.


\item \hypertarget{Akkus}{} Akkus, O., J.A. Cookson, and A. Horta\c{c}su (2016): ``The Determinants of Bank Mergers: A Revealed Preference Analysis,'' {\it Management Science}, 62(8), 2241--2258.


\item \hypertarget{AS}{ } Andrews, D.W.K., and G. Soares (2010): ``Inference for Parameters Defined by Moment Inequalities Using Generalized Moment Selection,'' {\itshape Econometrica}, 78(1), 119--157.

\item \hypertarget{Baccara}{} Baccara, M., A. \.{I}mrohoro\u{g}lu, A.J. Wilson, and L. Yariv (2012): ``A Field Study on Matching with Network Externalities,'' {\it American Economic Review}, 102(5), 1773--1804.

\item \hypertarget{Banal}{} Banal-Esta\~{n}ol, A., I. Macho-Stadler, D. P\'{e}rez-Castrillo (2018): ``Endogenous Matching in University-Industry Collaboration: Theory and Empirical Evidence from the United Kingdom,'' {\it Management Science}, 64(4), 1591--1608.

\item \hypertarget{becker}{ } Becker, G.S. (1973):  ``A Theory of Marriage: Part I,'' {\it Journal of Political Economy}, 81(4), 813--846.

\item \hypertarget{Bertsimas}{} Bertsimas, D., and J.N. Tsitsiklis (1997): {\it Introduction to Linear Optimisation}, Athena Scientific, Belmont, Massachusetts.

\item \hypertarget{Bisin}{ } Bisin, A., and G. Tura (2020): ``Marriage, Fertility, and Cultural Integration in Italy,'' NBER Working Paper No. 26303.


\item \hypertarget{Botticini}{ } Botticini, M, and A. Siow (2011): ``Are There Increasing Returns to Scale in Marriage Markets?,'' Working Paper.
 
\item \hypertarget{Brandt}{} Brandt, L., A. Siow, and C. Vogel (2016): ``Large Demographic Shocks and Small Changes in the Marriage Market,'' {\it Journal of the European Economic Association}, 14(6), 1437--1468.

\item \hypertarget{Bruze}{} Bruze, G., M.  Svarer, and Y. Weiss (2015): ``The Dynamics of Marriage and Divorce,'' {\it Journal of Labor Economics}, 33,(1) 123--170.

\item \hypertarget{Chen}{} Chen, L. (2017): ``Compensation, Moral Hazard, and Talent Misallocation in the Market for CEOs,'' SSRN Working Paper.

\item \hypertarget{CHT} Chernozhukov, V., H. Hong, and E. Tamer (2007): ``Estimation and Confidence Regions for Parameter Sets in Econometric Models,'' {\it Econometrica}, 75(5), 1243--1284.

\item \hypertarget{Chiappori3}{ } Chiappori, P.-A. (2017): {\itshape Matching with Transfers: 
The Economics of Love and Marriage}, Princeton University Press.

\item \hypertarget{Chiappori6}{ } Chiappori, P.-A., M. Costa-Dias, C. Meghir (2018): ``The Marriage Market, Labor Supply, and Education Choice,'' {\it Journal of Political Economy}, 126(S1), S26-S72. 

\item \hypertarget{Chiappori_Meghir2}{} Chiappori, P.-A., M. Costa-Dias, S. Crossman, and C. Meghir (2020): ``Changes in Assortative Matching and Inequality in Income: Evidence for the UK,'' {\it Fiscal Studies}, 41(1), 39--63. 

\item \hypertarget{Chiappori_Meghir}{ } Chiappori, P.-A., M. Costa-Dias, C. Meghir (2020): ``Changes in Assortative Matching: Theory and Evidence for the US,'' NBER Working Papers 26932. 

\item \hypertarget{Chiappori_Meghir3}{ } Chiappori, P.-A., M. Costa-Dias, C. Meghir (2021): `` The Measuring of Assortativeness in Marriage: A Comment,'' Cowles Foundation Discussion Paper  2316.

\item \hypertarget{Chiappori5}{ } Chiappori, P.-A., M. Iyigun, Y. Weiss (2009): ``Investment in Schooling and the Marriage Market,'' {\it American Economic Review}, 99(5), 1689--1713.

\item \hypertarget{Nesheim}{} Chiappori, P.-A., R.J. McCann, and L.P. Nesheim (2010): ``Hedonic Price Equilibria,
Stable Matching, and Optimal Transport: Equivalence, Topology, and Uniqueness,'' {\it Economic Theory}, 42(2), 317--354.

\item \hypertarget{Chiappori_McCann_Pass}{} Chiappori, P.-A., R.J. McCann, and B. Pass (2020): ``Multidimensional matching: theory and empirics,'' Working Paper.


\item \hypertarget{Chiappori}{ } Chiappori, P.-A., and B. Salani\'e (2016): ``The Econometrics of Matching Models,'' {\itshape Journal of Economic Literature}, 54(3), 832--861.

\item \hypertarget{CSW}{ } Chiappori, P.-A., B. Salani\'e, and Y. Weiss (2017): ``Partner Choice, Investment in Children, and the Marital College Premium,'' {\itshape American Economic Review}, 107(8), 2109--2167.

\item \hypertarget{Chiappori_UV}{ } Chiappori, P.-A., B. Salani\'e, A. Tillman, and Y. Weiss (2008): ``Assortative Matching on the Marriage Market:
A Structural Investigation,'' slides available at \url{http://adres.ens.fr/IMG/pdf/09022009.pdf}.

\item \hypertarget{Choo}{ } Choo, E. (2015): ``Dynamic Marriage Matching: An Empirical Framework,'' {\it Econometrica}, 83(4), 1373--1423. 

\item \hypertarget{Choo2}{ } Choo, E., and S. Seitz (2013): ``The Collective Marriage Matching Model: Identification, Estimation, and Testing,'' {\it Structural Econometric Models (Advances in Econometrics)}, 31, 291--336.

\item \hypertarget{Choo_Siow}{ } Choo, E., and A. Siow (2006): ``Who Marries Whom and Why,'' {\itshape Journal of Political Economy}, 114(1), 175--201.

\item \hypertarget{CC}{} Christensen, T., and B. Connault (2022): ``Counterfactual Sensitivity and Robustness,''  arXiv:1904.00989.


\item \hypertarget{Ciscato}{ } Ciscato, E., A. Galichon, and M. Gouss\'e (2020): ``Like Attract Like? A Structural Comparison of Homogamy Across Same-Sex and Different-Sex Households,'' {\it Journal of Political Eocnomy}, 128(2), 740--781.

\item \hypertarget{Ciscato_Weber}{} Ciscato, E., and S. Weber (2020) ``The Role of Evolving Marital Preferences in Growing Income Inequality,'' {\it Journal of Population Economics}, 33(1), 307--347.


\item \hypertarget{Dupuy_Galichon}{ } Dupuy, A., and A. Galichon (2014): ``Personality traits and the marriage market,'' {\itshape Journal of Political Economy}, 122(6), 1271--1319.

\item \hypertarget{Dupuy_Weber}{} Dupuy, A., and S. Weber (2020): ``Marital Patterns and Income Inequality,'' SSRN Working Paper.

\item \hypertarget{Eika}{} Eika, L., M. Mogstad, and B. Zafar (2019): ``Educational Assortative Mating and Household Income Inequality,'' {\it Journal of Political Economy}, 127(6), 2795--2835.


\item \hypertarget{Fernandez1}{} Fern\'andez, R., N. Guner, and J. Knowles (2005): ``Love and Money: A Theoretical and Empirical Analysis of Household Sorting and Inequality,'' {\it Quarterly Journal of Economics}, 120 (1), 273--344.

\item \hypertarget{Fernandez2}{} Fern\'andez, R. and E. Rogerson (2001): ``Sorting and Long-Run Inequality,'' {\it Quarterly Journal of Economics}, 116 (4), 1305--1341.

\item \hypertarget{Fox_2010}{ } Fox, J.T. (2010): ``Identification in Matching Games,'' {\itshape Quantitative Economics}, 1(2), 203--254.

\item \hypertarget{Fox_2018}{ } Fox, J.T. (2018): ``Estimating Matching Games with Transfers,'' {\itshape Quantitative Economics}, 9(1), 1--38.

\item \hypertarget{Fox_Yang_Hsu}{ } Fox, J.T., C. Yang, and D.H. Hsu (2018): ``Unobserved Heterogeneity in Matching Games,'' {\itshape Journal of Political Economy}, 126(4), 1339--1373.

\item \hypertarget{Galichon_Kominers_Weber}{} Galichon, A., S.D. Kominers, and S. Weber (2019): ``Costly Concessions: An Empirical Framework for Matching with Imperfectly Transferable Utility,'' {\it Journal of Political Economy}, 127(6), 2875--2925.

\item \hypertarget{Galichon_Salanie2}{ } Galichon, A., and B. Salani\'e (2019):  ``IIA in Separable Matching Markets,” Columbia University mimeo.

\item \hypertarget{Galichon_Salanie}{ } Galichon, A., and B. Salani\'e (2021):  ``Cupid’s Invisible Hand: Social Surplus and Identification in Matching Models,'' forthcoming in the {\it Review of Economic Studies}.

\item \hypertarget{Gayle}{ } Gayle, G.-L., and A. Shephard (2019): ``Optimal Taxation, Marriage, Home Production, and Family Labor Supply,'' {\it Econometrica}, 87(1), 291--326.

\item \hypertarget{Graham_2011}{} Graham, B. (2011): ``Econometric Methods for the Analysis of Assignment Problems in the Presence of Complementarity and Social Spillovers,'' in {\it Handbook of Social Economics}, ed. by J. Benhabib, M.O. Jackson, A. Bisin, 1B, 965--1052.

\item \hypertarget{Graham}{ } Graham, B. (2013a): ``Comparative Static and Computational Methods for an Empirical One-To-One Transferable Utility Matching Model,'' {\it Advances in Econometrics: Structural Econometric Models}, 31(1), 151--179.

\item \hypertarget{Graham_errata}{} Graham, B. (2013b): ``Errata in Econometric Methods for the Analysis of Assignment Problems in the Presence of Complementarity and Social Spillovers,'' Unpublished.



\item \hypertarget{Greenwood1}{} Greenwood, J., N. Guner, and J.A. Knowles (2003): ``More on Marriage, Fertility, and the Distribution of Income,'' {\it International Economic Review}, 44(3), 827--862.

\item \hypertarget{Greenwood2}{} Greenwood, J., N. Guner, G. Kocharkov, and C. Santos (2014): ``Marry your Like: Assortative Mating and Income Inequality,'' {\it American Economic Review: Papers and Proceedings}, 104(5), 348--353.

\item \hypertarget{Gretsky}{} Gretsky, N.E., J.M. Ostroy, and W.R. Zame (1992): ``The Nonatomic Assignment Model,'' {\it Economic Theory}, 2(1), 103--127.

\item \hypertarget{HHK}{} Haile, P.A., A. Horta\c{c}su,  and G. Kosenok (2008): ``On the Empirical Content of Quantal Response Equilibrium,'' {\it American Economics Review}, 98(1), 180--200.

\item \hypertarget{Heckman}{} Heckman, J.J., and S. Mosso (2014): ``The Economics of Human Development and Social Mobility,'' {\it Annual Review of Economics}, 6, 689--733.


\item \hypertarget{Kremer}{} Kremer, M. (1997): ``How Much Does Sorting Increase Inequality,'' {\it Quarterly Journal of Economics}, 112 (1), 115--139.


\item \hypertarget{Liu}{} Liu, H. and J. Lu (2006) ``Measuring the Degree of Assortative Mating,'' {\it Economics Letters}, 92(3), 317--322.

\item \hypertarget{McFadden}{} McFadden, D. (1974): ``Conditional Logit Analysis of Qualitative Choice Behavior,'' in {\it Frontiers in Econometrics}, ed. by  Paul Zarembka,  Newark: Academic Press, 105--142.

 
 \item \hypertarget{Mindruda}{} Mindruda, D. (2013): ``Value Creation in University‐Firm Research Collaborations: A Matching Approach,'' {\it Strategic management journal}, 34(6), 644--665.

 \item \hypertarget{Mindruda2}{} Mindruda, D., M. Mohen, R. Agarwal (2016): ``A Two-sided Matching Approach for Partner Selection and Assessing Complementarities in Inter-firm Alliances,'' {\it Strategic Management Journal}, 37(1), 206--231.


\item \hypertarget{Mourifie}{} Mourifi\'e, I., and A. Siow (2021): ``The Cobb Douglas Marriage Matching Function: Marriage Matching with Peer and Scale Effects,''  {\it Journal of Labor Economics}, 39(S1), S239--S274.

\item \hypertarget{Nelsen}{} Nelsen, R. (2006): {\it An Introduction to Copulas}, Springer, New York.




 \item \hypertarget{Shapley_Shubik}{ }  Shapley, L., and M. Shubik (1972): ``The Assignment Game I: The Core'', {\itshape International Journal
of Game Theory}, 1(1), 111--130.

\item \hypertarget{Shen}{} Shen, J. (2019): “(Non-)Marital Assortative Mating and the Closing of the Gender Gap in Education,” Working Paper.

\item \hypertarget{Sinha}{ } Sinha, S. (2018): ``Identification in One-to-One Matching Models with Nonparametric Unobservables,'' TSE Working Paper 18-897.

 \item \hypertarget{Siow}{ }  Siow, A. (2015): ``Testing Becker’s Theory of Positive Assortative Matching'', {\itshape Journal of Labor Economics}, 33(2), 409-441.
 
  \item \hypertarget{Sklar1}{ }  Sklar, A. (1959): ``Fonctions de R\'epartition \'a $n$ Dimensions et Leurs Marges,'' {\it Publications de l’Institut Statistique de l’Universit\'e de Paris}, 8, 229--231.
  
    \item \hypertarget{Sklar2}{ } Sklar, A. (1996): ``Random Variables, Distribution Functions, and Copulas: A Personal Look Backward and Forward,'' {\it Institute of Mathematical Statistics Lecture Notes-Monograph Series}, 28, 1--14.


\item \hypertarget{Torgo}{ } Torgovitsky, A. (2019): ``Partial Identification by Extending Subdistributions,'' {\itshape Quantitative Economics}, 10(1), 105--144.

\item \hypertarget{Villani}{} Villani C., (2009): {\it Optimal Transport. Old and new.}, Grundlehren der Mathematischen Wissenschaften (Fundamental Principles of Mathematical Sciences), 338. Springer.

\end{description}

\newpage
\begin{appendix}
\counterwithin{figure}{section}
\counterwithin{table}{section}
\counterwithin{equation}{subsection}
\counterwithin{theorem}{section}
\counterwithin{corollary}{section}
\counterwithin{proposition}{section}
\counterwithin{lemma}{section}
\counterwithin{definition}{section}
\counterwithin{remarkex}{section}
\counterwithin{assumptionex}{section}

\counterwithin{equation}{subsection}
\section{Further details on Sections \MYref{first_step} and \MYref{second_step}}

\subsection{Characterisation of $\pmb{\mathcal{B}^c}$ for $\pmb{r=2}$}
\label{box}
When $r=2$ (hence, $d=3$),  recall that $\mathcal{B}$ is  the plane
$$
\mathcal{B}\equiv \{(b_1,b_2,b_3)\in \mathbb{R}^3: b_1=b_2+b_3\}.
$$
Given $(b_1,b_2, b_3)\in \mathcal{B}$, let $\mathcal{B}_{b_1,b_2,b_3}$ be a $3$-box of {\it any} of these forms: 
\vspace{-0.5cm}
\par\nobreak
{\footnotesize 
\begin{align} 
&(b_1,+\infty]\times [-\infty,b_2]\times [-\infty,b_3],\quad  && [-\infty, b_1)\times [b_2,+\infty]\times [b_3,+\infty],\nonumber \\
&(b_1,+\infty]\times [-\infty,b_2)\times [-\infty,b_3),\quad  && [-\infty, b_1)\times (b_2,+\infty]\times (b_3,+\infty],\nonumber \\
&(b_1,+\infty]\times [-\infty,b_2]\times [-\infty,b_3),\quad  &&[-\infty, b_1)\times [b_2,+\infty]\times (b_3,+\infty],\nonumber \\
&(b_1,+\infty]\times [-\infty,b_2)\times [-\infty,b_3],\quad  &&[-\infty, b_1)\times (b_2,+\infty]\times [b_3,+\infty],\nonumber \\
&[b_1,+\infty]\times [-\infty,b_2]\times [-\infty,b_3),\quad  &&[-\infty, b_1]\times [b_2,+\infty]\times (b_3,+\infty],\nonumber \\
&[b_1,+\infty]\times [-\infty,b_2)\times [-\infty,b_3],\quad  &&[-\infty, b_1]\times (b_2,+\infty]\times [b_3,+\infty],\nonumber \\
&[b_1,+\infty]\times [-\infty,b_2)\times [-\infty,b_3),\quad  &&[-\infty, b_1]\times (b_2,+\infty]\times (b_3,+\infty].\nonumber 
\end{align}}
Then, $\mathcal{B}^c=\cup_{(b_1,b_2,b_3)\in \mathcal{B}} \mathcal{B}_{b_1,b_2,b_3}$.

\subsection{A linear program (generic $\pmb{r}$)}
\label{deg_prop_general_section}
In this section, we generalise the discussion of Section \MYref{first_step} to any $r$. As in Section \MYref{first_step}, we  illustrate the result in the   case where   $\Delta \mathcal{F}^{\dagger}=\Delta \mathcal{F}$. Hence, (\MYref{step1_general}) becomes
\begin{equation}
\label{existence2_general}
\begin{aligned}
&\exists \text{ } \{\Delta F_x\}_{x\in \mathcal{X}}\in \Delta \mathcal{F} \text{ s.t. } \forall x \in \mathcal{X},\\
&p_{1|x}=\kappa(U, \Delta F_x,1),p_{2|x}=\kappa(U, \Delta F_x,2),...,p_{r|x}=\kappa(U, \Delta F_x,r),p_{0|x}=\kappa(U, \Delta F_x,0).
\end{aligned}
\end{equation}
It is straightforward to explicitly express  $\kappa$ as in (\MYref{existence2}), although notationally cumbersome. Once this is done, we can see that, for each $x\in \mathcal{X}$, (\MYref{existence2_general}) depends on the values of $\Delta F_x$ at a finite number of $d$-tuples. We collect such $d$-tuples in the sets $\mathcal{A}_{x,1,U},..., \mathcal{A}_{x,d,U}$, as in (\MYref{A_sets}), where  $\mathcal{A}_{x,1,U}$ collects the elements at which $\Delta F_x$  is evaluated  along the first dimension,  ..., $\mathcal{A}_{x,d,U}$ collects the elements at which $\Delta F_x$  is evaluated along the $d$-th dimension.   Further, we define $\mathcal{A}_{x,U}\equiv \times_{l=1}^d \mathcal{A}_{x,l,U}$.  Thus, the infinite-dimensional existence problem (\MYref{existence2_general}) is equivalent to verifying whether there exists a    finite-domain function $ \Delta \bar{F}_x:  \mathcal{A}_{x,U} \rightarrow [0,1]$ that satisfy the equations in (\MYref{existence2_general}) and that can be  extended to a proper CDF $\Delta F_x: \bar{\mathbb{R}}^d\rightarrow [0,1]$, for every $x\in \mathcal{X}$. That is, (\MYref{existence2_general}) is equivalent to
\begin{align}
& \exists \text{ }   \Delta \bar{F}_x:  \mathcal{A}_{x,U} \rightarrow [0,1] \text{ s.t. } \forall x \in \mathcal{X}, \nonumber\\
&p_{1|x}=\kappa(U, \Delta \bar{F}_x,1),   p_{2|x}= \kappa(U, \Delta \bar{F}_x,2), ...,  p_{r|x}=\kappa(U, \Delta \bar{F}_x,r), p_{0|x}=\kappa(U, \Delta \bar{F}_x,0), \label{LP_1_general}\\\
&\text{and $\{ \Delta \bar{F}_x\}_{x\in \mathcal{X}}$ can be extended to a proper family of conditional CDFs in $\Delta \mathcal{F}$ }. \label{extension_naive_general}  
\end{align}
Importantly, observe that (\MYref{LP_1_general}) is a collection of $r+1$ equations that are linear in $ \Delta \bar{F}_x$. Further, we  show below that (\MYref{extension_naive_general}) can be expressed as  a  finite collection of equations and inequalities that are also  linear in $ \Delta \bar{F}_x$. Therefore, we can  transform (\MYref{existence2_general}) into a linear program.
 
 We now explain how to write (\MYref{extension_naive_general})  as  a finite collection of linear equations and inequalities. It is clear that $ \Delta \bar{F}_x$ can be extended to a proper  CDF $\Delta F_x$  only if   
(\MYref{ext_statement}) holds. 
Consider first  the  case where   (\MYref{support_restriction}) is ignored. Then,  Lemma 2 of \hyperlink{Torgo}{Torgovitsky (2019)} shows that (\MYref{ext_statement}) is also  sufficient for extendibility. In particular, the defining properties of  CDFs are:\\
 (i) $ \Delta \bar{F}_x(a_1,..., A_d)=0$ for every $(a_1,..., a_d)\in\mathcal{A}_{x,U}$ that has at least one component equal to $-\infty$. That is,
 \begin{equation}
 \label{subdistribution_ext_0_general}
\begin{aligned}
& \Delta \bar{F}_x(-\infty,a_2,..., a_d)=0  && \forall (a_2,a_3,..., a_d)\in  \times_{l=2}^d \mathcal{A}_{x,l,U},\\
& \Delta \bar{F}_x(a_1,-\infty,a_3,..., a_d)=0 && \forall (a_1,a_3,..., a_d)\in  \times_{l\neq 2}^d \mathcal{A}_{x,l,U}, \\
& \vdots    \\
& \Delta \bar{F}_x(a_1,a_2,..., a_{d-1},-\infty)=0&&  \forall (a_1,a_2,..., a_{d-1})\in  \times_{l=1}^{d-1} \mathcal{A}_{x,l,U}.  \\
\end{aligned}
\end{equation}
(ii) $ \Delta \bar{F}_x(a_1...,a_d)=1$  when $a_l=+\infty$ for every $l\in\{1,..., d\}$. That is,
\begin{equation}
 \label{subdistribution_ext_1_general}
 \Delta \bar{F}_x(+\infty,..., +\infty)=1.
\end{equation}
(iii) $ \Delta \bar{F}_x$ is $d$-increasing. Formally, given a pair of   $d$-tuples, $ (a_1,...,a_d),(a_1',...,a_d')$ in $\mathcal{A}_{x,U}$ with $(a_1,...,a_d)\leq (a_1',...,a_d')$, let $\mbox{Vol}_{ \Delta \bar{F}_x}(\times_{l=1}^d[a_l,a_l'])$ denote the volume of the $d$-box $\times_{l=1}^d[a_l,a_l']$.  $ \Delta \bar{F}_x$ is called $d$-increasing if 
 \begin{equation}
  \label{prop_incr_general}
 \begin{aligned}
& \mbox{Vol}_{ \Delta \bar{F}_x}(\times_{l=1}^d[a_l,a_l']) \geq 0,\\
& \text{for every }  (a_1,..., a_d), (a_1',...,a_d')\in \mathcal{A}_{x,U},  \\
&  \text{s.t. } \hspace{1cm} (a_1,...,a_d)\leq (a_1',...,a_d').\footnotemark\\
 \end{aligned}
\end{equation}
\footnotetext{Take  $(a_1,...,a_d), (a_1',...,a_d')\in \mathcal{A}_{x,U}$ with $ (a_1,...,a_d)\leq (a_1',...,a_d')$. Then, $$\mbox{Vol}_{ \Delta \bar{F}_x}(\times_{l=1}^d[a_l,a_l']) \equiv \sum_{v\in \text{vert}((a_1,...,a_d), (a_1',...,a_d'))} \Delta \bar{F}_x(v)*\mbox{sgn}(v),$$ where  $\text{vert}((a_1,...,a_d), (a_1',...,a_d'))$  is the set of the box's vertices, $v\equiv (v_1,...,v_d)$ denotes a generic vertex, $\mbox{sign}(v)$ is equal to $1$ if $v_l=a_l$ for an even number of $l\in \{1,..., d\}$, and equal to $-1$ otherwise; 0 is considered even.}%
 Note that (\MYref{subdistribution_ext_0_general})-(\MYref{prop_incr_general}) constitute a   finite collection of  equations and inequalities that are linear in $ \Delta \bar{F}_x$.
Therefore, if (\MYref{support_restriction}) was absent, we could rewrite (\MYref{existence2_general}) as a linear program by  direct application of Lemma 2 of \hyperlink{Torgo}{Torgovitsky (2019)}.

Next, we discuss how to incorporate (\MYref{support_restriction}). As in Section \MYref{first_step}, note that assuming  $\lambda_{\Delta F_x}(\mathcal{B})=1$ is equivalent to assuming  $\lambda_{\Delta F_x}(\mathcal{B}^c)=0$ where $\mathcal{B}^c$ is the complement of $\mathcal{B}$ in $\bar{\mathbb{R}}^d$. Also observe that $\mathcal{B}^c$ can be written as the union of an infinite collection of  $d$-boxes. 
In fact, given $(\gamma,\delta)\in\mathbb{R}^2$, let $ t \in \mathcal{T} \equiv \{1,...,r-1\}$,  $p \in \mathcal{P}_t   \equiv \{t+1,...,r\}$,  and $q\in \mathcal{Q}_{t,p}$ where $\mathcal{Q}_{t,p} \equiv  \binom{r-t}{2}-(r-p) $ if $r-t\geq 2$ and $\mathcal{Q}_{t,p}\equiv d-(r-p)$ otherwise. Let $\mathcal{B}^{t,p,q}_{\gamma+\delta, \gamma,\delta}$ be a $d$-box of any of these forms: 
\vspace{-0.5cm}
\par\nobreak
{\scriptsize  
\begin{align} 
 \{(z_1,...,z_d)\in \mathbb{R}^d \text{: } z_t>\gamma+\delta, \text{ } z_p \leq \gamma, \text{ }z_q\leq \delta \},\quad \quad 
 \{(z_1,...,z_d)\in \mathbb{R}^d \text{: } z_t<\gamma+\delta, \text{ } z_p \geq \gamma, \text{ }z_q\geq \delta \},\nonumber\\
\{(z_1,...,z_d)\in \mathbb{R}^d \text{: } z_t>\gamma+\delta, \text{ } z_p < \gamma, \text{ }z_q < \delta \}, \quad \quad 
 \{(z_1,...,z_d)\in \mathbb{R}^d \text{: } z_t<\gamma+\delta, \text{ } z_p > \gamma, \text{ }z_q > \delta \},\nonumber\\
\{(z_1,...,z_d)\in \mathbb{R}^d \text{: } z_t>\gamma+\delta, \text{ } z_p \leq \gamma, \text{ }z_q < \delta \}, \quad \quad 
 \{(z_1,...,z_d)\in \mathbb{R}^d \text{: } z_t<\gamma+\delta, \text{ } z_p \geq \gamma, \text{ }z_q > \delta \},\nonumber\\
 \{(z_1,...,z_d)\in \mathbb{R}^d \text{: } z_t>\gamma+\delta, \text{ } z_p < \gamma, \text{ }z_q\leq \delta \}, \quad \quad 
  \{(z_1,...,z_d)\in \mathbb{R}^d \text{: } z_t<\gamma+\delta, \text{ } z_p > \gamma, \text{ }z_q\geq \delta \},\nonumber\\
 \{(z_1,...,z_d)\in \mathbb{R}^d \text{: } z_t\geq \gamma+\delta, \text{ } z_p \leq \gamma, \text{ }z_q <  \delta \}, \quad \quad 
  \{(z_1,...,z_d)\in \mathbb{R}^d \text{: } z_t\leq \gamma+\delta, \text{ } z_p \geq \gamma, \text{ }z_q > \delta \},\nonumber\\
\{(z_1,...,z_d)\in \mathbb{R}^d \text{: } z_t\geq \gamma+\delta, \text{ } z_p <\gamma, \text{ }z_q\leq \delta \}, \quad \quad 
 \{(z_1,...,z_d)\in \mathbb{R}^d \text{: } z_t\leq \gamma+\delta, \text{ } z_p > \gamma, \text{ }z_q\geq \delta \},\nonumber\\
\{(z_1,...,z_d)\in \mathbb{R}^d \text{: } z_t\geq \gamma+\delta, \text{ } z_p < \gamma, \text{ }z_q <  \delta \}, \quad \quad 
 \{(z_1,...,z_d)\in \mathbb{R}^d \text{: } z_t\leq \gamma+\delta, \text{ } z_p > \gamma, \text{ }z_q >\delta \}. \nonumber
\end{align}}
\hspace{-0.2cm}Then, $\mathcal{B}^c=\cup_{t\in \mathcal{T}, p\in \mathcal{P}_t, q\in \mathcal{Q}_{t,p}} \cup_{(\gamma,\delta)\in\mathbb{R}^2} \mathcal{B}^{t,p,q}_{\gamma+\delta, \gamma,\delta}$. In turn,  assuming $\lambda_{\Delta F_x}(\mathcal{B}^c)=0$ is equivalent to assuming 
\begin{equation}
\label{deg_general}
\begin{aligned}
&\mbox{Vol}_{\Delta F_x}(\mathcal{B}^{t,p,q}_{\gamma+\delta, \gamma,\delta})=0 \quad \text{$\forall (\gamma,\delta)\in\mathbb{R}^2$,  $t \in \mathcal{T}$, $p\in \mathcal{P}_t$, and $q\in \mathcal{Q}_{t,p}$. }\\
\end{aligned}
\end{equation}

Therefore, it is clear that  $ \Delta \bar{F}_x$ can be extended to a  proper CDF $\Delta F_x$  satisfying (\MYref{deg_general}) only if the increasingness condition  (\MYref{prop_incr_general}) holds as  equality for every  pair of $d$-tuples, $ (a_1,..., a_d),(a_1',..., a_d')$ in $\mathcal{A}_{x,U}$ with $(a_1,...,a_d)< (a_1',...,a_d')$, such that the box $\times_{l=1}^d[a_l,a_l']$  is  contained in a box $\mathcal{B}^{t,p,q}_{\gamma+\delta, \gamma,\delta}$ for some $(\gamma,\delta)\in\mathbb{R}^2$, $ t \in  \mathcal{T}$,  $p\in \mathcal{P}_t$,  and $q\in \mathcal{Q}_{t,p}$. Proposition \MYref{deg_prop_general} proves that such a condition is also sufficient for extendibility when combined with (\MYref{subdistribution_ext_0_general})-(\MYref{prop_incr_general}).

\begin{prop}{\normalfont({\itshape Extendibility})}
\label{deg_prop_general}
Given $U\in \mathcal{U}$ and $x\in \mathcal{X}$, let $\mathcal{A}_{x,U}\equiv \times_{l=1}^d \mathcal{A}_{x,l,U} $, where $ \mathcal{A}_{x,l,U}$ is a finite subset of $\bar{\mathbb{R}}$ and contains $\{+\infty,-\infty\}$ for each $l\in \{1,...,d\}$. Let $ \Delta \bar{F}_x:  \mathcal{A}_{x,U} \rightarrow[0,1]$ be a function satisfying 
(\MYref{subdistribution_ext_0_general})-(\MYref{prop_incr_general}), and   
\begin{align}
\label{supp_constraint_general}
&&&\mbox{Vol}_{ \Delta \bar{F}_x}( \times_{l=1}^d[a_l,a_l'] )=0,\\
 &&& \text{for every } (a_1,...,a_d), (a_1',...,a_d')\in \mathcal{A}_{x,U}, \nonumber\\
&&& \text{s.t. } \hspace{1cm} (a_1,...,a_d)< (a_1',...,a_d'), \nonumber\\
&&& \hspace{1.8cm}\text{and } \times_{l=1}^d[a_l,a_l'] \subset\mathcal{B}^{t,p,q}_{\gamma+\delta, \gamma,\delta}\text{ for some $(\gamma,\delta)\in\mathbb{R}^2$, $ t \in  \mathcal{T}$,  $p\in \mathcal{P}_t$,  and $q\in \mathcal{Q}_{t,p}$}. \nonumber 
\end{align}
Then, there exists a proper CDF $\Delta F_x: \bar{\mathbb{R}}^d\rightarrow [0,1]$  such that: (i)   $ \Delta \bar{F}_x$ can be extended to $\Delta F_x$, i.e.,  $\Delta F_x(a_1,..., a_d)=   \Delta \bar{F}_x(a_1,..., a_d)$ for each $(a_1,..., a_d)\in \mathcal{A}_{x,U}$; (ii) $ \lambda_{\Delta F_x}(\mathcal{B})=1$. 
\end{prop}

In addition, Lemma \MYref{deg_lemma} also applies to verify if a box $\times_{l=1}^d[a_l,a_l']$  is  contained in a box $\mathcal{B}^{t,p,q}_{\gamma+\delta, \gamma,\delta}$ for some $(\gamma,\delta)\in\mathbb{R}^2$, $ t \in  \mathcal{T}$,  $p\in \mathcal{P}_t$,  and $q\in \mathcal{Q}_{t,p}$

Lastly, as   in Section \MYref{first_step}, the above methodology remains valid under various classes of nonparametric restrictions on $\{\Delta F_x\}_{x\in \mathcal{X}}$, which can be simply imposed on  $\{ \Delta \bar{F}_x\}_{x\in \mathcal{X}}$ as linear constraints.

\subsection{Theorem 1 in Torgovitsky (2019)}
\label{Torgo_formal}

In this section, we provide a formal statement of Theorem 1 in \hyperlink{Torgo}{Torgovitsky (2019)} within our framework. 
 We refer the reader to Definitions 1-5, Lemmas 1-2, and Corollary 1 in \hyperlink{Torgo}{Torgovitsky (2019)},  which are the other key results and definitions used by Theorem 1. In what follows, $ \Delta F_x|_{\mathcal{C}}$ denotes the restriction of $\Delta F_x$ to a subset, $\mathcal{C}$, of its domain.

\paragraph{Assumption A, Torgovitsky (2019)}
$\Delta \mathcal{F}^{\dagger}$ satisfies the following properties: for each $\{\Delta F_x\}_{x\in \mathcal{X}}\in \Delta \mathcal{F}^{\dagger}$, it holds that 
\begin{enumerate}[1.]
\item $\Delta F_x(a)=\Delta F_{\tilde{x}}(a)$ $\forall x,\tilde{x}\in \{\mathcal{X}\cap \mathcal{X}^{\dagger}_{0,m_0}\}$, $\forall a\in \bar{\mathbb{R}}^d$, $\forall m_0\in \{1,...,M_0\}$, where each $\mathcal{X}^{\dagger}_{0,m_0}$ is a known (possibly empty) subset of $\mathcal{X}$.
\item $\Delta F_{x,l}(a)=\Delta F_{\tilde{x},l}(a)$ $\forall x,\tilde{x}\in \{\mathcal{X}\cap \mathcal{X}^{\dagger}_{l,m_l}\}$, $\forall a\in \bar{\mathbb{R}}$, $\forall m_l\in \{1,...,M_L\}$, $\forall l\in \{1,...,d\}$, where each $\mathcal{X}^{\dagger}_{l,m_l}$ is a known (possibly empty) subset of $\mathcal{X}$.
\item  $\{\Delta F_{x,l}\}_{x\in \mathcal{X}}\in \Delta \mathcal{F}^{\dagger}_l$ $\forall l\in \{1,...,d\}$, where $\Delta \mathcal{F}^{\dagger}_l$ is a known collection of families of one-dimensional conditional CDFs.
\item $\rho(U,\{\Delta F_x\}_{x\in \mathcal{X}})\geq 0$ for some known vector-valued function $\rho$, where the inequality is interpreted component wise.
\end{enumerate}

\noindent In Assumption A, Conditions 1 and 2 are independence restrictions on $\{\Delta F_x\}_{x\in \mathcal{X}}$  and $\{\Delta F_{x,l}\}_{x\in \mathcal{X}}$, respectively. Condition 3 requires $\{\Delta F_{x,l}\}_{x\in \mathcal{X}}$ to be extendible in the sense described below in Theorem 1. Condition 4 allows for miscellaneous restrictions, represented here by a function $\rho$ chosen by the researcher. Any of the Conditions 1-4 can be made non-restrictive by using specific choices of $\mathcal{X}^{\dagger}_{0,m_0}$, $\mathcal{X}^{\dagger}_{l,m_l}$, $\Delta \mathcal{F}^{\dagger}_l$, or $\rho$. The restrictions listed in Assumption \MYref{nonpar_assumption} of Section \MYref{first_step} can be written in terms of 1-4.

\paragraph{Condition U, Torgovitsky (2019)}
Suppose that $\Delta \mathcal{F}^{\dagger}$ satisfies Assumption A. A collection of sets, $\{\mathcal{A}_{x,U}\}_{x\in \mathcal{X}}$, satisfies Condition U if the following properties hold:
\begin{enumerate}
\item $\forall x \in \mathcal{X}$, $\mathcal{A}_{x,U}\equiv \times_{l=1}^d\mathcal{A}_{x,l,U}$, where $\mathcal{A}_{x,l,U}\subseteq \bar{\mathbb{R}}$ is closed and such that $\{+\infty, -\infty\}\subseteq \mathcal{A}_{x,l,U}$ $\forall l\in \{1,...,d\}$. 
\item There exists functions $\bar{\kappa}$ and $\bar{\rho}$ such that, $\forall \{\Delta F_x\}_{x\in \mathcal{X}}\in \Delta \mathcal{F}^\dagger$, 
$$
\begin{aligned}
& \kappa(U, \Delta F_x,y)=\bar{\kappa}(U,\Delta F_x|_{\mathcal{A}_{x,U}},y) \text{ }\forall (x,y) \in \mathcal{X}\times \mathcal{Y}_0,\\
& \rho(U,\{\Delta F_x\}_{x\in \mathcal{X}})= \bar{\rho}(U,\{ \Delta F_x|_{\mathcal{A}_{x,U}}\}_{x\in \mathcal{X}}).
\end{aligned}
$$
\item $\forall l\in \{1,...,d\}$, there exists a collection of families of conditional subsdistributions, $\Delta \bar{\mathcal{F}}^{\dagger}_l$, such that
$$
\begin{aligned}
& \text{$\Delta \bar{\mathcal{F}}^{\dagger}_l$ is extendible to $\Delta \mathcal{F}^{\dagger}_l$},\\
& \text{$\Delta {\mathcal{F}}^{\dagger}_l$ is reducible to $\Delta \bar{\mathcal{F}}^{\dagger}_l$},\\
& \text{$\forall \{\Delta \bar{F}_{x,l}\}_{x\in \mathcal{X}}\in \Delta \bar{\mathcal{F}}^{\dagger}_l$, every $\Delta \bar{F}_{x,l}$ has common domain $\mathcal{A}_{x,l,U}$.}
\end{aligned}
$$
\item $\mathcal{A}_{x,U}=\mathcal{A}_{\tilde{x},U}$  $\forall x,\tilde{x}\in \{\mathcal{X}\cap \mathcal{X}^{\dagger}_{0,m_0}\}$ and $\forall m_0\in \{1,...,M_0\}$.
\item $\mathcal{A}_{x,l,U}=\mathcal{A}_{\tilde{x},l,U}$  $\forall x,\tilde{x}\in \{\mathcal{X}\cap \mathcal{X}^{\dagger}_{l,m_l}\}$, $\forall m_l\in \{1,...,M_L\}$, and $\forall l\in \{1,...,d\}$.
\end{enumerate}

\paragraph{Theorem 1, Torgovitsky (2019)}
Suppose that $\Delta \mathcal{F}^{\dagger}$ can be represented as in Assumption A. Take any $U\in \mathcal{U}$. Let $\{\mathcal{A}_{x,U}\}_{x\in \mathcal{X}}$ be any collection of subsets of $\bar{\mathbb{R}}^{d}$ that satisfy Condition U. $U\in \mathcal{U}^*$ if and only if, for each $ x \in \mathcal{X}$, there exists $ \Delta \bar{F}_x:  \mathcal{A}_{x,U} \rightarrow [0,1]$ such that:
$$\begin{aligned}
&  p_{y|x}=\bar{\kappa}(U, \Delta \bar{F}_x,y) \text{ }\forall (x,y) \in \mathcal{X}\times \mathcal{Y}_0,\\
& \text{$ \Delta \bar{F}_x$ is a $d$-dimensional subdistribution $\forall x \in \mathcal{X}$}, \\
& \Delta \bar{F}_x(a)=\Delta \bar{F}_{\tilde{x}}(a)\text{ }\forall x,\tilde{x}\in \{\mathcal{X}\cap \mathcal{X}^{\dagger}_{0,m_0}\}, \text{ }\forall a\in \mathcal{A}_{x,U}, \text{ }\forall m_0\in \{1,...,M_0\},\\
& \Delta \bar{F}_{x,l}(a)=\Delta \bar{F}_{\tilde{x},l}(a)\text{ }\forall x,\tilde{x}\in \{\mathcal{X}\cap \mathcal{X}^{\dagger}_{l,m_l}\}, \text{ }\forall a\in \mathcal{A}_{x,l,U}, \text{ }\forall m_l\in \{1,...,M_L\}, \text{ }\forall l\in \{1,...,d\}, \\
& \{\Delta \bar{F}_{x,l}\}_{x\in \mathcal{X}}\in \Delta \bar{\mathcal{F}}^{\dagger}_l \text{ }\forall l\in \{1,...,d\}, \\
& \bar{\rho}(U, \{ \Delta \bar{F}_x\}_{x\in \mathcal{X}})\geq 0.
\end{aligned}
$$

\subsection{Example of a linear program}
\label{example_LP}
Let $r=2$ (hence, $d=3$). Fix  $U\in \mathcal{U}$. Impose, for instance, Assumption 5.2. Therefore,
$$
\begin{aligned}
& \mathcal{A}_{x,1,U}\equiv \{-U_{x1}, U_{x1}, 0,+\infty, -\infty\},\\
&  \mathcal{A}_{x,2,U}\equiv \{-U_{x2}, U_{x2}, 0,  +\infty, -\infty\},\\
& \mathcal{A}_{x,3,U}\equiv \{U_{x2}-U_{x1},-U_{x2}+U_{x1}, 0,+\infty, -\infty\},
\end{aligned}
$$
for every $ x \in \mathcal{X}$. 
By the arguments of Sections \MYref{first_step}, $U\in \mathcal{U}^*$  if and only if the following linear program admits a solution with respect to $ \Delta \bar{F}_x:\mathcal{A}_{x,U}\rightarrow [0,1]$ for every $x\in \mathcal{X}$:

{\footnotesize 
 \begin{align}
&p_{1|x}=1+ \Delta \bar{F}_x(-U_{x1}, +\infty, U_{x2}-U_{x1})- \Delta \bar{F}_x(+\infty, +\infty, U_{x2}-U_{x1})- \Delta \bar{F}_x(-U_{x1}, +\infty, +\infty), \label{eq0} \\
&p_{2|x}= \Delta \bar{F}_x(+\infty, +\infty, U_{x2}-U_{x1})- \Delta \bar{F}_x(+\infty, -U_{x2}, U_{x2}-U_{x1}), \label{eq1} \\
&p_{0|x}= \Delta \bar{F}_x(-U_{x1}, -U_{x2}, +\infty), \label{eq2}\\
& \Delta \bar{F}_x(-U_{x1}, +\infty, +\infty)=1- \Delta \bar{F}_x(-U_{x1}, +\infty, +\infty), \label{eq3}\\
& \Delta \bar{F}_x(+\infty, -U_{x2}, +\infty)=1- \Delta \bar{F}_x(+\infty, -U_{x2}, -\infty), \label{eq4}\\
& \Delta \bar{F}_x( +\infty, +\infty, U_{x2}-U_{x1})=1- \Delta \bar{F}_x( +\infty, +\infty, -U_{x2}+U_{x1}), \label{eq5}\\
& \Delta \bar{F}_x(0, +\infty, +\infty)=1/2,\label{eq6} \\
& \Delta \bar{F}_x(+\infty, 0, +\infty)=1/2,\label{eq7} \\
& \Delta \bar{F}_x( +\infty, +\infty, 0)=1/2, \label{eq8}\\
& \Delta \bar{F}_x(-\infty,a_2,a_3)=0 \text{ } \forall (a_2,a_3)\in \mathcal{A}_{x,2,U}\times \mathcal{A}_{x,3,U}, \label{eq19}\\
& \Delta \bar{F}_x(a_1,-\infty,a_3)=0 \text{ } \forall (a_1,a_3)\in \mathcal{A}_{x,1,U}\times \mathcal{A}_{x,3,U}, \label{eq20}\\
& \Delta \bar{F}_x(a_1,a_2,-\infty)=0 \text{ } \forall (a_1,a_2)\in \mathcal{A}_{x,1,U}\times \mathcal{A}_{x,2,U},\label{eq21}\\
& \Delta \bar{F}_x(+\infty, +\infty, +\infty)=1, \label{eq22}\\
&\mbox{Vol}_{ \Delta \bar{F}_x}([a_1,a_1']\times[a_2,a_2']\times[a_3,a'_3]) \geq 0  \text{ } \forall (a_1,a_2,a_3), (a_1',a_2',a_3')\in \mathcal{A}_{x,U}, \nonumber\\
&\hspace{6.2cm}\text{ s.t. } (a_1,a_2,a_3)\leq (a_1',a_2',a_3'), \label{eq24}\\
&\mbox{Vol}_{ \Delta \bar{F}_x}([a_1,a_1']\times[a_2,a_2']\times[a_3,a'_3]) = 0  \text{ }   \forall (a_1,a_2,a_3), (a_1',a_2',a_3')\in \mathcal{A}_{x,U}, \nonumber \\
&\hspace{6.2cm}\text{ s.t. } (a_1,a_2,a_3)<(a_1',a_2',a_3') \text{ and } a_1>a_2'+a_3' \text{ or } a'_1<a_2+a_3. \label{eq25}
\end{align} }

\noindent In the linear program above: (\MYref{eq0})-(\MYref{eq2}) match the model-implied conditional choice probabilities with the empirical ones; (\MYref{eq3})-(\MYref{eq8}) impose Assumption 5.2 on  $ \Delta \bar{F}_x$; (\MYref{eq19})-(\MYref{eq24}) ensure that   $ \Delta \bar{F}_x$ can be extended to a proper  CDF; (\MYref{eq25}) impose that   $ \Delta \bar{F}_x$ concentrates its mass within $\mathcal{B}$.

\counterwithin{equation}{subsection}
\section{Proofs}
\label{proofs_identification}

\subsection{Proof of Proposition \MYref{deg_prop}}
The proof    revisits the proof of the multidimensional case of Lemma 2 of \hyperlink{Torgo}{Torgovitsky (2019)}  to accommodate   the support restriction (\MYref{support_restriction}). It is organised in the following steps. In Step 0, we report some useful  definitions and results from copula theory.  In Step 1, we introduce a subcopula. In Steps 2 and 3, we    extend this subcopula to a proper copula. In Step 4, we construct a proper CDF from such a copula and show that it satisfies (\MYref{support_restriction}). The proof of Proposition \MYref{deg_prop_general} follows exactly the same steps, but becomes notationally more complicated.

\paragraph{Step 0} In this step, we report some definitions and results that are used below.  We follow the discussion in Appendix A of \hyperlink{Torgo}{Torgovitsky (2019)}.
See \hyperlink{Sklar1}{Sklar (1959}; \hyperlink{Sklar2}{1996)} and \hyperlink{Nelsen}{Nelsen (2006)} for more details.

\paragraph{Definition of subdistribution}  Let  $\mathcal{A}\equiv  \times_{l=1}^d \mathcal{A}_l$, where $\mathcal{A}_l\subseteq \bar{\mathbb{R}}$ and $\{-\infty,+\infty\}\subseteq \mathcal{A}_l$ for each $l\in \{1,..., d\}$. A $d$-dimensional subdistribution is a function $\bar{F}: \mathcal{A}\rightarrow [0,1]$ such that:
\begin{enumerate}[1.]
\item $\bar{F}(a_1,..., a_d)=0$ for every $(a_1,..., a_d)\in \mathcal{A}$ that has at least one component equal to $-\infty$.
\item For each $l\in \{1,..., d\}$, $\bar{F}(a_1,..., a_d)=a_l$ for every $(a_1,..., a_d)\in \mathcal{A}$ that has all components equal to $+\infty$.
\item $\bar{F}$ is $d$-increasing. That is, $\mbox{Vol}_{\bar{F}}( \times_{l=1}^d[a_l,a_l'])\geq 0$ for every pair of $d$-tuples, $(a_1,...,a_d), (a'_1, ..., a'_d)\in \mathcal{A}$ with $(a_1,...,a_d)\leq  (a'_1, ..., a'_d)$. 
\end{enumerate}
\vspace{0.3cm}
A  $d$-dimensional CDF is a subdistribution with domain $\bar{\mathbb{R}}^d$.

\paragraph{Definition of subcopula} Let  $\mathcal{T}\equiv  \times_{l=1}^d \mathcal{T}_l$, where $\mathcal{T}_l\subseteq [0,1]$ and $\{0,1\}\subseteq \mathcal{T}_l$ for each $l\in \{1,..., d\}$. A $d$-dimensional subcopula is a function $\bar{C}: \mathcal{T}\rightarrow [0,1]$ such that:
\begin{enumerate}[1.]
\item $\bar{C}(t_1,..., t_d)=0$ for every $(t_1,..., t_d)\in \mathcal{T}$ that has at least one component equal to $0$.
\item For each $l\in \{1,..., d\}$, $\bar{C}(t_1,..., t_d)=t_l$ for every $(t_1,..., t_d)\in \mathcal{T}$ that has all components, except the $l$-th one, equal to $1$.
\item $\bar{C}$ is $d$-increasing. That is, $\mbox{Vol}_{\bar{C}}( \times_{l=1}^d[t_l,t_l'])\geq 0$ for every pair of $d$-tuples, $(t_1,...,t_d), (t'_1, ..., t'_d)\in \mathcal{T}$ with $(t_1,...,t_d)\leq  (t'_1, ..., t'_d)$. 
\end{enumerate}
\vspace{0.3cm}
A  $d$-dimensional copula is a subcopula with domain $[0,1]^d$.

\paragraph{Sklar's Theorem} 1. Let $F: \bar{\mathbb{R}}^d\rightarrow [0,1]$ be a $d$-dimensional CDF with margins $F_l: \bar{\mathbb{R}}\rightarrow [0,1]$ for each $l\in \{1,...,d\}$. Then, there exists a $d$-dimensional copula $C: [0,1]^d\rightarrow [0,1]$ such that $F(a_1,...,a_d)=C(F_1(a_1), ..., F_d(a_d))$ for every $(a_1,...,a_d)\in \bar{\mathbb{R}}^d$. If $F_l$ is continuous on $\bar{\mathbb{R}}$ for every $l\in \{1,...,d\}$, then $C$ is unique. Otherwise, $C$ is uniquely determined on $\times _{l=1}^d\{F_l(a_l): a_l\in \bar{\mathbb{R}}\}$.

2. If $C: [0,1]^d\rightarrow [0,1]$ is a $d$-dimensional copula and $F_l: \bar{\mathbb{R}}\rightarrow [0,1]$ is a one-dimensional CDF for each $l\in \{1,...,d\}$, then the function $F: \bar{\mathbb{R}}^d\rightarrow [0,1]$ such that $F(a_1,...,a_d)= C(F_1(a_1), ..., F_d(a_d))$ for every $(a_1,...,a_d)\in \bar{\mathbb{R}}^d$ is a $d$-dimensional CDF with margins $F_l $ for each $l\in \{1,...,d\}$.\\

\vspace{0.1cm}
\noindent As for Lemma 2 of \hyperlink{Torgo}{Torgovitsky (2019)}, the proof of Proposition \MYref{deg_prop} uses the second part of Sklar’s Theorem. Further, it uses a lemma that is part of the proof of the first part of Sklar’s Theorem.

\paragraph{Sklar's Lemma}
Let $\bar{C}:\mathcal{T}\rightarrow [0,1]$ be a $d$-dimensional copula. Then, there exists a $d$-dimensional copula $C: [0,1]^d\rightarrow [0,1]$ such that $C(t_1,..., t_d)=\bar{C}(t_1,..., t_d)$ for every $(t_1,..., t_d)\in \mathcal{T}$.

\paragraph{Step 1}  
In this step, we introduce a subdistribution and a subcopula that will be useful below. Hereafter, we focus on the case where $r=2$ (hence, $d=3$), as considered by Proposition \MYref{deg_prop}.
 
 Given $U\in \mathcal{U}$ and $x\in \mathcal{X}$, let $\mathcal{A}_{x,U}\equiv  \mathcal{A}_{x,1,U}\times \mathcal{A}_{x,2,U}\times \mathcal{A}_{x,3,U} $, where $ \mathcal{A}_{x,l,U}$ is a finite subset of $\bar{\mathbb{R}}$ and contains $\{+\infty,-\infty\}$ for each $l\in \{1,2,3\}$. Let $ \Delta \bar{F}_x:\mathcal{A}_{x,U}\rightarrow [0,1]^3$ be a function satisfying 
(\MYref{subdistribution_ext_0})-(\MYref{prop_incr}) and (\MYref{supp_constraint}). As per the definition given in Step 0, $ \Delta \bar{F}_x$ is a $3$-dimensional subdistribution. 
For each $l\in \{1,2,3\}$, let $\Delta \bar{F}_{x,l}$ be the $l$-th margin of $ \Delta \bar{F}_x$. That is,
 $$
 \begin{aligned}
 &\Delta \bar{F}_{x,1}: \mathcal{A}_{x,1, U}\rightarrow [0,1] \quad \text{s.t.} \quad  \Delta \bar{F}_{x,1}(a_1)\equiv  \Delta \bar{F}_{x}(a_1,+\infty,+\infty) \quad \forall a_1\in \mathcal{A}_{x,1,U},\\
 &  \Delta \bar{F}_{x,2}: \mathcal{A}_{x,2, U}\rightarrow [0,1] \quad \text{s.t.} \quad  \Delta \bar{F}_{x,2}(a_2)\equiv  \Delta \bar{F}_{x}( +\infty,a_2,+\infty)\quad \forall a_2\in \mathcal{A}_{x,2,U},\\
  &  \Delta \bar{F}_{x,3}: \mathcal{A}_{x,3, U}\rightarrow [0,1] \quad \text{s.t.} \quad  \Delta \bar{F}_{x,3}(a_3)\equiv  \Delta \bar{F}_{x}( +\infty,+\infty,a_3)\quad \forall a_3\in \mathcal{A}_{x,3,U}.
   \end{aligned}
 $$
 By Lemma 1 of \hyperlink{Torgo}{Torgovitsky (2019)},   $\Delta \bar{F}_{x,l}$ is itself a one-dimensional subdistribution for each $l\in \{1,2,3\}$. 
 
 Next,  define the set
 $$
 \mathcal{T}\equiv \mathcal{T}_1\times \mathcal{T}_2\times \mathcal{T}_3\equiv \{\Delta \bar{F}_{x,1}(a_1): a_1\in \mathcal{A}_{x,1,U}\}\times  \{\Delta \bar{F}_{x,2}(a_2): a_2\in \mathcal{A}_{x,2,U}\}\times  \{\Delta \bar{F}_{x,3}(a_3): a_3\in \mathcal{A}_{x,3,U}\},
 $$
 and define the function  
 $$
 \bar{C}:\mathcal{T}\rightarrow [0,1] \quad \text{s.t.} \quad  \bar{C}( \Delta \bar{F}_{x,1}(a_1), \Delta \bar{F}_{x,2}(a_2), \Delta \bar{F}_{x,3}(a_3))= \Delta \bar{F}_x(a_1,a_2,a_3) \quad \forall (a_1,a_2,a_3)\in \mathcal{A}_{x,U}.
 $$
 As shown by \hyperlink{Torgo}{Torgovitsky (2019)}, $ \bar{C}$ is a $3$-dimensional subcopula with domain $\mathcal{T}$. 
 
 \paragraph{Step 2} In this step, we apply the one-dimensional case of Lemma 2 of \hyperlink{Torgo}{Torgovitsky (2019)} to extend every marginal subdistribution $\Delta \bar{F}_{x,l}$ to a proper  CDF.

Formally, Lemma 2 of \hyperlink{Torgo}{Torgovitsky (2019)}  shows that, for each $l\in \{1,2,3\}$, there exists 
 $$
 \Delta F_{x,l}:\bar{\mathbb{R}}\rightarrow [0,1]\quad \text{s.t.} \quad \Delta F_{x,l}(a_l)= \Delta \bar{F}_{x,l}(a_l)\quad \forall a_l\in  \mathcal{A}_{x,l,U}. 
 $$

\paragraph{Step 3}  In this step, we prove that there exists a proper $3$-dimensional copula $C : [0,1]^3\rightarrow [0,1]$  such that:
\begin{align}
&\text{$ \bar{C}$ can be extended to $C$, i.e.,  $C(t)=\bar{C}(t)$ for every $t\in \mathcal{T}$}, \label{copula_ext}\\
&\mbox{Vol}_C(\mathcal{B}_{\Delta F_{x,1}(b_1), \Delta F_{x,2}(b_2),\Delta F_{x,3}(b_3)})=0 \quad \forall (b_1,b_2,b_3)\in\mathcal{B}\equiv\{(b_1,b_2,b_3)\in \mathbb{R}^3: b_1=b_2+b_3\}, \label{deg_copula}
\end{align}
where $\mathcal{B}_{\Delta F_{x,1}(b_1), \Delta F_{x,2}(b_2),\Delta F_{x,3}(b_3)}$ is a $3$-box of any of these forms:
\vspace{-0.3cm}
\par\nobreak
{\scriptsize
\begin{equation}
\label{boxes}
\begin{aligned} 
&(\Delta F_{x,1}(b_1),+\infty]\times [-\infty,\Delta F_{x,2}(b_2)]\times [-\infty,\Delta F_{x,3}(b_3)],\quad  && [-\infty, \Delta F_{x,1}(b_1))\times [\Delta F_{x,2}(b_2),+\infty]\times [\Delta F_{x,3}(b_3),+\infty],  \\
&(\Delta F_{x,1}(b_1)+\infty]\times [-\infty,\Delta F_{x,2}(b_2))\times [-\infty,\Delta F_{x,3}(b_3)),\quad  && [-\infty, \Delta F_{x,1}(b_1))\times (\Delta F_{x,2}(b_2),+\infty]\times (\Delta F_{x,3}(b_3),+\infty],  \\
&(\Delta F_{x,1}(b_1),+\infty]\times [-\infty,\Delta F_{x,2}(b_2)]\times [-\infty,\Delta F_{x,3}(b_3)),\quad  &&[-\infty, \Delta F_{x,1}(b_1))\times [\Delta F_{x,2}(b_2),+\infty]\times (\Delta F_{x,3}(b_3),+\infty],  \\
&(\Delta F_{x,1}(b_1),+\infty]\times [-\infty,\Delta F_{x,2}(b_2))\times [-\infty,\Delta F_{x,3}(b_3)],\quad  &&[-\infty,\Delta F_{x,1}(b_1))\times (\Delta F_{x,2}(b_2),+\infty]\times [\Delta F_{x,3}(b_3),+\infty],  \\
&[\Delta F_{x,1}(b_1),+\infty]\times [-\infty,\Delta F_{x,2}(b_2)]\times [-\infty,\Delta F_{x,3}(b_3)),\quad  &&[-\infty, \Delta F_{x,1}(b_1)]\times [\Delta F_{x,2}(b_2),+\infty]\times (\Delta F_{x,3}(b_3),+\infty], \\
&[\Delta F_{x,1}(b_1),+\infty]\times [-\infty,\Delta F_{x,2}(b_2))\times [-\infty,\Delta F_{x,3}(b_3)],\quad  &&[-\infty, \Delta F_{x,1}(b_1)]\times (\Delta F_{x,2}(b_2),+\infty]\times [\Delta F_{x,3}(b_3),+\infty],  \\
&[\Delta F_{x,1}(b_1),+\infty]\times [-\infty,\Delta F_{x,2}(b_2))\times [-\infty,\Delta F_{x,3}(b_3)),\quad  &&[-\infty, \Delta F_{x,1}(b_1)]\times (\Delta F_{x,2}(b_2),+\infty]\times (\Delta F_{x,3}(b_3),+\infty]. 
\end{aligned}
\end{equation}}

By Sklar's Lemma,  there exists a proper $3$-dimensional copula $C^\dagger : [0,1]^3\rightarrow [0,1]$  such that $ \bar{C}$ can be extended to $C^\dagger$, i.e.,  $C^\dagger(t)=\bar{C}(t)$ for every $t\in \mathcal{T}$. 
In particular,  $C^\dagger$ can be constructed by multilinear interpolation, as shown by  \hyperlink{Nelsen}{Nelsen (2006)}  for $d=2$ and  \hyperlink{Sklar2}{Sklar (1996)} for a generic $d$.
In what follows, we show that $C^\dagger$ can be tweaked into another $3$-dimensional copula $C:[0,1]^3\rightarrow [0,1]$ such that both (\MYref{copula_ext}) and (\MYref{deg_copula}) hold. This adjustment of $C^\dagger$ is essentially a ``volume swapping/redistributing'' procedure that appropriately introduces ``holes'' in $C^\dagger$ so as to satisfy both (\MYref{copula_ext}) and (\MYref{deg_copula}).

We start with two remarks, (\MYref{union_boxes}) and (\MYref{supp_constraint_copula}). First,  consider any $ (t_1,t_2,t_3), (t_1',t_2',t_3')\in \mathcal{T}$ with $(t_1,t_2,t_3)<(t_1',t_2',t_3')$ and take the  $3$-box  $  [t_1,t_1']\times[t_2,t_2']\times[t_3,t'_3]$. We say that such a box is ``atomic'' if $t_1,t_1'$ are consecutive elements of $\mathcal{T}_1$, $t_2,t_2'$ are consecutive elements of $\mathcal{T}_2$, and $t_3,t_3'$ are consecutive elements of $\mathcal{T}_3$. Let  $\mathcal{A}_{\text{tomic}}$ be the (finite) collection of these atomic boxes. Let $\tilde{\mathcal{A}}_{\text{tomic}}$ be the collection of every  element of $\mathcal{A}_{\text{tomic}}$ that is contained in a box $\mathcal{B}_{\Delta F_{x,1}(b_1), \Delta F_{x,2}(b_2),\Delta F_{x,3}(b_3)}$ for some $(b_1,b_2,b_3)\in\mathcal{B}$. Let $\mathcal{A}^c_{\text{tomic}}$ be the complement of $\tilde{\mathcal{A}}_{\text{tomic}}$ in $\mathcal{A}_{\text{tomic}}$. 
By continuity arguments,   there is no element of $\mathcal{A}_{\text{tomic}}^c$ that is covered by the union of some boxes of the form (\MYref{boxes}).  That is,
$$
\nexists \quad \mathcal{H}\in \mathcal{A}^c_{\text{tomic}} \quad \text{s.t.} \quad \mathcal{H}\subseteq \mathcal{Z}\equiv \cup_{(b_1,b_2,b_3)\in\mathcal{B}} \mathcal{B}_{\Delta F_{x,1}(b_1), \Delta F_{x,2}(b_2),\Delta F_{x,3}(b_3)}.
$$
Therefore,
\begin{equation}
\label{union_boxes}
\mathcal{H} \cap \mathcal{Z}^c \neq \emptyset \quad \forall \mathcal{H}\in \mathcal{A}^c_{\text{tomic}}.
\end{equation}
Second,  by (\MYref{supp_constraint}), $C^\dagger$ is such that
$$
\begin{aligned}
& \mbox{Vol}_{C^\dagger}([t_1,t_1']\times[t_2,t_2']\times[t_3,t'_3])=0,\\
 &  \text{for every } (t_1,t_2,t_3), (t_1',t_2',t_3')\in \mathcal{T},  \\
&  \text{s.t. } \hspace{1cm} (t_1,t_2,t_3)< (t_1',t_2',t_3'),  \\
& \hspace{1.8cm}\text{and } [t_1,t_1']\times [t_2,t_2']\times [t_3,t'_3] \subset \mathcal{B}_{\Delta F_{x,1}(b_1), \Delta F_{x,2}(b_2),\Delta F_{x,3}(b_3)} \text{ for some $(b_1,b_2,b_3)\in\mathcal{B}$}.
\end{aligned} 
$$
Therefore, 
\begin{equation}
\label{supp_constraint_copula}
\mbox{Vol}_{C^\dagger}(\mathcal{H})=0 \quad \forall \mathcal{H}\in \tilde{\mathcal{A}}_{\text{tomic}}.
\end{equation}

Next, let $\lambda_{C^\dagger}$ be the probability measure associated with $C^\dagger$. Let $\lambda$ denote the Lebesgue measure.
Define the probability measure $\lambda_C$ such that
$$
\lambda_C(\mathcal{S}) = \sum_{\mathcal{H}\in  \mathcal{A}^c_{\text{tomic}}} \lambda_{C^\dagger}(\mathcal{S} \cap \mathcal{H}) \frac{\lambda(\mathcal{S} \cap \mathcal{H} \cap \mathcal{Z}^c)}{\lambda( \mathcal{H} \cap \mathcal{Z}^c) }   \quad \forall \mathcal{S}\subseteq [0,1]^3,
$$
where $\lambda(\mathcal{H} \cap \mathcal{Z}^c) \neq 0$ for every $\mathcal{H}\in \mathcal{A}^c_{\text{tomic}}$ by (\MYref{union_boxes}).  The   CDF associated with $\lambda_C$ is a $3$-dimensional copula $C : [0,1]^3\rightarrow [0,1]$.

 For every $\mathcal{H}\in \mathcal{A}_{\text{tomic}}$, observe that $C$ and $C^\dagger$ agree on each atomic box, i.e., $\mbox{Vol}_{C}(\mathcal{H})=\mbox{Vol}_{C^\dagger}(\mathcal{H})$. Indeed,
\begin{equation}
\label{first_prop}
\begin{aligned}
& \forall \mathcal{H}\in \mathcal{A}^c_{\text{tomic}} &&\mbox{Vol}_{C}(\mathcal{H})\equiv  \lambda_C(\mathcal{H})=\lambda_{C^\dagger}(  \mathcal{H}) \frac{\lambda( \mathcal{H} \cap \mathcal{Z}^c)}{\lambda( \mathcal{H} \cap \mathcal{Z}^c)} = \lambda_{C^\dagger}(  \mathcal{H}) \equiv \mbox{Vol}_{C^\dagger}(\mathcal{H}),\\
& \forall \mathcal{H}\in \tilde{\mathcal{A}}_{\text{tomic}} && \mbox{Vol}_{C}(\mathcal{H})\equiv  \lambda_C(\mathcal{H})=0  =   \mbox{Vol}_{C^\dagger}(\mathcal{H}), \\
\end{aligned}
\end{equation}
where the last equality in the second line uses (\MYref{supp_constraint_copula}).

Given that $C^\dagger$ is an extension of $\bar{C}$ by construction and $C$ is equal to $C^\dagger$ on each atomic box as highlighted by (\MYref{first_prop}), it follows that also $C$  is an extension of $\bar{C}$. Therefore, $C$ satisfies (\MYref{copula_ext}).
Further,
$$
\begin{aligned}
\lambda_C(\mathcal{Z}) & = \sum_{\mathcal{H}\in  \mathcal{A}^c_{\text{tomic}}} \lambda_{C^\dagger}(\mathcal{Z} \cap \mathcal{H}) \frac{\lambda(\mathcal{Z} \cap \mathcal{H} \cap \mathcal{Z}^c)}{\lambda( \mathcal{H} \cap \mathcal{Z}^c)  } =  \sum_{\mathcal{H}\in  \mathcal{A}^c_{\text{tomic}}} \lambda_{C^\dagger}(\mathcal{Z} \cap \mathcal{H}) \frac{0}{\lambda( \mathcal{H} \cap \mathcal{Z}^c)   }=0.
\end{aligned}
$$
Therefore, $C$ satisfies (\MYref{deg_copula}).


   \paragraph{Step 4} This step concludes the proof by constructing a proper $d$-dimensional CDF from $C$. 
   
   Define the function 
  $$
  \Delta F_x: \bar{\mathbb{R}}^d\rightarrow [0,1] \quad \text{s.t.} \quad  \Delta F_x(a_1,a_2,a_3)=C(\Delta F_{x,1}(a_1), \Delta F_{x,2}(a_2),\Delta F_{x,3}(a_3)) \quad \forall (a_1,a_2,a_3)\in  \bar{\mathbb{R}}^d.
  $$ 
By Sklar’s Theorem, $ \Delta F_x$ is a $d$-dimensional CDF. Following the proof of Lemma 2 of \hyperlink{Torgo}{Torgovitsky (2019)}, $\Delta F_x$ is an extension of $ \Delta \bar{F}_x$. In fact, for every $(a_1,a_2,a_3)\in \mathcal{A}_{x,U}$,
$$
\begin{aligned}
\Delta F_x(a_1,a_2,a_3)& \equiv C(\Delta F_{x,1}(a_1), \Delta F_{x,2}(a_2),\Delta F_{x,3}(a_3)),\\
& = C(\Delta \bar{F}_{x,1}(a_1), \Delta \bar{F}_{x,2}(a_2),\Delta \bar{F}_{x,3}(a_3)),\\
&= \bar{C}(\Delta \bar{F}_{x,1}(a_1), \Delta \bar{F}_{x,2}(a_2),\Delta \bar{F}_{x,3}(a_3)),\\
&=\Delta \bar{F}_x(a_1,a_2,a_3).
\end{aligned}
$$
Further, by (\MYref{deg_copula}), $\Delta F_x$ satisfies (\MYref{deg})  or, equivalently, (\MYref{support_restriction}).

\subsection{Proof of Lemma \MYref{deg_lemma}}
Given $U\in \mathcal{U}$ and $x\in \mathcal{X}$, let $\mathcal{A}_{x,U}\equiv  \mathcal{A}_{x,1,U}\times \mathcal{A}_{x,2,U}\times \mathcal{A}_{x,3,U} $, where $ \mathcal{A}_{x,l,U}$ is a finite subset of $\bar{\mathbb{R}}$ and contains $\{+\infty,-\infty\}$ for each $l\in \{1,2,3\}$. Take $ (a_1,a_2,a_3),(a_1',a_2',a_3')$ in $\mathcal{A}_{x,U}$ with $(a_1,a_2,a_3)< (a_1',a_2',a_3')$. Define the $3$-box $\mathcal{H}\equiv [a_1,a_1']\times [a_2,a_2']\times [a_3,a'_3]$. 

First, we show that if $\mathcal{H} \subset \mathcal{B}_{b_1,b_2,b_3}$ for some $(b_1,b_2,b_3)\in \mathcal{B}\equiv \{(b_1,b_2,b_3)\in \mathbb{R}^3: b_1=b_2+b_3\}$, then (\MYref{deg_lemma_condition}) holds. For instance, suppose 
$$
\mathcal{B}_{b_1,b_2,b_3}\equiv (b_1,+\infty]\times [-\infty,b_2]\times [-\infty,b_3].
$$
Since $\mathcal{H} \subset \mathcal{B}_{b_1,b_2,b_3}$, it holds that
\begin{align}
&b_1<a_1, \label{dir1_1}\\
& b_2\geq a_2', \label{dir1_2}\\
&b_3\geq a_3'. \label{dir1_3}
\end{align}
By (\MYref{dir1_2}) and (\MYref{dir1_3}), it holds that $b_1\geq a_2'+a_3'$. By combining this with (\MYref{dir1_1}), it holds that $a_2'+a_3'\leq b_1<a_1$.  Therefore, (\MYref{deg_lemma_condition}) is verified because $a_1>a_2'+a_3'$.
As another example, suppose 
$$
\mathcal{B}_{b_1,b_2,b_3}\equiv [-\infty, b_1]\times (b_2,+\infty]\times [b_3,+\infty].
$$
Since $\mathcal{H} \subset \mathcal{B}_{b_1,b_2,b_3}$, it holds that
\begin{align}
&b_1> a_1', \label{dir1_11}\\
& b_2< a_2, \label{dir1_21}\\
&b_3\leq  a_3. \label{dir1_31}
\end{align}
By (\MYref{dir1_21}) and (\MYref{dir1_31}), it holds that $b_1< a_2+a_3$. By combining this with (\MYref{dir1_11}),  it holds that $a_1'< b_1<a_2+a_3$. herefore, (\MYref{deg_lemma_condition}) is verified because $a_1'<a_2+a_3$.
We can proceed similarly for the other forms of $\mathcal{B}_{b_1,b_2,b_3}$.

Second,  we show that if (\MYref{deg_lemma_condition}) holds, then $\mathcal{H} \subset \mathcal{B}_{b_1,b_2,b_3}$ for some $(b_1,b_2,b_3)\in\mathcal{B}$. For instance, suppose that $a_1>a_2'+a_3'$. Consider the box
$$
\mathcal{B}_{b_1,b_2,b_3}\equiv (b_1, \infty]\times [-\infty,b_2]\times [-\infty, b_3],
$$
with $b_2\equiv a_2'$ and $b_3\equiv a_3'$.
Observe that $\mathcal{H}\subset \mathcal{B}_{b_1,b_2,b_3}$.
As another example, suppose that  $a_1'<a_2+a_3$. Consider the box
$$
\mathcal{B}_{b_1,b_2,b_3}\equiv [-\infty, b_1]\times (b_2,+\infty]\times [b_3,+\infty],
$$
with $b_2\equiv a_2$ and $b_3\equiv a_3$.
Observe that $\mathcal{H}\subset \mathcal{B}_{b_1,b_2,b_3}$. We can obtain similar conclusions by using the other forms of $\mathcal{B}_{b_1,b_2,b_3}$.

\subsection{Proof of Proposition \MYref{part2}}

The proof is organised in the following steps. In Step 0, we recall the notation introduced in Section \MYref{second_step} and introduce some new one.   In Step 1, we present the notion of an equivalence class for every $U\in \mathcal{U}$ and prove that if $\tilde{U},\hat{U}\in \mathcal{U}$ belong to the same equivalence class, then they induce the same set of solutions of the linear program. In Step 2, we show how such equivalence classes are related to the notion of $\pi$-ordering used in Proposition \MYref{part2}. In Step 3, we conclude. Remark \MYref{practice} explains how Proposition \MYref{part2} is used in practice.

For simplicity of exposition, we provide the proof of Proposition \MYref{part2} for the case $r=2$ (hence, $d=3$) and Assumption 5.2.  The proof for a generic case follows exactly the same steps, but becomes notationally more complicated. In the case considered, we have that $\mathcal{A}_{x,1,U}\equiv \{-U_{x1}, U_{x1},0,+\infty, -\infty\}$, $\mathcal{A}_{x,2,U}\equiv \{-U_{x2}, U_{x2},0,+\infty, -\infty\}$, and $\mathcal{A}_{x,3,U}\equiv \{U_{x2}-U_{x1},-U_{x2}+U_{x1}, 0, +\infty, -\infty\}$, for every $x\in \mathcal{X}$ and $U\in \mathcal{U}$.
Therefore, for any given $U\in \mathcal{U}$ and by following Section \MYref{first_step},  $U\in \mathcal{U}^*$ if and only if the linear program (\MYref{eq0})-(\MYref{eq25}) has  a solution with respect to $ \Delta \bar{F}_x: \mathcal{A}_{x,U}\rightarrow [0,1]$ for each $x\in \mathcal{X}$.

 \paragraph{Step 0}
In this step, we recall the notation introduced in Section \MYref{second_step} and introduce some new ones.   

 Fix $U\in \mathcal{U}$ and $x\in \mathcal{X}$.
In the example considered, $\mathcal{A}_{x,l,U}$ has cardinality $5$ for every $l\in \{1,2,3\}$. $\mathcal{A}_{x,U}$ has cardinality $5^3$. The image set of $ \Delta \bar{F}_x$, which we denote by $ \Delta \bar{F}_x(\mathcal{A}_{x,U})$, has cardinality $5^3$. {\it Importantly}, in all such sets,   repetitions of elements are kept. 

For every $l\in \{1,2,3\}$, fix an order of the $5$ elements in $\mathcal{A}_{x,l,U}$ and list them in   a $5\times 1$ vector, $\alpha_{x,l,U}$. For instance, 
$$
 \begin{aligned}
&\alpha_{x,1,U}\equiv (-U_{x1}, U_{x1},0,+\infty, -\infty)^\top,\\
&\alpha_{x,2,U}\equiv (-U_{x2}, U_{x2},0,+\infty, -\infty)^\top,\\
& \alpha_{x,3,U}\equiv (U_{x2}-U_{x1},-U_{x2}+U_{x1}, 0, +\infty, -\infty)^\top.\\
 \end{aligned}
 $$ 
Similarly, fix an order of the $5^3$ $3$-tuples in $\mathcal{A}_{xU}$ and list them in   a $5^3\times 3$ matrix, $\alpha_{x,U}$. Using the same order, list the $5^3$ elements of $ \Delta \bar{F}_x(\mathcal{A}_{x,U})$ in a $5^3\times 1$ vector, $f_{x,U}$.
Lastly, construct a $(5+5^2)\times 1$ vector, $\beta_{x,U}$, listing $ \alpha_{x,1,U}$ and the sum of every possible element of $\alpha_{x,2,U}$ with every possible element of $\alpha_{x,3,U}$.

Define the functions $\iota:  \Delta \bar{F}_x(\mathcal{A}_{x,U})\to \{1,2,...,5^3\}$, where $\iota(k)$ is  the row index of scalar $k$ in the vector $f_{x,U}$, and  $\tau: \mathcal{A}_{x,U}\to \{1,2,...,5^3\}$, where $\tau(k)$ is the row index of $3$-tuple $k$ in the matrix $\alpha_{x,U}$.  
 Define $\pi_1$ and $\pi_2$  as in Section \MYref{second_step}.

Finally, by using the formula of $\mbox{Vol}_{ \Delta \bar{F}_x}$, it is useful to write (\MYref{eq24}) and (\MYref{eq25}) in the following more explicit way:
\begin{align}
& -\Delta\bar{F}_x(a_1,a_2,a_3)
+\Delta\bar{F}_x(a_1', a_2, a_3)
+\Delta\bar{F}_x(a_1, a_2', a_3)
-\Delta\bar{F}_x(a_1', a_2', a_3) \nonumber\\
&+ \Delta\bar{F}_x(a_1, a_2, a_3')
-\Delta\bar{F}_x(a_1', a_2, a_3')
-\Delta\bar{F}_x(a_1, a_2', a_3')
+\Delta\bar{F}_x(a_1', a_2', a_3' )\geq 0, \nonumber\\
&\hspace{1cm} \forall (a_1,a_2,a_3), (a_1',a_2',a_3')\in \mathcal{A}_{x,U} \text{ s.t. } (a_1,a_2,a_3)\leq (a_1',a_2',a_3'), \label{eq24_1}\\
& -\Delta\bar{F}_x(a_1,a_2,a_3)
+\Delta\bar{F}_x(a_1', a_2, a_3)
+\Delta\bar{F}_x(a_1, a_2', a_3)
-\Delta\bar{F}_x(a_1', a_2', a_3) \nonumber\\
&+ \Delta\bar{F}_x(a_1, a_2, a_3')
-\Delta\bar{F}_x(a_1', a_2, a_3')
-\Delta\bar{F}_x(a_1, a_2', a_3') 
+\Delta\bar{F}_x(a_1', a_2', a_3' )=0, \nonumber\\
&\hspace{1cm}  \forall (a_1,a_2,a_3), (a_1',a_2',a_3')\in \mathcal{A}_{x,U} \text{ s.t. } (a_1,a_2,a_3)<(a_1',a_2',a_3') \text{ and } a_1>a_2'+a_3' \text{ or } a'_1<a_2+a_3. \label{eq25_1}
\end{align}


\paragraph{Step 1} 
In this step, we present the notion of an equivalence class for every $U\in \mathcal{U}$ and prove that if $\tilde{U},\hat{U}\in \mathcal{U}$ belong to the same equivalence class, then they induce the same set of solutions of the linear program (\MYref{eq0})-(\MYref{eq25}). We add superscripts $\tilde{U}$ or $\hat{U}$ to the 
function $\Delta\bar{F}_x$ to clearly distinguish between a potential solution to the linear program for $\tilde{U}$ and a potential solution to the linear program for $\hat{U}$.

Let $x\in \mathcal{X}$ and $\tilde{U}, \hat{U}\in \mathcal{U}$. Define
\par\nobreak
\vspace{-0.7cm}
{\small $$
\begin{aligned}
&\mathcal{C}_x(\tilde{U})\equiv \Big\{\{(\tilde{t}, \tilde{q}, \tilde{r}),(\tilde{t}', \tilde{q}', \tilde{r}')\}:(\tilde{t}, \tilde{q}, \tilde{r}),(\tilde{t}', \tilde{q}', \tilde{r}') \in \mathcal{A}_{x,\tilde{U}}, (\tilde{t}, \tilde{q}, \tilde{r})\leq (\tilde{t}', \tilde{q}', \tilde{r}')\Big\},\\
&\mathcal{D}_x(\tilde{U})\equiv \Big\{\{(\tilde{t}, \tilde{q}, \tilde{r}),(\tilde{t}', \tilde{q}', \tilde{r}')\}:(\tilde{t}, \tilde{q}, \tilde{r}),(\tilde{t}', \tilde{q}', \tilde{r}') \in \mathcal{A}_{x,\tilde{U}}, (\tilde{t}, \tilde{q}, \tilde{r})< (\tilde{t}', \tilde{q}', \tilde{r}'), \text{ and }\tilde{t}> \tilde{q}'+ \tilde{r}' \text{ or } \tilde{t}'< \tilde{q}+ \tilde{r}\Big\},\\
&\mathcal{C}_x(\hat {U})\equiv \Big\{ \{( \hat{t}, \hat{q}, \hat{r}), (\hat{t}', \hat{q}', \hat{r}')\}:( \hat{t}, \hat{q}, \hat{r}), (\hat{t}', \hat{q}', \hat{r}')  \in  \mathcal{A}_{x,\hat{U}}, (\hat{t}, \hat{q}, \hat{r}) \leq (\hat{t}', \hat{q}', \hat{r}') \Big \},\\
&\mathcal{D}_x(\hat{U})\equiv \Big\{\{(\hat{t}, \hat{q}, \hat{r}),(\hat{t}', \hat{q}', \hat{r}')\}:(\hat{t}, \hat{q}, \hat{r}),(\hat{t}', \hat{q}', \hat{r}') \in \mathcal{A}_{x,\hat{U}}, (\hat{t}, \hat{q}, \hat{r})< (\hat{t}', \hat{q}', \hat{r}'), \text{ and }\hat{t}> \hat{q}'+ \hat{r}' \text{ or } \hat{t}'< \hat{q}+\hat{r}\Big\}.\\
\end{aligned}
$$ }
\begin{def2}{\normalfont({\itshape Equivalence class})}
\label{equiv}
Let $\tilde{U}, \hat{U}\in \mathcal{U}$. $\hat{U}$   belongs to the equivalence class of $\tilde{U}$ at $x\in \mathcal{X}$ if the following conditions hold: 
\begin{enumerate}[1.]
\item For every $\{(\tilde{t}, \tilde{q}, \tilde{r}),(\tilde{t}', \tilde{q}', \tilde{r}')\}\in \mathcal{C}_x(\tilde{U})$, there exists $\{( \hat{t}, \hat{q}, \hat{r}), (\hat{t}', \hat{q}', \hat{r}')\} \in\mathcal{C}_x(\hat {U})$ such that 
\begin{equation}
\label{iota1}
\begin{aligned}
&\iota(\Delta\bar{F}^{\tilde{U}}_x(\tilde{t}, \tilde{q}, \tilde{r}))=\iota(\Delta\bar{F}^{\hat{U}}_x(\hat{t}, \hat{q}, \hat{r})),\\
&\iota(\Delta\bar{F}^{\tilde{U}}_x(\tilde{t}', \tilde{q}, \tilde{r}))=\iota(\Delta\bar{F}^{\hat{U}}_x(\hat{t}', \hat{q}, \hat{r})),\\
&\iota(\Delta\bar{F}^{\tilde{U}}_x(\tilde{t}, \tilde{q}', \tilde{r}))=\iota(\Delta\bar{F}^{\hat{U}}_x(\hat{t}, \hat{q}', \hat{r})),\\
&\iota(\Delta\bar{F}^{\tilde{U}}_x(\tilde{t}', \tilde{q}', \tilde{r}))=\iota(\Delta\bar{F}^{\hat{U}}_x(\hat{t}', \hat{q}', \hat{r})),\\
&\iota(\Delta\bar{F}^{\tilde{U}}_x(\tilde{t}, \tilde{q}, \tilde{r}'))=\iota(\Delta\bar{F}^{\hat{U}}_x(\hat{t}, \hat{q}, \hat{r}')),\\
&\iota(\Delta\bar{F}^{\tilde{U}}_x(\tilde{t}', \tilde{q}, \tilde{r}'))=\iota(\Delta\bar{F}^{\hat{U}}_x(\hat{t}', \hat{q}, \hat{r}')),\\
&\iota(\Delta\bar{F}^{\tilde{U}}_x(\tilde{t}, \tilde{q}', \tilde{r}'))=\iota(\Delta\bar{F}^{\hat{U}}_x(\hat{t}, \hat{q}', \hat{r}')),\\
&\iota(\Delta\bar{F}^{\tilde{U}}_x(\tilde{t}', \tilde{q}', \tilde{r}' ))=\iota(\Delta\bar{F}^{\hat{U}}_x(\hat{t}', \hat{q}', \hat{r}' )),
\end{aligned}
\end{equation}
and vice-versa.
\item For every $\{(\tilde{t}, \tilde{q}, \tilde{r}),(\tilde{t}', \tilde{q}', \tilde{r}')\}\in \mathcal{D}_x(\tilde{U})$, there exists $\{( \hat{t}, \hat{q}, \hat{r}), (\hat{t}', \hat{q}', \hat{r}')\} \in\mathcal{D}_x(\hat {U})$ such that 
\begin{equation}
\label{iota2}
\begin{aligned}
&\iota(\Delta\bar{F}^{\tilde{U}}_x(\tilde{t}, \tilde{q}, \tilde{r}))=\iota(\Delta\bar{F}^{\hat{U}}_x(\hat{t}, \hat{q}, \hat{r})),\\
&\iota(\Delta\bar{F}^{\tilde{U}}_x(\tilde{t}', \tilde{q}, \tilde{r}))=\iota(\Delta\bar{F}^{\hat{U}}_x(\hat{t}', \hat{q}, \hat{r})),\\
&\iota(\Delta\bar{F}^{\tilde{U}}_x(\tilde{t}, \tilde{q}', \tilde{r}))=\iota(\Delta\bar{F}^{\hat{U}}_x(\hat{t}, \hat{q}', \hat{r})),\\
&\iota(\Delta\bar{F}^{\tilde{U}}_x(\tilde{t}', \tilde{q}', \tilde{r}))=\iota(\Delta\bar{F}^{\hat{U}}_x(\hat{t}', \hat{q}', \hat{r})),\\
&\iota(\Delta\bar{F}^{\tilde{U}}_x(\tilde{t}, \tilde{q}, \tilde{r}'))=\iota(\Delta\bar{F}^{\hat{U}}_x(\hat{t}, \hat{q}, \hat{r}')),\\
&\iota(\Delta\bar{F}^{\tilde{U}}_x(\tilde{t}', \tilde{q}, \tilde{r}'))=\iota(\Delta\bar{F}^{\hat{U}}_x(\hat{t}', \hat{q}, \hat{r}')),\\
&\iota(\Delta\bar{F}^{\tilde{U}}_x(\tilde{t}, \tilde{q}', \tilde{r}'))=\iota(\Delta\bar{F}^{\hat{U}}_x(\hat{t}, \hat{q}', \hat{r}')),\\
&\iota(\Delta\bar{F}^{\tilde{U}}_x(\tilde{t}', \tilde{q}', \tilde{r}' ))=\iota(\Delta\bar{F}^{\hat{U}}_x(\hat{t}', \hat{q}', \hat{r}' )),
\end{aligned}
\end{equation}
and vice-versa.
\item $ \pi_2(\alpha_{x,l,\tilde{U}})=\pi_2(\alpha_{x,l,\hat{U}})$ for every $l\in \{1,2,3\}$. 
\end{enumerate}
Let  $[\tilde{U}]_x$ denote the equivalence class of $\tilde{U}$ at $x\in \mathcal{X}$. 
\end{def2}
\begin{lm}
\label{equivlemma}
Let $x\in \mathcal{X}$ and $\tilde{U}, \hat{U}\in \mathcal{U}$. If $\hat{U}\in [\tilde{U}]_x$, then $\tilde{U}$ and $\hat{U}$ induce the same set of solutions of the linear program (\MYref{eq0})-(\MYref{eq25}).
\end{lm}

\begin{proof} Let $x\in \mathcal{X}$ and $\tilde{U}, \hat{U}\in \mathcal{U}$. As discussed in Section \MYref{second_step}, the only pieces of the linear program (\MYref{eq0})-(\MYref{eq25}) that might induce different sets of solutions for different values of $U$ are (\MYref{eq24}) and (\MYref{eq25}).
Therefore, if  (\MYref{eq24}) and (\MYref{eq25})  are identical under $\tilde{U}$ and $\hat{U}$, then $\tilde{U}$ and $\hat{U}$ induce the same set of solutions of the linear program (\MYref{eq0})-(\MYref{eq25}). In what follows, we show that  if  $\hat{U}\in [\tilde{U}]_x$, then (\MYref{eq24}) and (\MYref{eq25})  are identical under $\tilde{U}$ and $\hat{U}$. To do so, we use the equivalent representations of (\MYref{eq24}) and (\MYref{eq25}), which are (\MYref{eq24_1}) and (\MYref{eq25_1}), respectively.

Suppose Condition 3 of Definition \MYref{equiv} holds.
Take any $\{(\tilde{t}, \tilde{q}, \tilde{r}),(\tilde{t}', \tilde{q}', \tilde{r}')\}\in \mathcal{C}_x(\tilde{U})$ and a corresponding  $\{( \hat{t}, \hat{q}, \hat{r}), (\hat{t}', \hat{q}', \hat{r}')\} \in\mathcal{C}_x(\hat {U})$ such that (\MYref{iota1}) holds. 
Write  constraint (\MYref{eq24_1}) at $\{\tilde{U},(\tilde{t}, \tilde{q}, \tilde{r}),(\tilde{t}', \tilde{q}', \tilde{r}')\}$, where the terms of the form $\Delta\bar{F}^{\tilde{U}}_x(\cdot)$ are unknowns to be determined by solving the linear program. Relabel them as $\theta_{\iota(\Delta\bar{F}^{\tilde{U}}_x(\cdot))}$. 
Then, restate (\MYref{eq24_1}) as
\begin{equation}
\label{equiv1}
\begin{aligned}
&-\theta_{\iota(\Delta\bar{F}^{\tilde{U}}_x(\tilde{t}, \tilde{q}, \tilde{r}))}
+\theta_{\iota(\Delta\bar{F}^{\tilde{U}}_x(\tilde{t}', \tilde{q}, \tilde{r}))}
+\theta_{\iota(\Delta\bar{F}^{\tilde{U}}_x(\tilde{t}, \tilde{q}', \tilde{r}))}
-\theta_{\iota(\Delta\bar{F}^{\tilde{U}}_x(\tilde{t}', \tilde{q}', \tilde{r}))}\\
& +\theta_{\iota(\Delta\bar{F}^{\tilde{U}}_x(\tilde{t}, \tilde{q}, \tilde{r}'))}
-\theta_{\iota(\Delta\bar{F}^{\tilde{U}}_x(\tilde{t}', \tilde{q}, \tilde{r}'))}
-\theta_{\iota(\Delta\bar{F}^{\tilde{U}}_x(\tilde{t}, \tilde{q}', \tilde{r}'))}
+\theta_{\iota(\Delta\bar{F}^{\tilde{U}}_x(\tilde{t}', \tilde{q}', \tilde{r}' ))} \geq 0, 
\end{aligned}
\end{equation}
where $\theta$ is a $5^3\times 1$ vector of unknowns and $\theta_h$ denotes the $h$-th element of $\theta$. 
Do the same for $\hat{U}$, 
\begin{equation}
\label{equiv2}
\begin{aligned}
&-\theta_{\iota(\Delta\bar{F}^{\hat{U}}_x(\hat{t}, \hat{q}, \hat{r}))}
+\theta_{\iota(\Delta\bar{F}^{\hat{U}}_x(\hat{t}', \hat{q}, \hat{r}))}
+\theta_{\iota(\Delta\bar{F}^{\hat{U}}_x(\hat{t}, \hat{q}', \hat{r}))}
-\theta_{\iota(\Delta\bar{F}^{\hat{U}}_x(\hat{t}', \hat{q}', \hat{r}))}\\
& +\theta_{\iota(\Delta\bar{F}^{\hat{U}}_x(\hat{t}, \hat{q}, \hat{r}'))}
-\theta_{\iota(\Delta\bar{F}^{\hat{U}}_x(\hat{t}', \hat{q}, \hat{r}'))}
-\theta_{\iota(\Delta\bar{F}^{\hat{U}}_x(\hat{t}, \hat{q}', \hat{r}'))}
+\theta_{\iota(\Delta\bar{F}^{\hat{U}}_x(\hat{t}', \hat{q}', \hat{r}' ))} \geq 0. 
\end{aligned}
\end{equation}
By (\MYref{iota1}), the subscripts of $\theta$ in (\MYref{equiv1}) and (\MYref{equiv2}) are identical. Further, observe that  if some or all of the components of $(\tilde{t}, \tilde{q}, \tilde{r})$ are equal to $(\tilde{t}', \tilde{q}', \tilde{r}')$, then (\MYref{equiv1}) becomes an {\it equality}.   Condition 3 of Definition \MYref{equiv} ensures that if some or all of the components of $(\tilde{t}, \tilde{q}, \tilde{r})$ are equal to $(\tilde{t}', \tilde{q}', \tilde{r}')$, then the same holds for $(\hat{t}, \hat{q}, \hat{r}),(\hat{t}', \hat{q}', \hat{r}')$. 
Therefore,   (\MYref{equiv1}) and (\MYref{equiv2}) are identical. In turn, if Conditions 1 and 3 of Definition \MYref{equiv} hold, then (\MYref{eq24_1}) is identical under $\tilde{U}$ and $\hat{U}$.

Next, take any $\{(\tilde{t}, \tilde{q}, \tilde{r}),(\tilde{t}', \tilde{q}', \tilde{r}')\}\in \mathcal{D}_x(\tilde{U})$ and a corresponding  $\{( \hat{t}, \hat{q}, \hat{r}), (\hat{t}', \hat{q}', \hat{r}')\} \in\mathcal{D}_x(\hat {U})$ such that (\MYref{iota2})   holds. 
Analogously to above, write  constraint (\MYref{eq25_1}) at $\{\tilde{U},(\tilde{t}, \tilde{q}, \tilde{r}),(\tilde{t}', \tilde{q}', \tilde{r}')\}$ as
\begin{equation}
\label{equiv1_1}
\begin{aligned}
&-\theta_{\iota(\Delta\bar{F}^{\tilde{U}}_x(\tilde{t}, \tilde{q}, \tilde{r}))}
+\theta_{\iota(\Delta\bar{F}^{\tilde{U}}_x(\tilde{t}', \tilde{q}, \tilde{r}))}
+\theta_{\iota(\Delta\bar{F}^{\tilde{U}}_x(\tilde{t}, \tilde{q}', \tilde{r}))}
-\theta_{\iota(\Delta\bar{F}^{\tilde{U}}_x(\tilde{t}', \tilde{q}', \tilde{r}))}\\
& +\theta_{\iota(\Delta\bar{F}^{\tilde{U}}_x(\tilde{t}, \tilde{q}, \tilde{r}'))}
-\theta_{\iota(\Delta\bar{F}^{\tilde{U}}_x(\tilde{t}', \tilde{q}, \tilde{r}'))}
-\theta_{\iota(\Delta\bar{F}^{\tilde{U}}_x(\tilde{t}, \tilde{q}', \tilde{r}'))}
+\theta_{\iota(\Delta\bar{F}^{\tilde{U}}_x(\tilde{t}', \tilde{q}', \tilde{r}' ))}= 0.
\end{aligned}
\end{equation}
Do the same for $\hat{U}$, 
\begin{equation}
\label{equiv2_1}
\begin{aligned}
&-\theta_{\iota(\Delta\bar{F}^{\hat{U}}_x(\hat{t}, \hat{q}, \hat{r}))}
+\theta_{\iota(\Delta\bar{F}^{\hat{U}}_x(\hat{t}', \hat{q}, \hat{r}))}
+\theta_{\iota(\Delta\bar{F}^{\hat{U}}_x(\hat{t}, \hat{q}', \hat{r}))}
-\theta_{\iota(\Delta\bar{F}^{\hat{U}}_x(\hat{t}', \hat{q}', \hat{r}))}\\
& +\theta_{\iota(\Delta\bar{F}^{\hat{U}}_x(\hat{t}, \hat{q}, \hat{r}'))}
-\theta_{\iota(\Delta\bar{F}^{\hat{U}}_x(\hat{t}', \hat{q}, \hat{r}'))}
-\theta_{\iota(\Delta\bar{F}^{\hat{U}}_x(\hat{t}, \hat{q}', \hat{r}'))}
+\theta_{\iota(\Delta\bar{F}^{\hat{U}}_x(\hat{t}', \hat{q}', \hat{r}' ))} =0. 
\end{aligned}
\end{equation}
By (\MYref{iota2}), the subscripts of $\theta$ in (\MYref{equiv1_1}) and (\MYref{equiv2_1}) are identical. Therefore,   (\MYref{equiv1_1}) and (\MYref{equiv2_1}) are identical.
In turn, if Condition 2 of  Definition \MYref{equiv} holds, then (\MYref{eq25_1}) is identical under $\tilde{U}$ and $\hat{U}$.

\end{proof}

\paragraph{Step 2} 
In this step, we show how the equivalence classes of Step 1 are related to the  notion of $\pi$-ordering used in Proposition \MYref{part2}. 

\begin{lm}
\label{sufflemma}
Let $x\in \mathcal{X}$ and $\tilde{U}, \hat{U}\in \mathcal{U}$. If 
\begin{enumerate}[i.]
\item  $ \pi_1(\alpha_{x,l,\tilde{U}})=\pi_1(\alpha_{x,l,\hat{U}})$ for every $l\in \{1,2,3\}$,
\item  $ \pi_2(\alpha_{x,l,\tilde{U}})=\pi_2(\alpha_{x,l,\hat{U}})$ for every $l\in \{1,2,3\}$,
\item $ \pi_1(\beta_{x,\tilde{U}})=\pi_2(\beta_{x,\hat{U}})$,
\item $ \pi_2(\beta_{x,\tilde{U}})=\pi_2(\beta_{x,\hat{U}})$,
\end{enumerate}
then $\hat{U}\in [\tilde{U}]_x$. 
\end{lm}
\begin{proof}
Condition ii of Lemma \MYref{sufflemma} coincides with Condition 3 of Definition \MYref{equiv}.  

Further, Condition i  of Lemma \MYref{sufflemma} implies Condition 1 of Definition \MYref{equiv}. Indeed,  let $x\in \mathcal{X}$. Take  $\tilde{U}, \hat{U}\in \mathcal{U}$ such that Condition  i   holds. Take any $\{(\tilde{t}, \tilde{q}, \tilde{r}),(\tilde{t}', \tilde{q}', \tilde{r}')\}\in \mathcal{C}_x(\tilde{U})$. Pick $( \hat{t}, \hat{q}, \hat{r}), (\hat{t}', \hat{q}', \hat{r}')\in \mathcal{A}_{x,\hat{U}}$ such that $\tau((\hat{t}, \hat{q}, \hat{r}))=\tau((\tilde{t}, \tilde{q}, \tilde{r}))$ and $\tau((\hat{t}', \hat{q}', \hat{r}'))=\tau((\tilde{t}', \tilde{q}', \tilde{r}'))$. By Condition i, it should be that $\{( \hat{t}, \hat{q}, \hat{r}), (\hat{t}', \hat{q}', \hat{r}')\}\in \mathcal{C}_x(\hat{U})$. Moreover, it holds that
$$
\begin{aligned}
&\tau((\tilde{t}', \tilde{q}, \tilde{r}))=\tau((\hat{t}', \hat{q}, \hat{r})),\\
&\tau((\tilde{t}, \tilde{q}', \tilde{r}))=\tau((\hat{t}, \hat{q}', \hat{r})),\\
&\tau((\tilde{t}', \tilde{q}', \tilde{r}))=\tau((\hat{t}', \hat{q}', \hat{r})),\\
&\tau((\tilde{t}, \tilde{q}, \tilde{r}'))=\tau((\hat{t}, \hat{q}, \hat{r}')),\\
&\tau((\tilde{t}', \tilde{q}, \tilde{r}'))=\tau((\hat{t}', \hat{q}, \hat{r}')),\\
&\tau((\tilde{t}, \tilde{q}', \tilde{r}'))=\tau((\hat{t}, \hat{q}', \hat{r}')).\\
\end{aligned}
$$
Therefore, (\MYref{iota1}) holds.

Lastly, Conditions i-iv of Lemma \MYref{sufflemma} imply Condition 2 of Definition \MYref{equiv}. Indeed, let $x\in \mathcal{X}$. Take  $\tilde{U}, \hat{U}\in \mathcal{U}$ such that Conditions i-iv  hold. Take any $\{(\tilde{t}, \tilde{q}, \tilde{r}),(\tilde{t}', \tilde{q}', \tilde{r}')\}\in \mathcal{D}_x(\tilde{U})$. Pick $( \hat{t}, \hat{q}, \hat{r}), (\hat{t}', \hat{q}', \hat{r}')\in \mathcal{A}_{x,\hat{U}}$ such that $\tau((\hat{t}, \hat{q}, \hat{r}))=\tau((\tilde{t}, \tilde{q}, \tilde{r}))$ and $\tau((\hat{t}', \hat{q}', \hat{r}'))=\tau((\tilde{t}', \tilde{q}', \tilde{r}'))$. By Conditions i-iv, it should be that $\{( \hat{t}, \hat{q}, \hat{r}), (\hat{t}', \hat{q}', \hat{r}')\}\in \mathcal{D}_x(\hat{U})$. 
Moreover, it holds that
$$
\begin{aligned}
&\tau((\tilde{t}', \tilde{q}, \tilde{r}))=\tau((\hat{t}', \hat{q}, \hat{r})),\\
&\tau((\tilde{t}, \tilde{q}', \tilde{r}))=\tau((\hat{t}, \hat{q}', \hat{r})),\\
&\tau((\tilde{t}', \tilde{q}', \tilde{r}))=\tau((\hat{t}', \hat{q}', \hat{r})),\\
&\tau((\tilde{t}, \tilde{q}, \tilde{r}'))=\tau((\hat{t}, \hat{q}, \hat{r}')),\\
&\tau((\tilde{t}', \tilde{q}, \tilde{r}'))=\tau((\hat{t}', \hat{q}, \hat{r}')),\\
&\tau((\tilde{t}, \tilde{q}', \tilde{r}'))=\tau((\hat{t}, \hat{q}', \hat{r}')).\\
\end{aligned}
$$
Therefore, (\MYref{iota2}) holds.

\end{proof}

\paragraph{Step 3} In this step, we combine Steps 1 and 2 and conclude. 
 Let $\tilde{U}, \hat{U}\in \mathcal{U}$. Lemmas \MYref{equivlemma} and \MYref{sufflemma} imply that if  if $\tilde{U}$ and $\hat{U}$ have the same $\pi$-ordering, then 
either both, $U$ and $\tilde{U}$, lie inside or outside the identified set, $\mathcal{U}^*$.

\begin{remark}{\normalfont({\itshape Proposition \MYref{part2} in practice})}
\label{practice}
In practice, we use Proposition \MYref{part2}  as follows. First, we generate a grid of points covering $\mathcal{U}$ as precisely as possible, depending on the available computational resources. We store the grid points in a matrix called \texttt{grid}. The number of columns of \texttt{grid} is equal to $r+1$. The number of rows of \texttt{grid} is equal to the number of values of $U$   considered. Second, we find the $\pi$-ordering of each row of \texttt{grid}.
Third, we collect the rows of \texttt{grid} producing the same $\pi$-ordering into the same equivalence class. In Matlab, steps 2 and 3 can be straightforwardly implemented by applying the pre-built function \texttt{sort}, without the necessity of solving an  optimisation routine.  
Fourth, we select a representative grid point from each equivalence class. 
Fifth, for each representative grid point, we solve the linear program of Section \MYref{first_step}. If the linear program has a solution, then all the rows of \texttt{grid}  belonging to the  representative grid point's equivalence class are   saved. Otherwise, all such rows   are discarded. The collection of  the rows of \texttt{grid} that are saved across different equivalence classes represents $\mathcal{U}^*$. 

Note that the overall procedure can be easily parallelised. For instance, if  Assumption 5.1 is not imposed, then  steps 1-5 are entirely separable across $x\in \mathcal{X}$, which substantially reduces the computational burden. 
Further, note that if $\Delta \mathcal{F}^\dagger=\Delta \mathcal{F}$, then there is only one equivalence class. Instead, if $\Delta \mathcal{F}^\dagger\subset\Delta \mathcal{F}$, then  the number of equivalence classes  increases with  $r$ and  the amount of nonparametric  restrictions imposed on  $\{\Delta F_x\}_{x\in \mathcal{X}}$. However, providing a general formula for the  number of equivalence classes  does not seem  possible to us.
Lastly, observe that if  \texttt{grid} does not granularly span $\mathcal{U}$, then one may obtain an imprecise approximation of $\mathcal{U}^*$, due to the risk of leaving unexplored some regions of the parameter space or neglecting potential disconnections inside the identified set.  This is a well-known issue in the partial identification literature, where ``gridding'' is still the most popular approach  to construct the sharp identified set for high-dimensional parameters. We discuss how we have carefully addressed this issue in Appendix \MYref{simulations}.
\end{remark}

\counterwithin{equation}{section}
\section{Simulations}
\label{simulations}

In this section, we implement the methodology described in Section \MYref{general_steps} using simulated data.  Given Assumption \MYref{nonpar_assumption}, we consider the six specifications of distributional assumptions summarised in Table \MYref{sim1_assumptions}. 
 \begin{table} [ht]
\centering
{\footnotesize \begin{tabular}{c| c c c c c c c }
\toprule
Assumptions     &  [1]                 &[2]                          &[3]                    &[4]                     & [5]               &[6]                   \\
\midrule
5.1                     &  \checkmark   & \checkmark          &                       &                        &                     &                        \\
5.2                     &                       & \checkmark           & \checkmark   & \checkmark     &                    &\checkmark        \\
5.3                      &                      & \checkmark           &                      & \checkmark     &                     &\checkmark       \\
5.4                      &                      &                              &                      &                         &\checkmark  &\checkmark     \\
 \bottomrule
 \end{tabular}}
\caption{Assumptions on the unobserved heterogeneity maintained in the different specifications.}
\label{sim1_assumptions}
\end{table}
In order to ensure that the volume of our identified sets is not improperly inflated relative to the point identified case, we impose some scale normalisations. In particular, for every $(x,y)\in \mathcal{X}\times \mathcal{Y}$, let  
$$
U^{\text{Logit}}_{xy}\equiv\log \frac{p_{y|x}}{p_{0|x}}, \quad V^{\text{Logit}}_{xy}\equiv\log \frac{p_{x|y}}{p_{0|y}}, \quad\text{and}\quad  \Phi^{\text{Logit}}_{xy}\equiv U^{\text{Logit}}_{xy}+V^{\text{Logit}}_{xy},
$$ be the values of $U_{xy}$, $V_{xy}$, and $\Phi_{xy}$, respectively, that are identified under the Logit  assumption (\hyperlink{Choo_Siow}{Choo and Siow, 2006}). When Assumption 5.1  is not imposed, we divide each element of $U_{x\cdot}\equiv (U_{xy}: y \in \mathcal{Y}_0)$ by  ${U_{x1}}/{U^{\text{Logit}}_{x1}}$ for every $x\in \mathcal{X}$ and each element of $V_{\cdot y} \equiv (V_{xy}: x \in \mathcal{X}_0)$ by  ${V_{1y}}/{V^{\text{Logit}}_{1y}}$ for every $y\in \mathcal{Y}$. Hence, the scale normalisations are $U_{x1}\equiv U^{\text{Logit}}_{x1}$  for every $x\in \mathcal{X}$ and $ V_{1y}\equiv V^{\text{Logit}}_{1y}$ for every $y \in \mathcal{Y}$.
Instead, when Assumption 5.1 is imposed, we divide  each element of $U$  by  $ {U_{11}}/{U^{\text{Logit}}_{11}}$ and each element of $V$ by  ${V_{11}}/{V^{\text{Logit}}_{11}}$. Hence, the scale normalisations are $U_{11}\equiv U^{\text{Logit}}_{11}$ and $   V_{11}\equiv V^{\text{Logit}}_{11}$. 
Note that, when Assumption 5.1 is not imposed, we include $|\mathcal{X}|+|\mathcal{Y}|$ scale normalisations. This is because  determining whether $U$ (resp. $V$) belongs to  $\mathcal{U}^*$ (resp. $\mathcal{V}^*$)   requires recovering $|\mathcal{X}|$ (resp. $|\mathcal{Y}|$) CDFs.  When Assumption 5.1 is imposed, we include  one scale normalisation on each side.  This is because  determining whether $U$  (resp. $V$) belongs to  $\mathcal{U}^*$ (resp. $\mathcal{V}^*$)  requires recovering one CDF.

Observe that Assumptions 5.1-5.4 are always satisfied under the Logit specification. Therefore, due to our choice of scale normalisations, $U^{\text{Logit}}=(U^{\text{Logit}}_{xy}: (x,y)\in \mathcal{X}\times \mathcal{Y}_0)$, $V^{\text{Logit}}=(V^{\text{Logit}}_{xy}: (x,y)\in \mathcal{X}_0\times \mathcal{Y})$, and $\Phi^{\text{Logit}}=(\Phi^{\text{Logit}}_{xy}: (x,y)\in \mathcal{X}\times \mathcal{Y})$  fall inside $\mathcal{U}^*$, $\mathcal{V}^*$, and $\Theta^*$, respectively, for each of specifications [1]-[6].

 As a first exercise, we fix $\mathcal{X}=\mathcal{Y}\equiv \{1,2\}$  and  investigate the identifying power of the 1to1TU model for each of specifications [1]-[6]. We simulate the data under three DGPs:
\begin{itemize}[itemindent=1.5em]
\item[(DGP1)] $\{\epsilon_{iy}\}_{y\in \mathcal{Y}_0}$ are i.i.d., where $\epsilon_{iy}$ is distributed independently from $X_i$, as  standard Extreme Value Type I. Analogous assumptions are 
imposed on the women's side. $\{p_x\}_{x\in \mathcal{X}}$ and  $\{p_y\}_{y\in \mathcal{Y}}$ are set equal to $\{p^{1950}_x\}_{x\in \mathcal{X}}$ and  $\{p^{1950}_y\}_{y\in \mathcal{Y}}$ from Section \MYref{empirical}.\footnote{\label{regroup}Since $r=2$ in this simulation, we regroup the 5 education types in 2 categories: \{HSD, HSG\} and \{SC, CG, CG+\}.}  $\Phi_{xy}$ is set equal to $\log  {p^{1950}_{y|x}}/{p^{1950}_{0|x}}+\log {p^{1950}_{x|y}}/{p^{1950}_{0|y}}$ for each $(x,y)\in \mathcal{X}\times \mathcal{Y}$. Hence, by construction, the simulated match probabilities are  almost equal (not exactly equal, due to simulation error) to   $p^{1950}_{y|x}$ and $p^{1950}_{x|y}$ for every $(x,y)\in \mathcal{Z}$.
\item[(DGP2)] $\epsilon_i$ is distributed independently of $X_i$ as a normal mixture, with $2$ equally weighted components. Every mixture component has mean zero. The two mixture components have the following variance-covariance matrices:
$$
\Sigma=\begin{pmatrix}
1 & 1 &1\\
1 & 1 & 1\\
1 & 1 & 1
\end{pmatrix}\quad\text{and}\quad \Sigma=\begin{pmatrix}
 50 &  -10   &-10\\
   -10&   50&   -10\\
   -10&   -10 & 50
\end{pmatrix}.
$$
Analogous assumptions are imposed on $\{G_y\}_{y\in \mathcal{Y}}$. $\{p_x\}_{x\in \mathcal{X}}$ and  $\{p_y\}_{y\in \mathcal{Y}}$ are set equal to $\{p^{1940}_x\}_{x\in \mathcal{X}}$ and  $\{p^{1940}_y\}_{y\in \mathcal{Y}}$ from Section \MYref{empirical}.  $\Phi_{xy}$ is calibrated so that  the simulated match probabilities are equal to   $p^{1940}_{y|x}$ and $p^{1940}_{x|y}$ for every $(x,y)\in \mathcal{Z}$.\footnote{In order to calibrate $\Phi_{xy}$, we use Proposition 2 in \hyperlink{Galichon_Salanie}{GS} showing that
$$
\Phi_{xy}=\frac{\partial F^*_x (\{p^{1940}_{y|x}\}_{y\in \mathcal{Y}_0})}{\partial p^{1940}_{y|x}} +\frac{\partial G^*_y(\{p^{1940}_{x|y}\}_{x\in \mathcal{X}_0})}{\partial p^{1940}_{y|x}},
$$
where $F^*_x(\{p^{1940}_{y|x}\}_{y\in \mathcal{Y}_0})$ is the Legendre-Fenchel transform of $F_x$ evaluated at $\{p^{1940}_{y|x}\}_{y\in \mathcal{Y}_0}$ and $G^*_y(\{p^{1940}_{x|y}\}_{x\in \mathcal{X}_0})$ is the Legendre-Fenchel transform of $G_y$ evaluated at $\{p^{1940}_{x|y}\}_{x\in \mathcal{X}_0}$. We compute the Legendre-Fenchel transforms by simulation and the derivatives numerically.}
\item[(DGP3)] $\epsilon_i$ is distributed as a normal mixture, with $2$ equally weighted components. When $X_i=1$, the first and second mixture components  have the following means and  variance-covariance matrices:
$$
\mu=\begin{pmatrix}
2\\
2
\end{pmatrix},   \Sigma=\begin{pmatrix}
1 & 1 &1\\
1 & 1 & 1\\
1 & 1 & 1
\end{pmatrix}\quad \text{and} \quad \mu=\begin{pmatrix}
0\\
0
\end{pmatrix},\Sigma=\begin{pmatrix}
 50 &  -20   &-20\\
   -20&   50&   -20\\
   -20&   -20 & 50
\end{pmatrix}.
$$
When $X_i=2$, the first and second mixture components  have the following means and  variance-covariance matrices:
$$
\mu=\begin{pmatrix}
0\\
0
\end{pmatrix},   \Sigma=\begin{pmatrix}
1 & 1 &1\\
1 & 1 & 1\\
1 & 1 & 1
\end{pmatrix}\quad \text{and} \quad \mu=\begin{pmatrix}
4\\
4
\end{pmatrix},\Sigma=\begin{pmatrix}
 40 &  0   &0\\
   0&   40&   0\\
   0&   0 & 40
\end{pmatrix}.
$$
Analogous assumptions are imposed on $\{G_y\}_{y\in \mathcal{Y}}$. $\{p_x\}_{x\in \mathcal{X}}$ and  $\{p_y\}_{y\in \mathcal{Y}}$ are set equal to $\{p^{1967}_x\}_{x\in \mathcal{X}}$ and  $\{p^{1967}_y\}_{y\in \mathcal{Y}}$ from Section \MYref{empirical}.  $\Phi_{xy}$ is calibrated so that  the simulated match probabilities are equal to   $p^{1967}_{y|x}$ and $p^{1967}_{x|y}$ for every $(x,y)\in \mathcal{Z}$. 
\end{itemize}

In Table \MYref{sim1_results}, we report the  true values and the  identified sets of $U$, $V$, $\Phi$, $D_{22,11}(\Phi)$, $C_{21}(U)$, and $C_{21}(V)$. Moreover, we report  $U^{\text{Logit}}$, $V^{\text{Logit}}$, $\Phi^{\text{Logit}}$, $D_{22,11}(\Phi^{\text{Logit}})$, $C_{21}(U^{\text{Logit}})$, and $C_{21}(V^{\text{Logit}})$. We distinguish between the case when Assumption 5.1 is imposed (``w/ 5.1'') and  the case when Assumption 5.1 is not imposed (``w/o 5.1'')  because, as highlighted earlier,  these two cases entail different scale normalisations.\footnote{Consequently, the corresponding identified sets are not necessarily nested.} 
Note that, in DGP3, we do not consider specifications [1] and [2] of Table \MYref{sim1_assumptions} because Assumption 5.1 does not hold.

We highlight a few facts from Table \MYref{sim1_results}. First, in each of the three DGPs considered, specifications [5] and [6] deliver the tightest bounds. This is consistent with the fact that specifications [5] and [6] impose the strongest restrictions on the unobserved heterogeneity among the six specifications considered.
Second, in none of the cases considered, the identified set of  $D_{22,11}(\Phi)$ is bounded on both sides. This is because there is always at least one component of $\Phi$ whose identified set is unbounded on at least one side. In particular, the upper bound for $D_{22,11}(\Phi)$ is always infinity. 
Third, in DGP1 and DGP2, the sign of $D_{22,11}(\Phi)$ is recovered  under specifications [5] and [6].   As discussed in Section \MYref{policy}, detecting the sign of $D_{22,11}(\Phi)$ is important in itself because it reveals the direction of assortativeness. \hyperlink{Graham_2011}{Graham (2011}; \hyperlink{Graham_errata}{2013b)} shows that if the taste shocks are i.i.d., then the sign of $D_{22,11}(\Phi)$ is  identified. Our simulations highlight that i.i.d.-ness is not a necessary condition. 
Fourth, the identified sets of $C_{21}(U)$ and $C_{21}(V)$  are always unbounded on at least one side, except for $C_{21}(U)$  in DGP3 under  specifications [5] and [6]. Such unboundedness implies that  
 the identified set of  the marital education premium will also be unbounded on at least one side (see Equation (\MYref{marital_our})). Further, the  signs of $C_{21}(U)$ and $C_{21}(V)$ are never identified. 
Lastly, in DGP2 and DGP3, the assumption that the  taste shocks are  i.i.d. standard Extreme Value Type I   is misspecified. This implies that $U^{\text{Logit}}$, $V^{\text{Logit}}$, $\Phi^{\text{Logit}}$, $D_{22,11}(\Phi^{\text{Logit}})$, $C_{21}(U^{\text{Logit}})$, and $C_{21}(V^{\text{Logit}})$ are different, sometimes quite significantly, from the true values of $U$, $V$, $\Phi$, $D_{22,11}(\Phi)$, $C_{21}(U)$, and $C_{21}(V)$, respectively. 

As a second exercise, we investigate how the identifying power of the 1to1TU model varies as the number of types, $r$,  increases. In particular, we simulate the data under three DGPs, featuring $r=3$, $r=4$, and $r=5$ for both sides of the market, respectively. In each DGP, $\{\epsilon_{iy}\}_{y\in \mathcal{Y}_0}$ are i.i.d., where $\epsilon_{iy}$ is distributed independently from $X_i$, as standard  Extreme Value Type I. Analogous assumptions are imposed on $\{G_y\}_{y\in \mathcal{Y}}$.  $\{p_x\}_{x\in \mathcal{X}}$ and  $\{p_y\}_{y\in \mathcal{Y}}$ are set equal to $\{p^{1950}_x\}_{x\in \mathcal{X}}$ and  $\{p^{1950}_y\}_{y\in \mathcal{Y}}$ from Section \MYref{empirical}.\footnote{\label{regroup}When $r=3$, we regroup the 5 education types in 3 categories: \{HSD\}, \{HSG\}, and \{SC, CG, CG+\}. When $r=4$, we regroup the 5 education types in 4 categories: \{HSD\}, \{HSG\},  \{SC\}, and \{CG, CG+\}.}  $\Phi_{xy}$ is set equal to $\log {p^{1950}_{y|x}}/{p^{1950}_{0|x}}+\log {p^{1950}_{x|y}}/{p^{1950}_{0|y}}$ for each $(x,y)\in \mathcal{X}\times \mathcal{Y}$, as in DGP1 above.\footnote{Note that we simulate new data for each $r=3,4,5$. This is why, for instance, $U^{\text{Logit}}_{11}$ is not exactly equal across Tables \MYref{sim2a_results}-\MYref{sim2f_results}.}
In Tables \MYref{sim2a_results}-\MYref{sim2f_results}, we report the true values and the  identified sets of $U$, $V$, $\Phi$, $D(\Phi)$, $C(U)$, and $C(V)$, under the three DGPs considered and for specifications [5] and [6] of Table \MYref{sim1_assumptions}. 
Overall, the findings of Table \MYref{sim1_results} are confirmed. In particular, note that in none of the cases considered, the identified sets of $D(\Phi)$, $C(U)$, and $C(V)$ are bounded on both sides. Further, the ability of the model to recover the sign of $D(\Phi)$, $C(U)$, and $C(V)$  seems to deteriorate as $r$ increases.

\begin{figure}
\begin{adjustwidth}{-1.5cm}{}
    \begin{subfigure}[t]{0.55\textwidth}          
            \includegraphics[width=\textwidth]{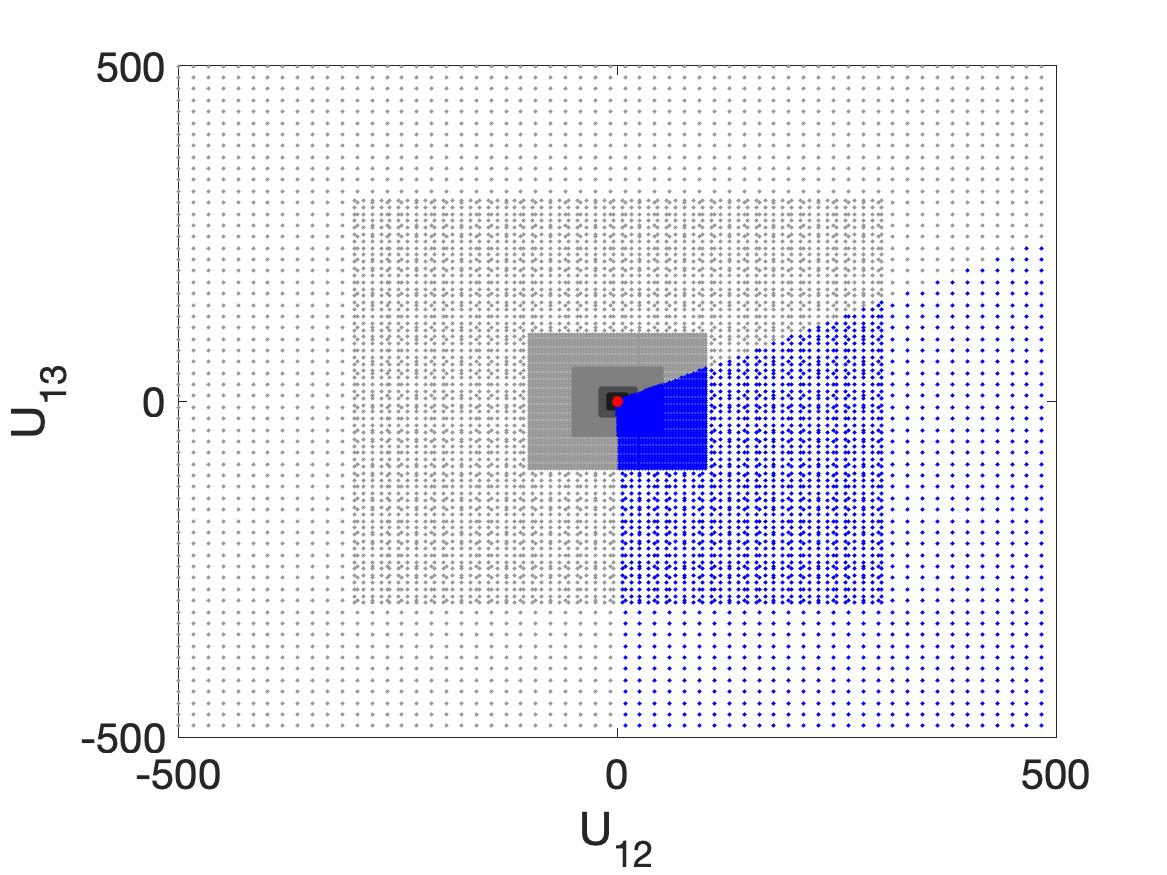}
            \caption{}
    \end{subfigure}%
    \begin{subfigure}[t]{0.55\textwidth}          
            \includegraphics[width=\textwidth]{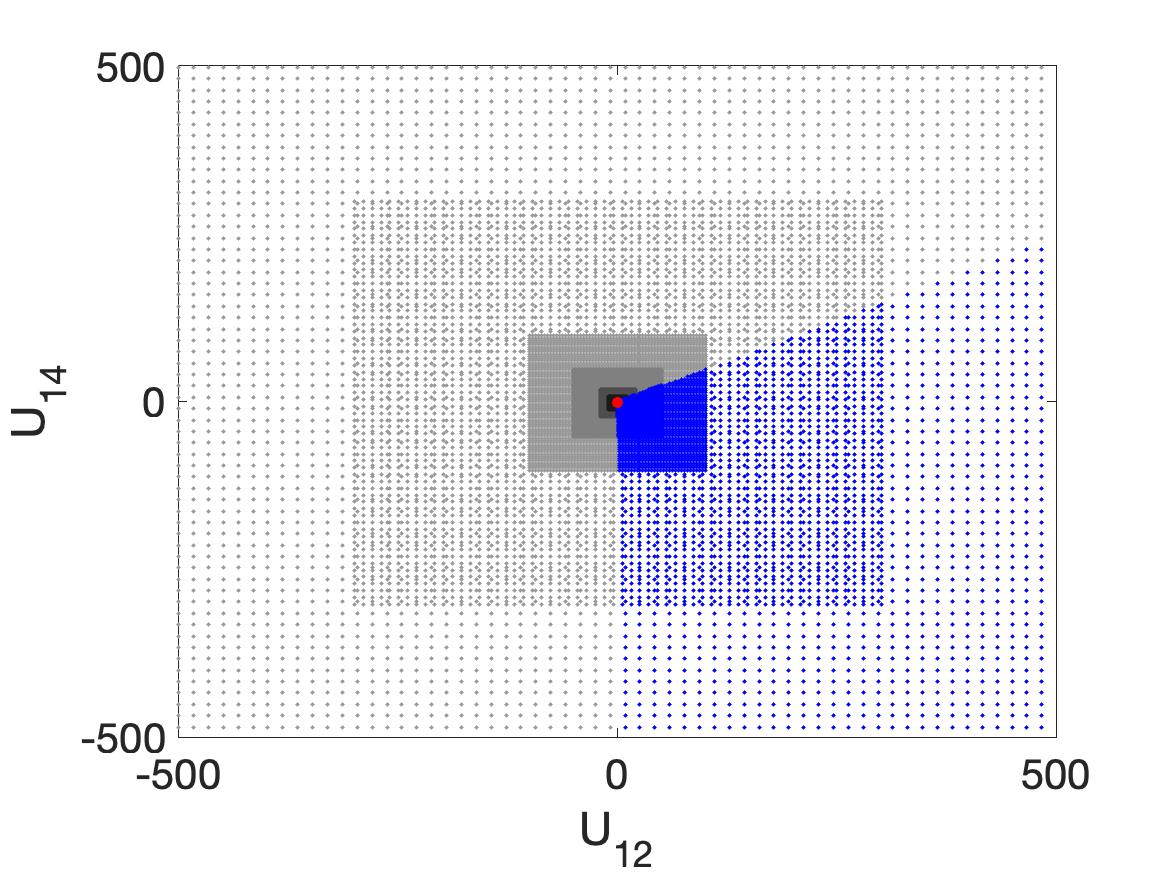}
            \caption{}
    \end{subfigure}%
\end{adjustwidth}
\caption{Projections of the six grids of points to be evaluated by the linear program  (in different shades of grey) and  of the six approximations of the identified set of $U_{1\cdot}$ (in blue). The red points represent the Logit estimates.}
\label{cloud}
\end{figure}

We conclude the section by discussing how we have obtained the grids of parameter values to be evaluated by the linear program.
For instance, consider the construction of the  identified set of $U$  in the  second simulation exercise with $r=4$.
Note that we can   construct the identified set of  $U_{x\cdot} \equiv (U_{xy}: y\in \mathcal{Y}_0)$  separately across $x\in \mathcal{X}$. Also observe that, for any given $x\in \mathcal{X}$, $U_{x0}=0$ (location normalisation) and $U_{x1}=U^\text{Logit}_{x1}$   (scale normalisation). Therefore, for any given $x\in \mathcal{X}$, we  have to span a  $3$-dimensional parameter space. For each $j\in \{500,300,100,50,20,10\}$, we construct   a $3$-dimensional grid by  evenly spacing 200 points between $U^{\text{Logit}}_{xy}-j$ and $U^{\text{Logit}}_{xy}+j$ in each dimension $y\in \{2,3,4\}$. Such a grid has $9\times 10^6$ rows, which can be feasibly evaluated  by combining    the ex-ante partitioning approach of Proposition 2 (as discussed in Remark \MYref{practice}), parallelisation, and cluster facilities. We  thus obtain six approximations of the identified set of $U_{x\cdot}$, one from each of the six grids evaluated. Next, for every pair of elements of $U_{x\cdot}$, we plot the two-dimensional projections of the    six approximated identified sets    in one graph. These graphs allow us to accurately determine the boundaries of the identified set of  $U_{x\cdot}$ and make sure that there are no neglected sources of non-sharpness.\footnote{In particular, we say that the lower (upper) bound of the identified set of $U_{xy}$ is equal to $-\infty$ ($+\infty$) if $U_{xy}$ can take value $U^{\text{Logit}}_{xy}-500$ ($U^{\text{Logit}}_{xy}+500$).}  Importantly, neither in the simulations nor empirical application  we have found a case featuring a disconnected two-dimensional projection. This   facilitates   the computation of the identified sets of  functions of    $U,V$. As an example, Figures  \MYref{cloud} (a) and (b)  display two two-dimensional projections of the identified set of $U_{1\cdot}\equiv (U_{1y}: y\in \mathcal{Y}_0)$. Each figure shows the projections of the six grids of points that are evaluated (in different shades of grey) and the projections of the six approximations of the identified set of $U_{1\cdot}$ (in blue). By construction, we  see a dense cloud of points around the Logit estimates, which gradually becomes sparser as we move towards the boundaries of the parameter space.

\begin{landscape}
\thispagestyle{empty}
 \begin{table} [!htbp]
\centering
\begin{adjustwidth}{-2.2cm}{}
\scalebox{.70}{\begin{tabular}{c|c| c c c c| c c c c| c c c c| c |c c }
\toprule
&Specifications                                      & $U_{11}$   &$U_{12}$          & $U_{21}$        & $U_{22}$          & $V_{11}$    &$V_{12}$           & $V_{21}$                 & $V_{22}$             & $\Phi_{11}$   &$\Phi_{12}$       &$\Phi_{21}$       & $\Phi_{22}$        & $D_{22,11}(\Phi)$   &$C_{21}(U)$      & $C_{21}(V)$ \\
&from Table \MYref{sim1_assumptions}&                  &                         &                         &                         &                    &                          &                                 &                            &                      &                          &                         &                            &                                &                          &                 \\
\toprule
& &&&&&&&&&&&&&&&\\                     
\multirowcell{3}[0pt][l]{w/ 5.1 \\ DGP1} &True \& Logit  &  $1.16$&$0.28$&$0.07$&$1.4$&$1.23$&$-0.55$&$0.95$&$1.36$&$2.39$&$-0.27$&$1.02$&$2.76$&$4.4$&$0.2$&$-0.03$\\
&$[1]$ & $1.16$&$(-\infty, +\infty)$&$(-\infty, +\infty)$&$(-\infty, +\infty)$&$1.23$&$(-\infty,1]$&$(-\infty, +\infty)$&$(-\infty, +\infty)$&$2.39$&$(-\infty, +\infty)$&$(-\infty, +\infty)$&$(-\infty, +\infty)$&$(-\infty, +\infty)$&$(-\infty, +\infty)$&$(-\infty, +\infty)$\\
&$ [2]$ &$1.16$&$(-\infty,1.1]$&$(-\infty, +\infty)$&$[0.1,+\infty)$&$1.23$&$(-\infty,1.2]$&$(-\infty, +\infty)$&$[0.1,+\infty)$&$2.38$&$(-\infty,2.3]$&$(-\infty,+\infty)$&$[0.2,+\infty)$&$(-\infty, +\infty)$&$(-\infty, +\infty)$&$(-\infty, +\infty)$\\
& &&&&&&&&&&&&&&&\\
\midrule
& &&&&&&&&&&&&&&&\\
\multirowcell{3}[0pt][l]{w/o 5.1 \\ DGP1} &True \& Logit  &  $1.16$&$0.28$&$0.07$&$1.4$&$1.23$&$-0.55$&$0.95$&$1.36$&$2.39$&$-0.27$&$1.02$&$2.76$&$4.4$&$0.2$&$-0.03$\\
&$[3]$ & $1.16$&$(-\infty,1.15]$&$0.07$&$[0.07, +\infty)$&$1.23$&$-0.55$&$(-\infty, +\infty)$&$[0.01, +\infty)$&$2.39$&$(-\infty,0.6]$&$(-\infty, +\infty)$&$[0.08, +\infty)$&$(-\infty, +\infty)$&$[-0.89, +\infty)$&$(-\infty, +\infty)$\\
&$[4]$ &$1.16$&$(-\infty,1.15]$&$0.07$&$[0.07, +\infty)$&$1.23$&$-0.55$&$(-\infty, +\infty)$&$[0.01, +\infty)$&$2.39$&$(-\infty,0.6]$&$(-\infty, +\infty)$&$[0.08, +\infty)$&$(-\infty, +\infty)$&$[-0.89, +\infty)$&$(-\infty, +\infty)$\\
&$[5]$ &$1.16$&$[-1.15,1.15]$&$0.07$&$[0.07, +\infty)$&$1.23$&$-0.55$&$[-1.22,1.22]$&$[-0.27, +\infty)$&$2.39$&$[-1.7,0.6]$&$[-1.15,1.29]$&$[-0.2, +\infty)$&$[0.3, +\infty)$&$[-0.89, +\infty)$&$[-1.3, +\infty)$\\
&$[6]$ &$1.16$&$[-1.15,1.15]$&$0.07$&$[0.07, +\infty)$&$1.23$&$-0.55$&$[-1.22,1.22]$&$[0.01, +\infty)$&$2.39$&$[-1.7,0.6]$&$[-1.15,1.29]$&$[0.08, +\infty)$&$[0.58, +\infty)$&$[-0.89, +\infty)$&$[-1.1,  \infty)$\\
 & &&&&&&&&&&&&&&&\\
\toprule
 & &&&&&&&&&&&&&&&\\
\multirowcell{4}[0pt][l]{w/ 5.1 \\ DGP2}  & True & $1.88$&$1.6$&$1.7$&$1.73$&$2$&$0.64$&$2.01$&$1.49$&$3.88$&$2.24$&$3.71$&$3.22$&$1.14$&$-0.07$&$-0.65$\\
& Logit    &$1.88$&$0.56$&$0.99$&$1.93$&$2$&$0.29$&$1.29$&$1.83$&$3.88$&$0.85$&$2.28$&$3.76$&$4.52$&$0.08$&$-0.24$\\
&$[1]$ &$1.88$&$(-\infty, +\infty)$&$(-\infty, +\infty)$&$(-\infty, +\infty)$&$2$&$(-\infty,2]$&$(-\infty, +\infty)$&$(-\infty, +\infty)$&$3.88$&$(-\infty, +\infty)$&$(-\infty, +\infty)$&$(-\infty, +\infty)$&$(-\infty, +\infty)$&$(-\infty, +\infty)$&$(-\infty, +\infty)$\\
&$[2]$ &$ 1.88$&$(-\infty,1.8]$&$(-\infty,+\infty)$&$[0.1,+\infty)$&$2$&$(-\infty,2]$&$(-\infty,2.01]$&$[0.1,+\infty)$&$3.88$&$(-\infty,3.8]$&$(-\infty,+\infty)$&$[0.2,+\infty)$&$(-\infty,+\infty)$&$(-\infty,+\infty]$&$(-\infty,+\infty]$\\
& &&&&&&&&&&&&&&&\\
\midrule
& &&&&&&&&&&&&&&&\\
\multirowcell{4}[0pt][l]{w/o 5.1 \\ DGP2}  & True & $1.88$&$1.6$&$0.99$&$1$&$2$&$0.29$&$2$&$0.68$&$3.88$&$1.89$&$2.99$&$1.68$&$0.67$&$-0.72$&$-1.3$\\
& Logit    &$1.88$&$0.56$&$0.99$&$1.93$&$2$&$0.29$&$1.29$&$1.83$&$3.88$&$0.85$&$2.28$&$3.76$&$4.52$&$0.08$&$-0.24$\\
&$[3]$&$1.88$&$(-\infty,1.87]$&$0.99$&$[0.99, +\infty)$&$2$&$0.29$&$(-\infty,2]$&$[0.3, +\infty)$&$3.88$&$(-\infty,2.16]$&$(-\infty,2.99]$&$[1.29, +\infty)$&$[0.02, +\infty)$&$[-0.78, +\infty)$&$[-1.57, +\infty)$\\
&$[4]$&$1.88$&$(-\infty,1.87]$&$0.99$&$[0.99, +\infty)$&$2$&$0.29$&$(-\infty,2]$&$[0.3, +\infty)$&$3.88$&$(-\infty,2.16]$&$(-\infty,2.99]$&$[1.29, +\infty)$&$[0.02, +\infty)$&$[-0.78, +\infty)$&$[-1.57, +\infty)$\\
&$[5]$&$1.88$&$[-1.87,1.87]$&$0.99$&$[0.99, +\infty)$&$2$&$0.29$&$[-2,2]$&$[0.3, +\infty)$&$3.88$&$[-1.58,2.16]$&$[-1.01,2.99]$&$[1.29, +\infty)$&$[0.02, +\infty)$&$[-0.78, +\infty)$&$[-1.57, +\infty)$\\
&$[6]$&$1.88$&$[-1.87,1.87]$&$0.99$&$[0.99, +\infty)$&$2$&$0.29$&$[-2,2]$&$[0.3, +\infty)$&$3.88$&$[-1.58,2.16]$&$[-1.01,2.99]$&$[1.29, +\infty)$&$[0.02, +\infty)$&$[-0.78, +\infty)$&$[-1.57, +\infty)$\\
 & &&&&&&&&&&&&&&&\\
\toprule
& &&&&&&&&&&&&&&&\\
\multirowcell{7}[0pt][l]{w/o 5.1 \\ DGP3}   & True &$-1.13$&$0$&$-2.41$&$-0.09$&$-1.06$&$-1.61$&$-1.15$&$0$&$-2.19$&$-1.61$&$-3.56$&$-0.09$&$2.89$&$0.01$&$0.22$\\
& Logit & $-1.13$&$-0.45$&$-2.41$&$-1.22$&$-1.06$&$-1.61$&$-1.21$&$-1.24$&$-2.19$&$-2.06$&$-3.62$&$-2.46$&$1.03$&$-0.08$&$-0.01$\\
&$[3]$&$-1.13$&$(-\infty,0]$&$-2.41$&$(-\infty,-0.01]$&$-1.06$&$-1.61$&$(-\infty,-0.01]$&$(-\infty,-0.01]$&$-2.19$&$(-\infty,-1.61]$&$(-\infty,-2.42]$&$(-\infty,-0.02]$&$(-\infty, +\infty)$&$(-\infty, +\infty)$&$(-\infty, +\infty)$\\
&$[4]$&$-1.13$&$(-\infty,0]$&$-2.41$&$(-\infty,-0.01]$&$-1.06$&$-1.61$&$(-\infty,-0.01]$&$(-\infty,-0.01]$&$-2.19$&$(-\infty,-1.61]$&$(-\infty,-2.42]$&$(-\infty,-0.02]$&$(-\infty, +\infty)$&$(-\infty, +\infty)$&$(-\infty, +\infty)$\\
&$[5]$&$-1.13$&$[-1.12,1.12]$&$-2.41$&$[-2.41,2.41]$&$-1.06$&$-1.61$&$(-\infty,-1.07]$&$[-1.61,1.61]$&$-2.19$&$[-2.73,-0.49]$&$(-\infty,-3.48]$&$[-4.02,4.02]$&$[-2.24, +\infty)$&$[-0.85, 0.91]$&$[-0.11, +\infty)$\\
&$[6]$&$-1.13$&$[-1.12,0]$&$-2.41$&$[-2.41,-0.01]$&$-1.06$&$-1.61$&$(-\infty,-1.07]$&$[-1.61,-0.01]$&$-2.19$&$[-2.73,-1.61]$&$(-\infty,-3.48]$&$[-4.02,-0.02]$&$[-1.12, +\infty)$&$[-0.48 ,0.39]$&$[-0.11, +\infty)$\\
& &&&&&&&&&&&&&&&\\
\bottomrule
\end{tabular}}
 \end{adjustwidth}
\caption{Projections of the identified sets of $U$, $V$, $\Phi$, $D_{22,11}(\Phi)$, $C_{21}(U)$, and $C_{21}(V)$ in the first simulation exercise when $r=2$.}
\label{sim1_results}
\end{table}
\raisebox{0cm}{\makebox[\linewidth]{\thepage}}
\end{landscape}
\begin{landscape}
\thispagestyle{empty}
 \begin{table}[!htbp]
\centering
\begin{adjustwidth}{-1cm}{}
\begin{tabular}{c|c|ccc|ccc|ccc}
\toprule
Specifications                                        &                               &\multicolumn{3}{c|}{$U$}  &\multicolumn{3}{c|}{$V$} &\multicolumn{3}{c}{$\Phi$} \\ 
from Table \MYref{sim1_assumptions}  &Wife $\rightarrow$ & $1$ &$2$  &$3$              &$1$ &$2$ &$3$             &$1$ &$2$  &$3$  \\ 
\midrule
                   & Husband $\downarrow$  & & & & & & & & & \\
True \& Logit & \multirow{3}{*}{$1$}          &  $0.13$&$0.65$&$-0.26$&$0.6$&$-0.47$&$-2.63$&$0.73$&$0.18$&$-2.89$\\
$[5] $         &                                          & $0.13$&$[0.2, +\infty)$&$(-\infty, +\infty)$&$0.6$&$-0.47$&$-2.63$&$0.73$&$[-0.27, +\infty)$&$(-\infty, +\infty)$\\
$[6]$          &                                          &$0.13$&$[0.2, +\infty)$&$(-\infty, +\infty)$&$0.6$&$-0.47$&$-2.63$&$0.73$&$[-0.27, +\infty)$&$(-\infty, +\infty)$\\
 \midrule      
True \& Logit & \multirow{3}{*}{$2$} &  $-1.11$&$1.08$&$0.56$&$0.55$&$1.15$&$-0.62$&$-0.56$&$2.23$&$-0.06$\\
 $[5] $        & &$-1.11$&$[-0.5, +\infty)$&$(-\infty, +\infty)$&$[-0.5,0.6]$&$[-0.2, +\infty)$&$[-2.6,2.6]$&$[-1.61,-0.51]$&$[-0.7, +\infty)$&$(-\infty, +\infty)$\\
$[6]$          &  &$-1.11$&$[-0.5, +\infty)$&$(-\infty, +\infty)$&$[-0.5,0.6]$&$[-0.2, +\infty)$&$[-2.6,2.6]$&$[-1.61,-0.51]$&$[-0.7, +\infty)$&$(-\infty, +\infty)$\\
\midrule       
True \& Logit &\multirow{3}{*}{$3$} &    $-2.78$&$-0.04$&$1.39$&$0$&$1.15$&$1.33$&$-2.78$&$1.11$&$2.72$\\
  $[5] $       & &$-2.78$&$[-2.7,2.7]$&$[-1.3, +\infty)$&$(-\infty,0.4]$&$[-0.1, +\infty)$&$[-1.2, +\infty)$&$(-\infty,-2.38]$&$[-2.8, +\infty)$&$[-2.5, +\infty)$\\
  $[6] $       & &$-2.78$&$[-2.7,2.7]$&$[0.1, +\infty)$&$(-\infty,0.4]$&$[-0.1, +\infty)$&$[0.1, +\infty)$&$(-\infty,-2.38]$&$[-2.8, +\infty)$&$[0.2, +\infty)$\\
\bottomrule
\end{tabular}
 \end{adjustwidth}
 \caption{Projections of the identified sets of $U$, $V$, and $\Phi$ in the second simulation exercise when $r=3$.}
\label{sim2a_results}
\end{table}
\vspace{1cm}
 \begin{table}[!htbp]
\centering
\begin{adjustwidth}{5cm}{}
\begin{tabular}{c|cc|cc|cc}
\toprule
Specifications &  \multicolumn{2}{c|}{Husband's payoff}  &\multicolumn{2}{c|}{Wife's payoff}&\multicolumn{2}{c}{Core} \\ 
from Table \MYref{sim1_assumptions}  & $C_{31}(U)$ & $C_{21}(U)$ &  $C_{31}(V)$ & $C_{21}(V)$ & $D_{33,11}(\Phi)$ & $D_{22,11}(\Phi)$  \\ 
\midrule
True \& Logit &  $0.64$&$0.38$&$0.47$&$0.51$&$9.12$&$3.34$\\
  $[5] $       & $(-\infty, +\infty)$&$(-\infty, +\infty)$&$[-1.56, +\infty)$&$[-0.58, +\infty)$&$(-\infty, +\infty)$&$(-\infty, +\infty)$\\
    $[6] $       &$(-\infty, +\infty)$&$(-\infty, +\infty)$&$[-0.65 , +\infty)$&$[-0.58, +\infty)$&$(-\infty, +\infty)$&$(-\infty, +\infty)$\\
\bottomrule
\end{tabular}
 \end{adjustwidth}
 \caption{Projections of the identified sets of  elements of $D(\Phi)$,  $C(U)$, and  $C(V)$ in the second simulation exercise when $r=3$. We take type  ``1'' as reference category.}
\label{sim2b_results}
\end{table}

\end{landscape}
\begin{landscape}
\thispagestyle{empty}
 \begin{table}[!htbp]
\centering
\begin{adjustwidth}{-2cm}{}
\scalebox{.80}{\begin{tabular}{c|c|cccc|cccc|cccc}
\toprule
Specifications                                        &                               &\multicolumn{4}{c|}{$U$}  &\multicolumn{4}{c|}{$V$} &\multicolumn{4}{c}{$\Phi$} \\ 
from Table \MYref{sim1_assumptions}  &Wife $\rightarrow$ & $1$ &$2$  &$3$  &$4$              &$1$ &$2$ &$3$  &$4$                  &$1$ &$2$  &$3$  &$4$       \\ 
\midrule
                   & Husband $\downarrow$  & & & & & & & & & & & & \\
True \& Logit & \multirow{3}{*}{$1$}       &  $0.35 $& $0.77 $& $-0.28 $& $-2.18  $                                         & $0.42 $& $-0.48 $& $-1.69 $& $-4.16   $                                                & $0.77 $& $0.29 $& $-1.96 $& $-6.34 $ \\
$[5] $         &                                          & $0.35$&$[0.76,+\infty)$&$(-\infty,+\infty)$&$(-\infty,+\infty)$         &$0.42$&$-0.48$&$-1.69$&$-4.16$           &$0.77$&$[0.28,+\infty)$&$(-\infty,+\infty)$&$(-\infty,+\infty)$\\
$[6]$          &                                          &$0.35$&$[0.76,+\infty)$&$(-\infty,+\infty)$&$(-\infty,+\infty)$         &$0.42$&$-0.48$&$-1.69$&$-4.16$           &$0.77$&$[0.28,+\infty)$&$(-\infty,+\infty)$&$(-\infty,+\infty)$\\
 \midrule      
True \& Logit & \multirow{3}{*}{$2$} &   $-1.15 $& $1.17 $& $0.2 $& $-0.69 $                                           & $0.19 $& $1.19 $& $0.06 $& $-1.4 $                                                     & $-0.96 $& $2.36 $& $0.26 $& $-2.09 $\\
 $[5] $        & &$-1.15$&$[-0.25,+\infty)$&$(-\infty,+\infty)$&$(-\infty,+\infty)$     &$[-0.25,0.25]$&$[0.76,+\infty)$&$[-0.76,+\infty)$&$[-3.81,3.31]$               &$[-1.40,-0.9]$&$[0.51,+\infty)$&$(-\infty,+\infty)$&$(-\infty,+\infty)$\\
$[6]$          &  &$-1.15$&$[0.25,+\infty)$&$(-\infty,+\infty)$&$(-\infty,+\infty)$      &$[-0.25,0.25]$&$[0.76,+\infty)$&$[-0.76,+\infty)$&$[-3.81,3.31]$               &$[-1.40,-0.9]$&$[1.02,+\infty)$&$(-\infty,+\infty)$&$(-\infty,+\infty)$\\
\midrule       
True \& Logit &\multirow{3}{*}{$3$} &     $-2.22 $& $0.51 $& $0.8 $& $0.09 $                                            & $-0.66 $& $0.75 $& $0.89 $& $-0.39 $                                                   & $-2.88 $& $1.26 $& $1.69 $& $-0.3 $\\
  $[5] $       & &$-2.22$&$[-0.76,+\infty)$&$[-0.25,+\infty)$&$(-\infty,+\infty)$     &$(-\infty,-0.25]$&$(-\infty,+\infty)$&$[-0.25,+\infty)$&$[-3.31,3.81]$           &$(-\infty,-2.47]$&$(-\infty,+\infty)$&$[-0.50,+\infty)$&$(-\infty,+\infty)$\\
  $[6] $       & &$-2.22$&$[-0.76,+\infty)$&$[-0.25,+\infty)$&$(-\infty,+\infty)$     &$(-\infty,-0.25]$&$(-\infty,+\infty)$&$[-0.25,+\infty)$&$[-3.31,3.81]$           &$(-\infty,-2.47]$&$(-\infty,+\infty)$&$[-0.50,+\infty)$&$(-\infty,+\infty)$\\
\midrule       
True \& Logit &\multirow{3}{*}{$4$} &     $-3.73 $& $-0.49 $& $0.22 $& $1.2 $                                           & $-1.76 $& $0.17 $& $0.71 $& $1.13 $                                                    & $-5.49 $& $-0.32 $& $0.93 $& $2.33 $\\
  $[5] $       & &$-3.73$&$[-3.31,3.31]$&$[-1.27,+\infty)$&$[-0.76,+\infty)$    &$(-\infty,-0.76]$&$(-\infty,+\infty)$&$(-\infty,+\infty)$&$[-1.27,+\infty)$                  &$(-\infty,-4.49]$&$(-\infty,+\infty)$&$(-\infty,+\infty)$&$[-2.03,+\infty)$\\
  $[6] $       & &$-3.73$&$[-3.31,3.31]$&$[-1.27,+\infty)$&$[0.25,+\infty)$     &$(-\infty,-0.76]$&$(-\infty,+\infty)$&$(-\infty,+\infty)$&$[0.25,+\infty)$                   &$(-\infty,-4.49]$&$(-\infty,+\infty)$&$(-\infty,+\infty)$&$[0.50,+\infty)$\\
\bottomrule
\end{tabular}}
 \end{adjustwidth}
 \caption{Projections of the identified sets of $U$, $V$, and $\Phi$ in the second simulation exercise when $r=4$.}
\label{sim2c_results}
\end{table}

\vspace{2cm}
 \begin{table}[!htbp]
\centering
\begin{adjustwidth}{1cm}{}
\scalebox{.95}{\begin{tabular}{c|ccc|ccc|ccc}
\toprule
Specifications &  \multicolumn{3}{c|}{Husband's payoff}  &\multicolumn{3}{c|}{Wife's payoff}&\multicolumn{3}{c}{Core} \\ 
from Table \MYref{sim1_assumptions}  & $C_{41}(U)$ & $C_{31}(U)$ & $C_{21}(U)$ & $C_{41}(V)$ & $C_{31}(V)$& $C_{21}(V)$ & $D_{44,11}(\Phi)$ & $D_{33,11}(\Phi)$& $D_{22,11}(\Phi)$  \\ 
\midrule
True \& Logit &  $0.31$&$0.1$&$0.21$&$0.51$&$0.45$&$0.61$&$14.93$&$7.3$&$3.8$\\
  $[5] $       & $(-\infty,+\infty)$&$(-\infty,+\infty)$&$(-\infty,+\infty)$&$[-1.58,+\infty)$&$(-\infty,+\infty)$&$(-\infty,+\infty)$&$(-\infty,+\infty)$&$(-\infty,+\infty)$&$(-\infty,+\infty)$\\
    $[6] $&      $(-\infty,+\infty)$&$(-\infty,+\infty)$&$(-\infty,+\infty)$&$[-0.64,+\infty)$&$(-\infty,+\infty)$&$(-\infty,+\infty)$&$(-\infty,+\infty)$&$(-\infty,+\infty)$&$(-\infty,+\infty)$\\
\bottomrule
\end{tabular}}
 \end{adjustwidth}
 \caption{Projections of the identified sets of  elements of $D(\Phi)$,  $C(U)$, and  $C(V)$ in the second simulation exercise when $r=4$. We take type  ``1'' as reference category.}
\label{sim2d_results}
\end{table}

\end{landscape}

\begin{landscape}
\thispagestyle{empty}
 \begin{table}[!htbp]
\centering
\begin{adjustwidth}{-2.2cm}{}
\scalebox{.69}{\begin{tabular}{c|c|ccccc|ccccc|ccccc}
\toprule
Specifications                                        &                               &\multicolumn{5}{c|}{$U$}  &\multicolumn{5}{c|}{$V$} &\multicolumn{5}{c}{$\Phi$} \\ 
from Table \MYref{sim1_assumptions}  &Wife $\rightarrow$ & $1$ &$2$  &$3$  &$4$    &$5$           &$1$ &$2$ &$3$  &$4$   &$5$                &$1$ &$2$  &$3$  &$4$  &$5$      \\ 
\midrule
                   & Husband $\downarrow$  & & & & & & & & & & & & & & & \\
True \& Logit & \multirow{3}{*}{$1$}       &  $0.30 $& $0.58 $& $-0.35 $& $-2.61 $ &  $-3.12$                         & $0.39 $& $-0.61 $& $-1.73 $& $-3.80 $& $-4.32$                                          & $0.69$ &  $-0.03$ & $ -2.08$ &  $-6.41$   &$-7.44$ \\
$[5] $         &                                          & $0.30$&$[0.52,+\infty)$&$(-\infty,+\infty)$&$(-\infty,+\infty)$&$(-\infty,+\infty)$         &$0.39$&$-0.61$&$-1.73$&$-3.80$& $-4.32$                                                    &$0.69$&$[-0.09,+\infty)$&$(-\infty,+\infty)$&$(-\infty,+\infty)$&$(-\infty,+\infty)$\\
$[6]$          &                                          &$0.30$&$[0.52,+\infty)$&$(-\infty,+\infty)$&$(-\infty,+\infty)$ &$(-\infty,+\infty)$        &$0.39$&$-0.61$&$-1.73$&$-3.80$ &  $-4.32$                                                    &$0.69$&$[-0.09,+\infty)$&$(-\infty,+\infty)$&$(-\infty,+\infty)$&$(-\infty,+\infty)$\\
 \midrule      
True \& Logit & \multirow{3}{*}{$2$} &   $-1.28 $& $1.03 $& $0.12 $& $-1.14 $& $-1.81$                                       & $0.09$  &  $1.11$   & $0.01$ &  $-1.05$ &  $-1.73$                                                                                        &   -$1.19$  &  $2.14$   & $0.13$  & $-2.19$ &  $-3.54$\\
 $[5] $        & &$-1.28$&$[0.52,+\infty)$&$(-\infty,+\infty)$&$(-\infty,+\infty)$&$(-\infty,+\infty)$                                                       &$[-3.32,3.32]$&$[-1.4,+\infty)$&$[-5.68,+\infty)$&$(-\infty,+\infty)$&$(-\infty,+\infty)$               &$[-4.6,2.04]$&$[-0.88,+\infty)$&$(-\infty,+\infty)$&$(-\infty,+\infty)$&$(-\infty,+\infty)$\\
$[6]$          &  &$-1.28$&$[0.52,+\infty)$&$(-\infty,+\infty)$&$(-\infty,+\infty)$& $(-\infty,+\infty)$                                                       &$[-3.32,3.32]$&$[-1.4,+\infty)$&$[-5.68,+\infty)$&$(-\infty,+\infty)$  & $(-\infty,+\infty)$            &$[-4.6,2.04]$&$[-0.88,+\infty)$&$(-\infty,+\infty)$&$(-\infty,+\infty)$&$(-\infty,+\infty)$\\
\midrule       
True \& Logit &\multirow{3}{*}{$3$} &     $-2.03 $& $0.51 $& $0.78 $& $-0.49 $  &$-0.96$                                          & $-0.52$  & $0.74$  &  $0.82$  &$-0.25$ &  $-0.74$                                               & $-2.55$  &  $1.25$  &  $1.60$ &  $-0.74$  &$ -1.70$\\
  $[5] $       & &$-2.03$&$[-1.53,+\infty)$&$[-2.36,+\infty)$&$(-\infty,+\infty)$& $(-\infty,+\infty)$                                                      &$(-\infty,1.66]$&$(-\infty,+\infty)$&$[-4.32,+\infty)$&$[-8.23,9.45]$  &$(-\infty,+\infty)$         &$(-\infty,0.37]$&$(-\infty,+\infty)$&$[-6.68+\infty)$&$(-\infty,+\infty)$&$(-\infty,+\infty)$\\
  $[6] $       & &$-2.03$&$[-1.53+\infty)$&$[-2.36,+\infty)$&$(-\infty,+\infty)$ &  $(-\infty,+\infty)$                                                    &$(-\infty,1.66]$&$(-\infty,+\infty)$&$[-4.32,+\infty)$&$[-8.23,9.45]$  &$(-\infty,+\infty)$         &$(-\infty,0.37]$&$(-\infty,+\infty)$&$[-6.68,+\infty)$&$(-\infty,+\infty)$&$(-\infty,+\infty)$\\
\midrule       
True \& Logit &\multirow{3}{*}{$4$} &     $-3.21 $& $ -0.51 $& $0.32 $& $0.43 $ & $-0.16$                                         & $-1.73$ & $-0.32$&  $0.32$  &  $0.63$ &   $0.03$                                                   &$ -4.94$   &$-0.83$  &  $0.64$   & $1.06$ &  $-0.13$\\
  $[5] $       & &$-3.21$&$(-\infty,+\infty)$&$(-\infty,+\infty)$&$(-\infty,+\infty)$&$(-\infty,+\infty)$                                                      &$(-\infty,2.89]$&$(-\infty,+\infty)$&$(-\infty,+\infty)$&$(-\infty,+\infty)$  &$(-\infty,+\infty)$                &$(-\infty,+\infty)$&$(-\infty,+\infty)$&$(-\infty,+\infty)$&$(-\infty,+\infty)$&$(-\infty,+\infty)$\\
  $[6] $       & &$-3.21$&$(-\infty,+\infty)$&$(-\infty,+\infty)$&$(-\infty+\infty)$&$(-\infty,+\infty)$                                                       &$(-\infty,2.89]$&$(-\infty,+\infty)$&$(-\infty,+\infty)$&$(-\infty,+\infty)$ &$(-\infty,+\infty)$                  &$(-\infty,+\infty)$&$(-\infty,+\infty)$&$(-\infty,+\infty)$&$(-\infty,+\infty)$&$(-\infty,+\infty)$\\
  \midrule       
True \& Logit &\multirow{3}{*}{$5$} &     $-4.25 $& $ -0.66 $& $0.27 $& $0.89 $ & $1.01$                                         & $-3.43$  & $-1.13$ &  $-0.39$  & $0.42$ &  $0.53$                                                 & $-7.68$ &  $-1.79$  & $-0.12$&   $1.31$  &  $1.54$\\
  $[5] $       & &$-4.25$&$(-\infty,+\infty)$&$(-\infty,+\infty)$&$(-\infty,+\infty)$&$(-\infty,+\infty)$                                                      &$(-\infty,+\infty)$&$(-\infty,+\infty)$&$(-\infty,+\infty)$&$(-\infty,+\infty)$   &$(-\infty,+\infty)$                &$(-\infty,+\infty)$&$(-\infty,+\infty)$&$(-\infty,+\infty)$&$(-\infty,+\infty)$&$(-\infty,+\infty)$\\
  $[6] $       & &$-4.25$&$(-\infty,+\infty)$&$(-\infty,+\infty)$&$(-\infty,+\infty)$&$(-\infty,+\infty)$                                                       &$(-\infty,+\infty)$&$(-\infty,+\infty)$&$(-\infty,+\infty)$&$(-\infty,+\infty)$ & $(-\infty,+\infty)$                 &$(-\infty,+\infty)$&$(-\infty,+\infty)$&$(-\infty,+\infty)$&$(-\infty,+\infty)$&$(-\infty,+\infty)$\\
\bottomrule
\end{tabular}}
 \end{adjustwidth}
 \caption{Projections of the identified sets of $U$, $V$, and $\Phi$ in the second simulation exercise when $r=5$.}
\label{sim2e_results}
\end{table}

\vspace{2cm}
 \begin{table}[!htbp]
\centering
\begin{adjustwidth}{-2cm}{}
\scalebox{.90}{\begin{tabular}{c|cccc|cccc|cccc}
\toprule
Specifications &  \multicolumn{4}{c|}{Husband's payoff}  &\multicolumn{4}{c|}{Wife's payoff}&\multicolumn{3}{c}{Core} \\ 
from Table \MYref{sim1_assumptions} & $C_{51}(U)$   & $C_{41}(U)$ & $C_{31}(U)$ & $C_{21}(U)$ & $C_{51}(V)$ & $C_{41}(V)$ & $C_{31}(V)$& $C_{21}(V)$ &  $D_{55,11}(\Phi)$ & $D_{44,11}(\Phi)$ & $D_{33,11}(\Phi)$& $D_{22,11}(\Phi)$  \\ 
\midrule
True \& Logit & $0.43$ & $-0.08$&$0.09$&$0.17$& $0.05$&$0.22$&$0.27$&$0.53$&$17.36$ &$13.11$&$6.92$&$4.05$\\
  $[5] $  &  $(-\infty,+\infty)$     & $(-\infty,+\infty)$&$(-\infty,+\infty)$&$(-\infty,+\infty)$&$(-\infty,+\infty)$ &$(-\infty,+\infty)$&$(-\infty,+\infty)$&$(-\infty,+\infty)$ &$(-\infty,+\infty)$ &$(-\infty,+\infty)$&$(-\infty,+\infty)$&$(-\infty,+\infty)$\\
    $[6] $&   $(-\infty,+\infty)$   & $(-\infty,+\infty)$&$(-\infty,+\infty)$&$(-\infty,+\infty)$&$(-\infty,+\infty)$ &$(-\infty,+\infty)$&$(-\infty,+\infty)$&$(-\infty,+\infty)$ &$(-\infty,+\infty)$ &$(-\infty,+\infty)$&$(-\infty,+\infty)$&$(-\infty,+\infty)$\\
\bottomrule
\end{tabular}}
 \end{adjustwidth}
 \caption{Projections of the identified sets of  elements of $D(\Phi)$,  $C(U)$, and  $C(V)$ in the second simulation exercise when $r=5$. We take type  ``1'' as reference category.}
\label{sim2f_results}
\end{table}

\end{landscape}

\counterwithin{equation}{section}


\section{Additional details on the empirical application}
\label{empirical_appendix}

\begin{figure}[!htbp]
\centering
\begin{adjustwidth}{-1.7cm}{}
    \begin{subfigure}[t]{0.43\textwidth}          
            \includegraphics[width=\textwidth, height=4.8cm]{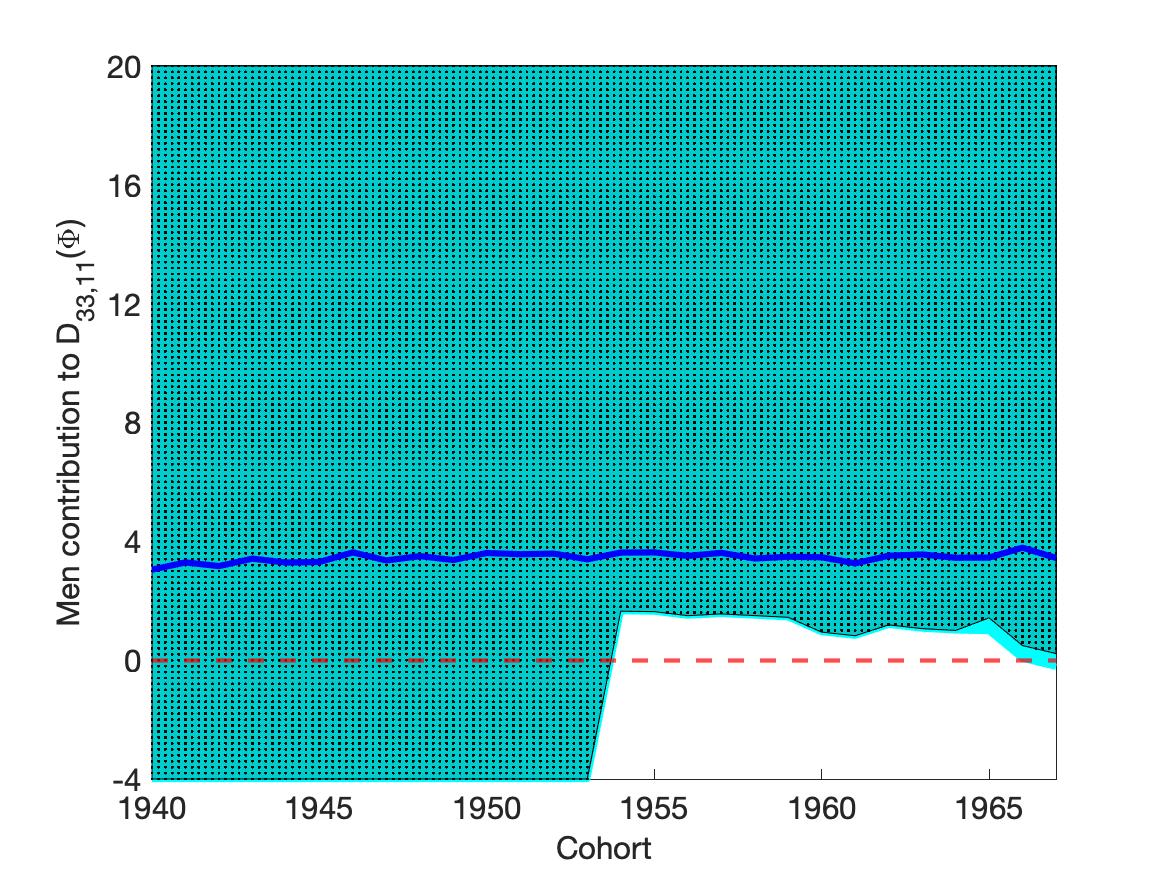}
            \subcaption{\hspace*{-1em}}
    \end{subfigure}%
    \begin{subfigure}[t]{0.43\textwidth}
        \hspace{-0.7cm}
            \includegraphics[width=\textwidth, height=4.8cm]{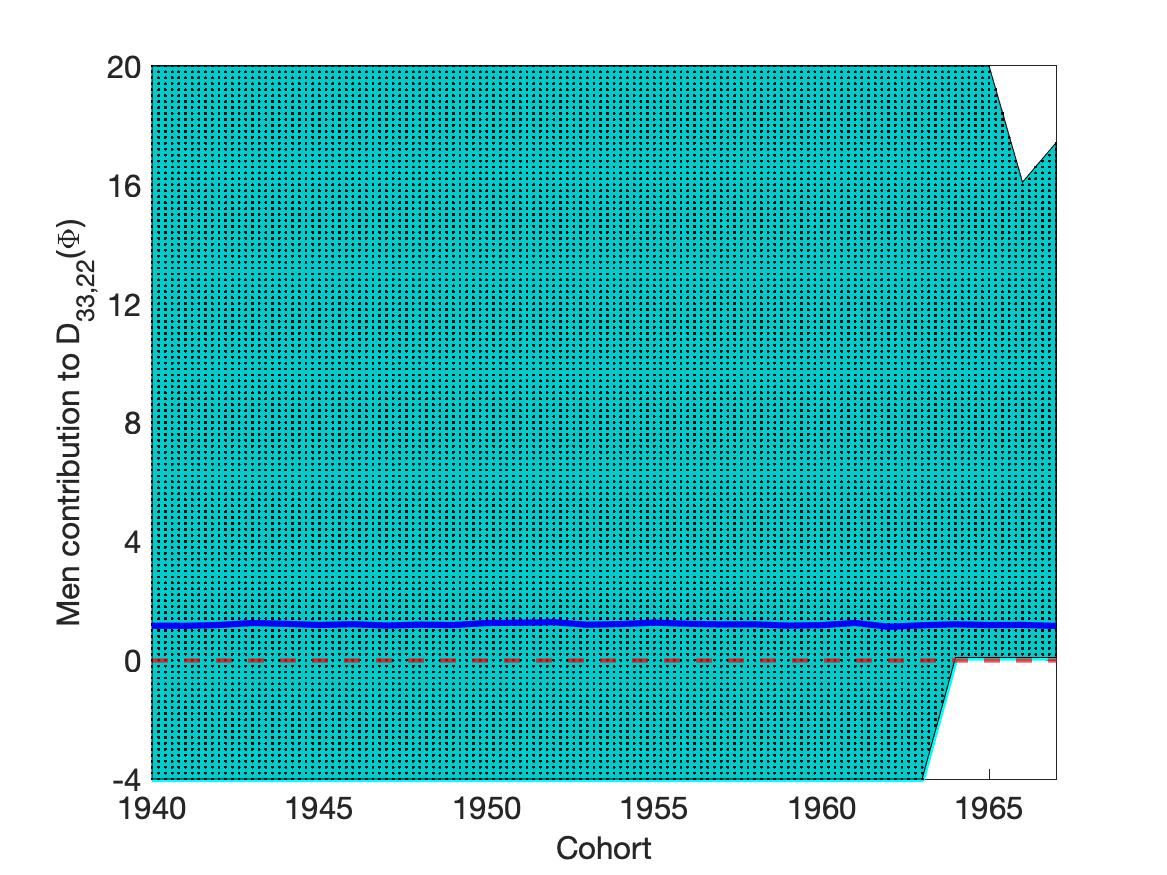}
           \subcaption{\hspace*{1.1em}}
    \end{subfigure}%
   \begin{subfigure}[t]{0.43\textwidth}
           \hspace{-1.3cm}
            \includegraphics[width=\textwidth, height=4.8cm]{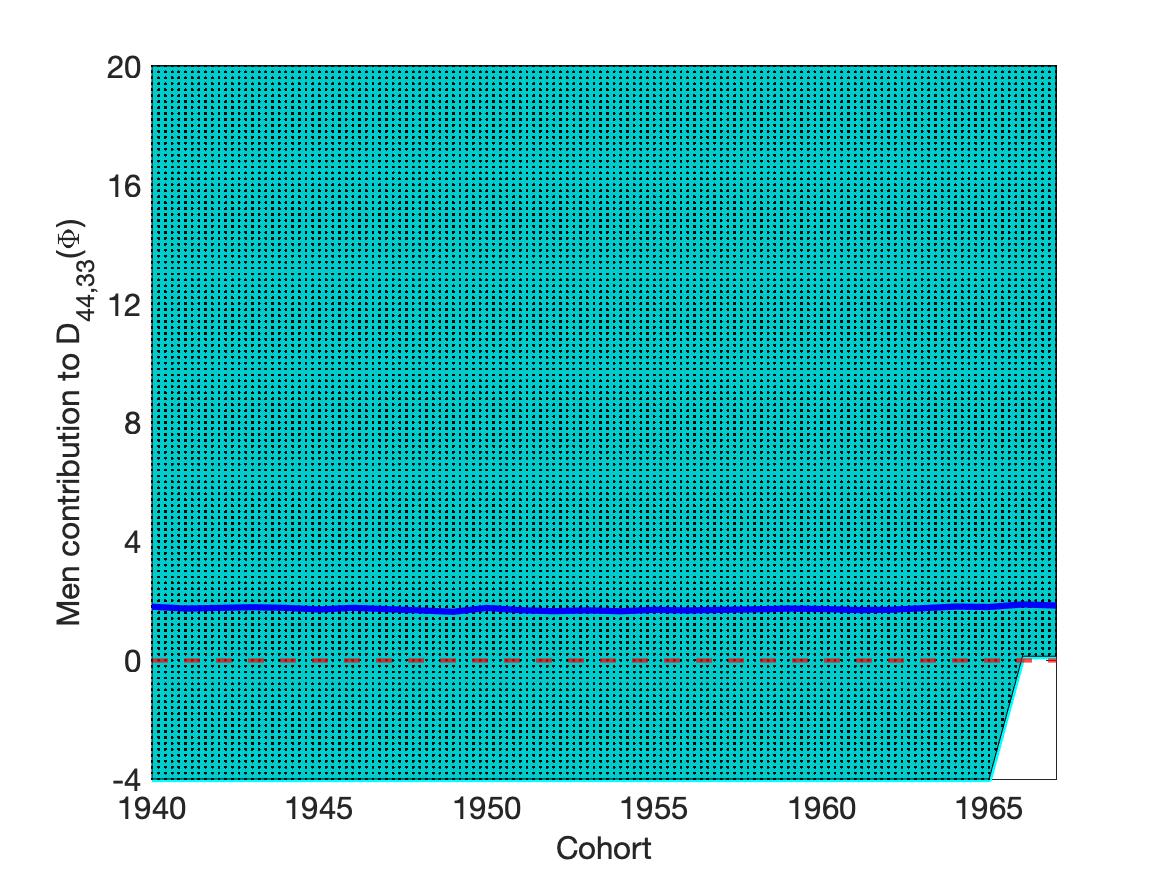}
             \subcaption{\hspace*{4.2em}}
    \end{subfigure}

    \begin{subfigure}[t]{0.43\textwidth}          
            \includegraphics[width=\textwidth, height=4.8cm]{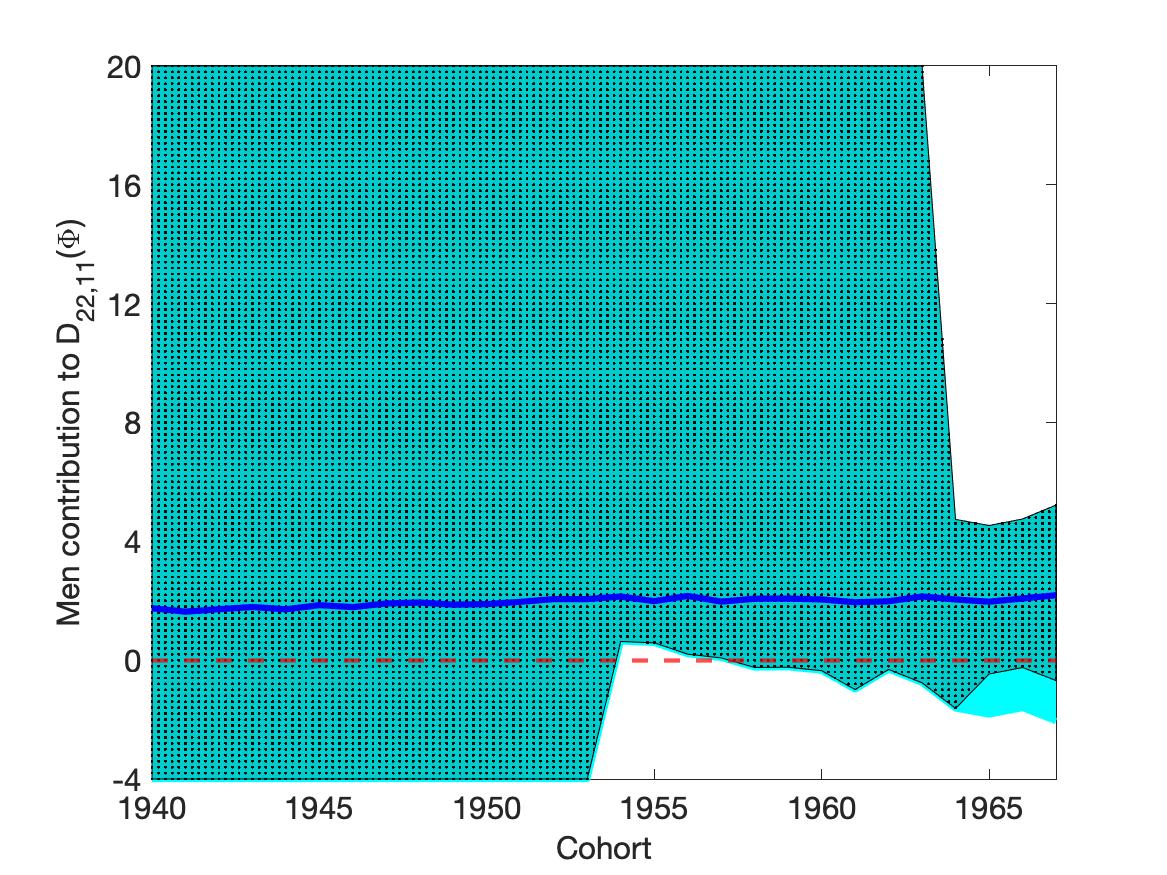}
             \subcaption{\hspace*{-1em}}
    \end{subfigure}%
    \begin{subfigure}[t]{0.43\textwidth}
            \hspace{-0.7cm}
            \includegraphics[width=\textwidth, height=4.8cm]{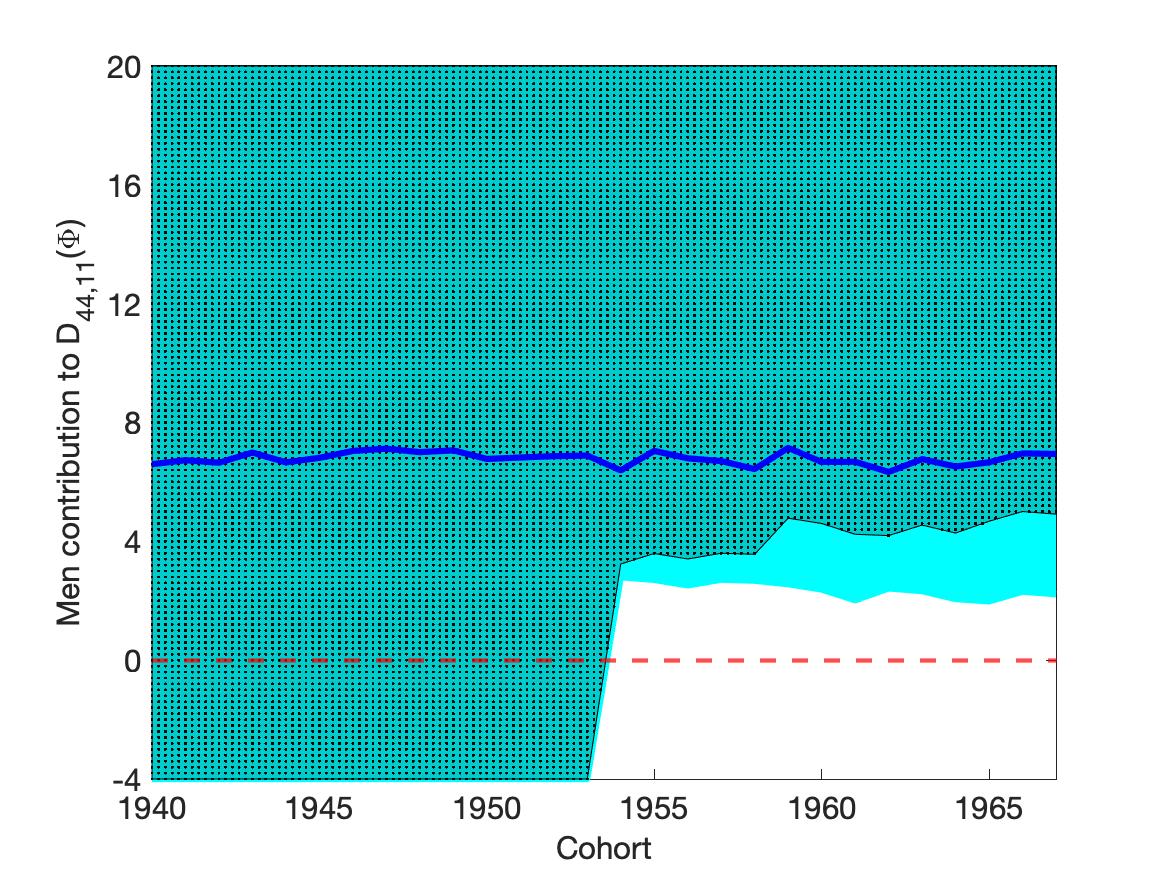}
              \subcaption{\hspace*{1.1em}}
    \end{subfigure}%
   \begin{subfigure}[t]{0.43\textwidth}
           \hspace{-1.3cm}
            \includegraphics[width=\textwidth, height=4.8cm]{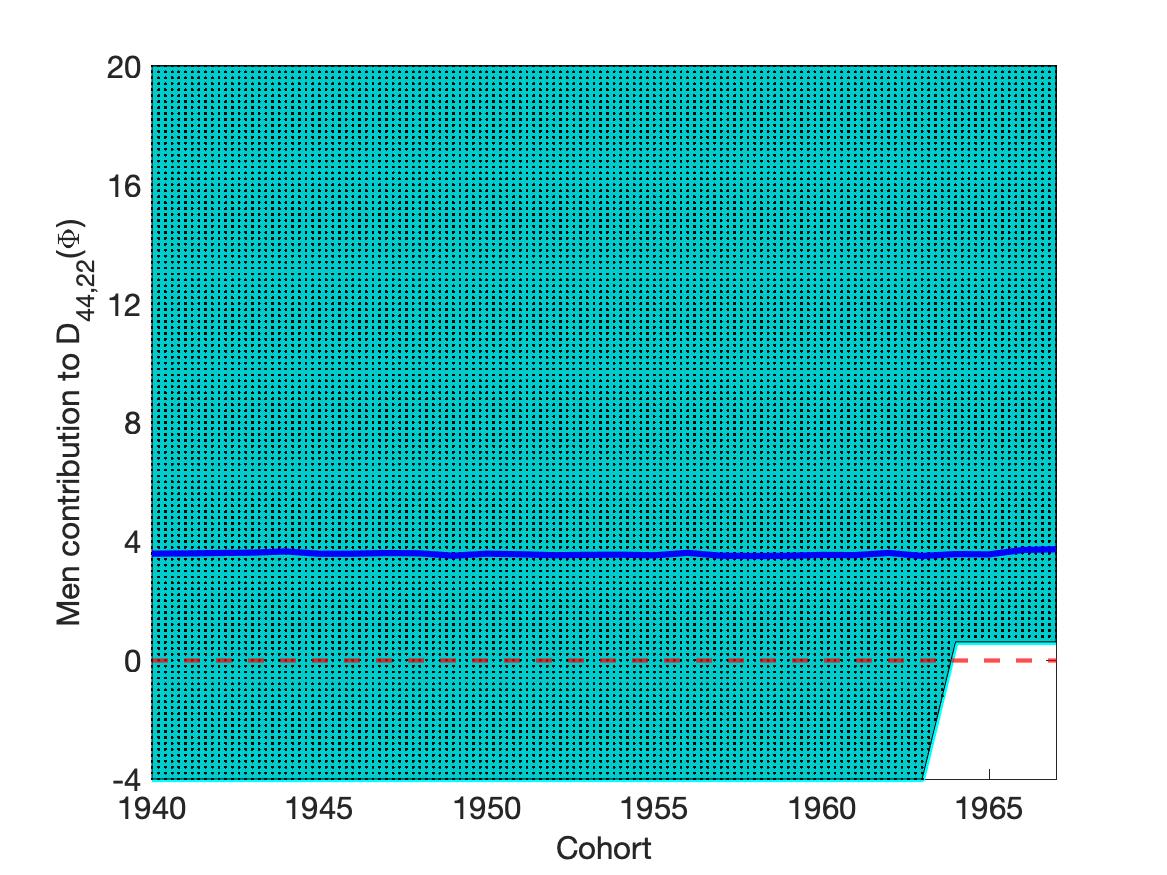}
            \subcaption{\hspace*{4.2em}}
    \end{subfigure}
    \end{adjustwidth}
\caption{The  blue and dotted regions are the estimated identified sets of $U_{xx}+U_{\tilde{x}\tilde{x}}-U_{x\tilde{x}}-U_{\tilde{x}x}$ for each $x,\tilde{x}\in \mathcal{X}$ with $x>\tilde{x}$, under specifications [\text{A}] and [B], respectively.   The dark blue line represents the Logit estimates.}
\label{core_ours_M}
\end{figure}

\begin{figure}[!htbp]
\centering
\begin{adjustwidth}{-1.7cm}{}
\vspace{-0.6cm}
    \begin{subfigure}[t]{0.43\textwidth}          
            \includegraphics[width=\textwidth, height=4.8cm]{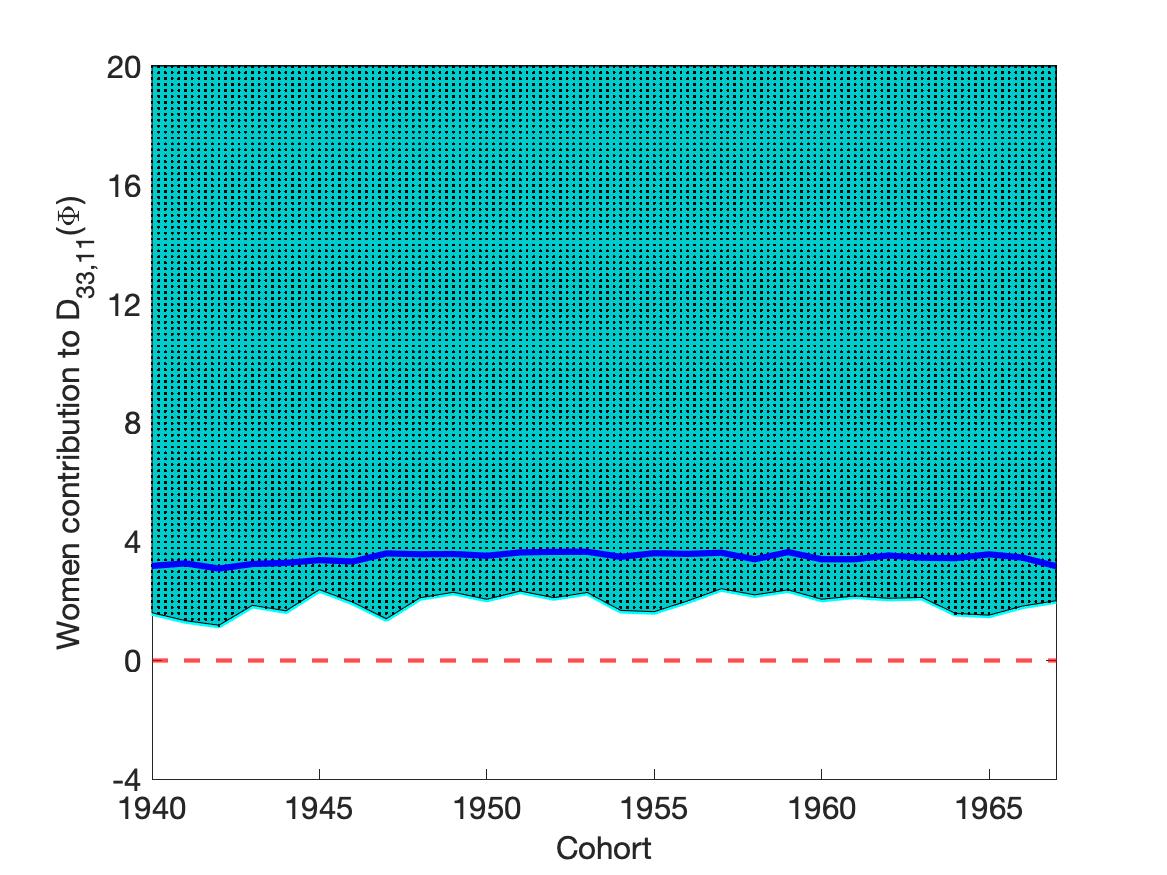}
            \subcaption{\hspace*{-1em}}
    \end{subfigure}%
    \begin{subfigure}[t]{0.43\textwidth}
        \hspace{-0.7cm}
            \includegraphics[width=\textwidth, height=4.8cm]{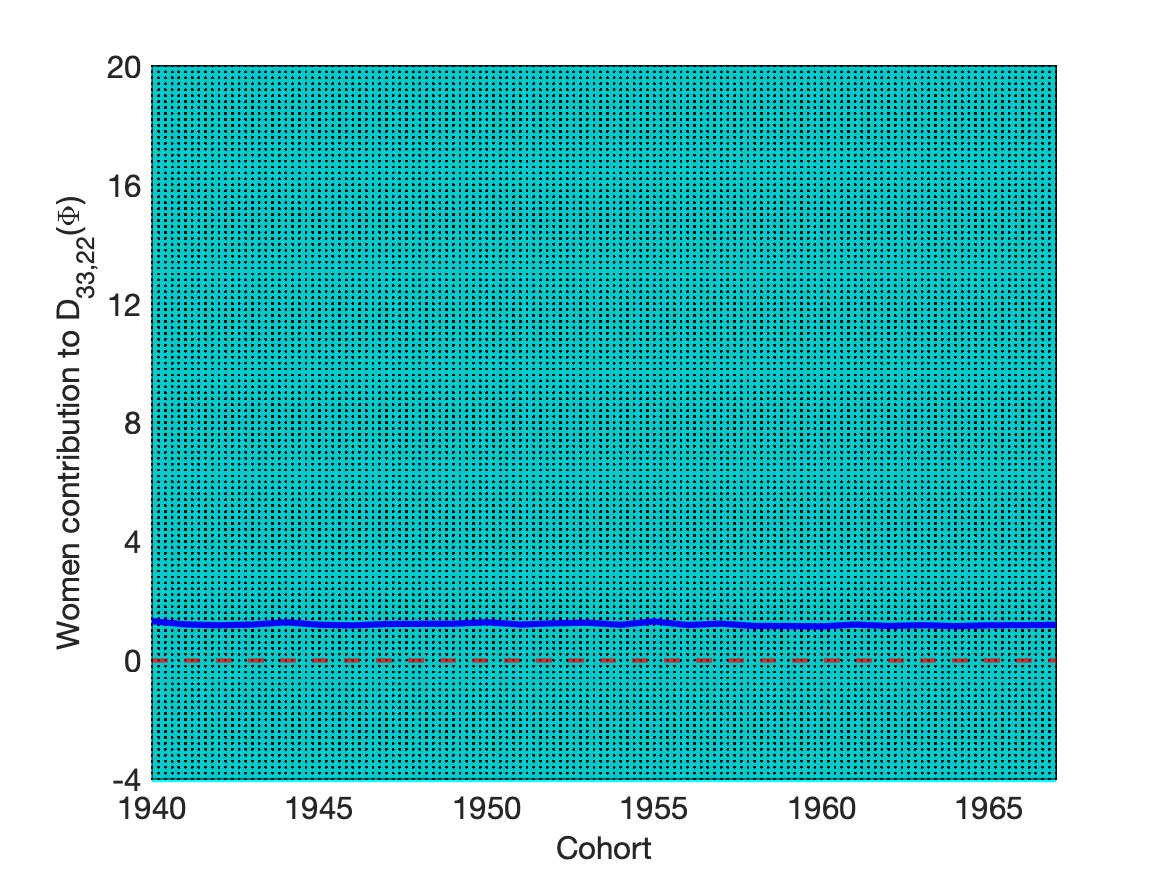}
           \subcaption{\hspace*{1.1em}}
    \end{subfigure}%
   \begin{subfigure}[t]{0.43\textwidth}
           \hspace{-1.3cm}
            \includegraphics[width=\textwidth, height=4.8cm]{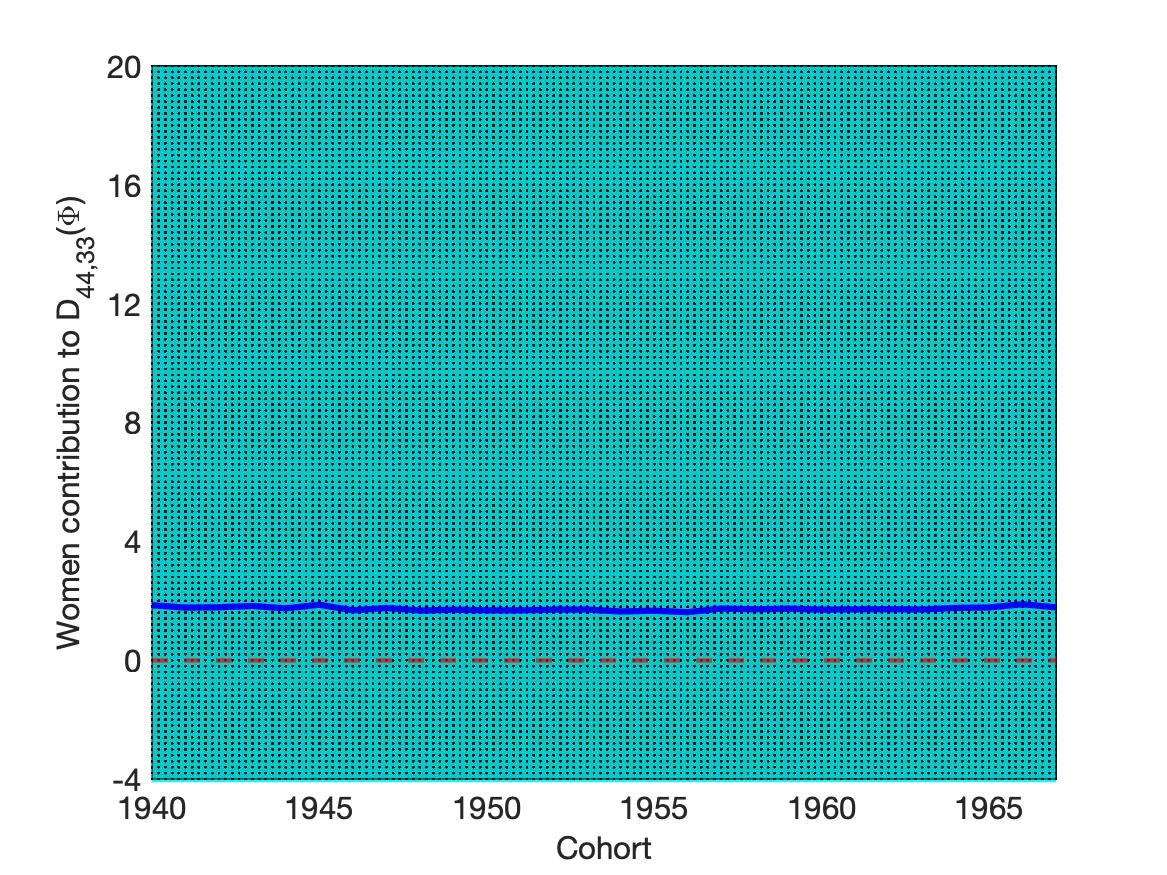}
             \subcaption{\hspace*{4.2em}}
    \end{subfigure}

    \begin{subfigure}[t]{0.43\textwidth}          
            \includegraphics[width=\textwidth, height=4.8cm]{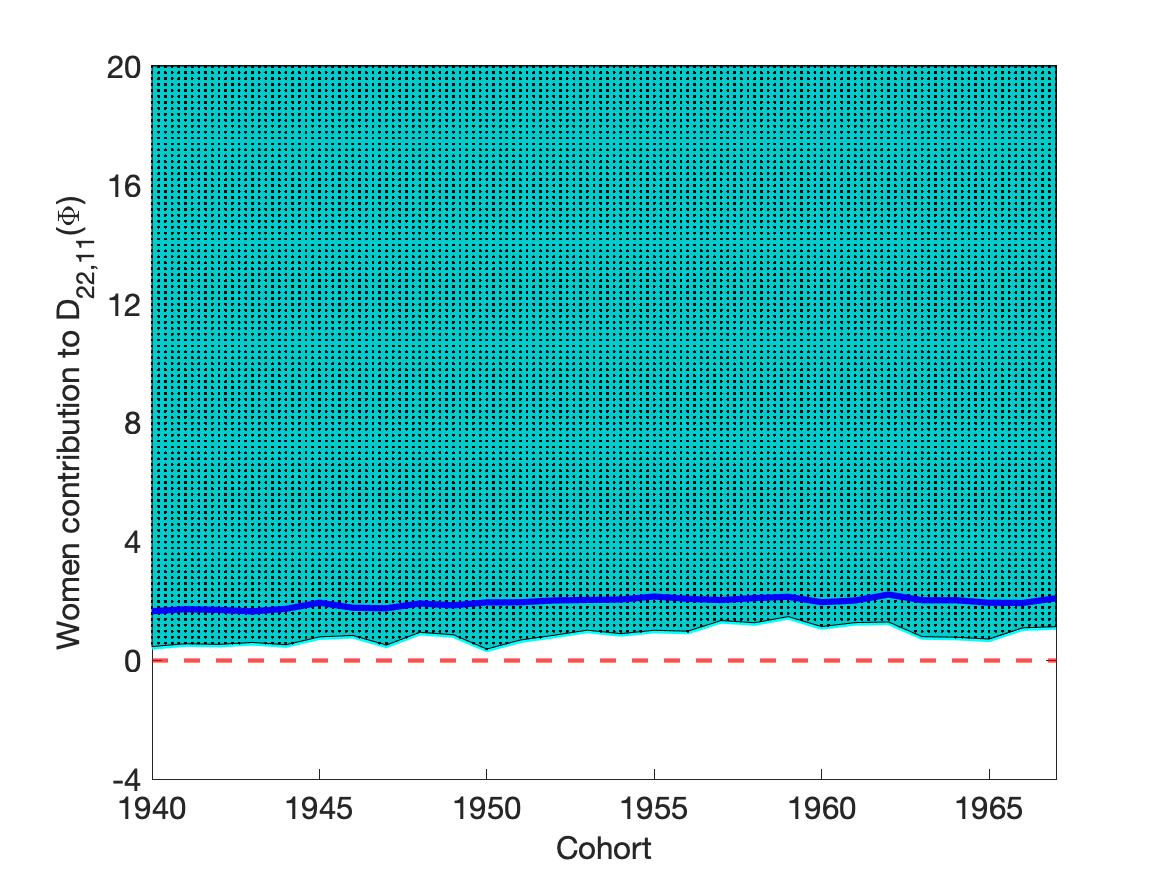}
             \subcaption{\hspace*{-1em}}
    \end{subfigure}%
    \begin{subfigure}[t]{0.43\textwidth}
            \hspace{-0.7cm}
            \includegraphics[width=\textwidth, height=4.8cm]{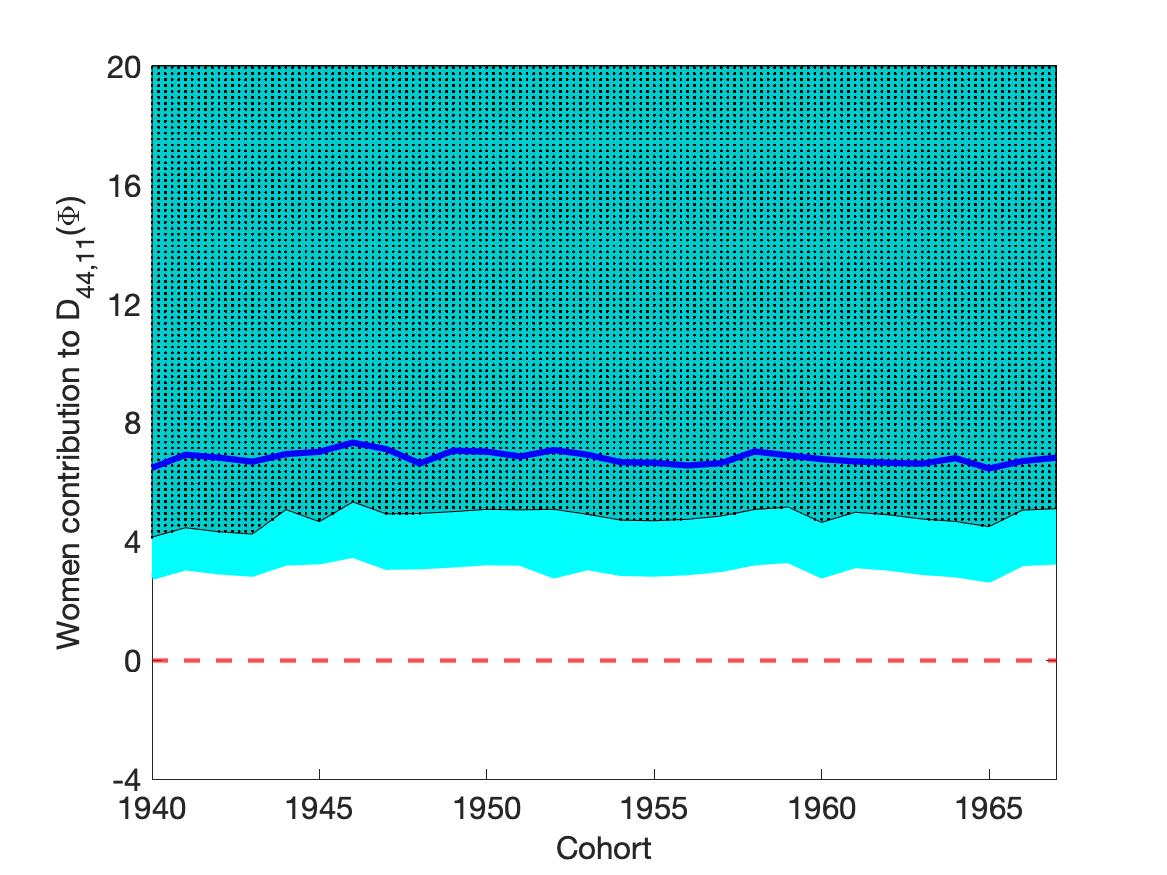}
              \subcaption{\hspace*{1.1em}}
    \end{subfigure}%
   \begin{subfigure}[t]{0.43\textwidth}
           \hspace{-1.3cm}
            \includegraphics[width=\textwidth, height=4.8cm]{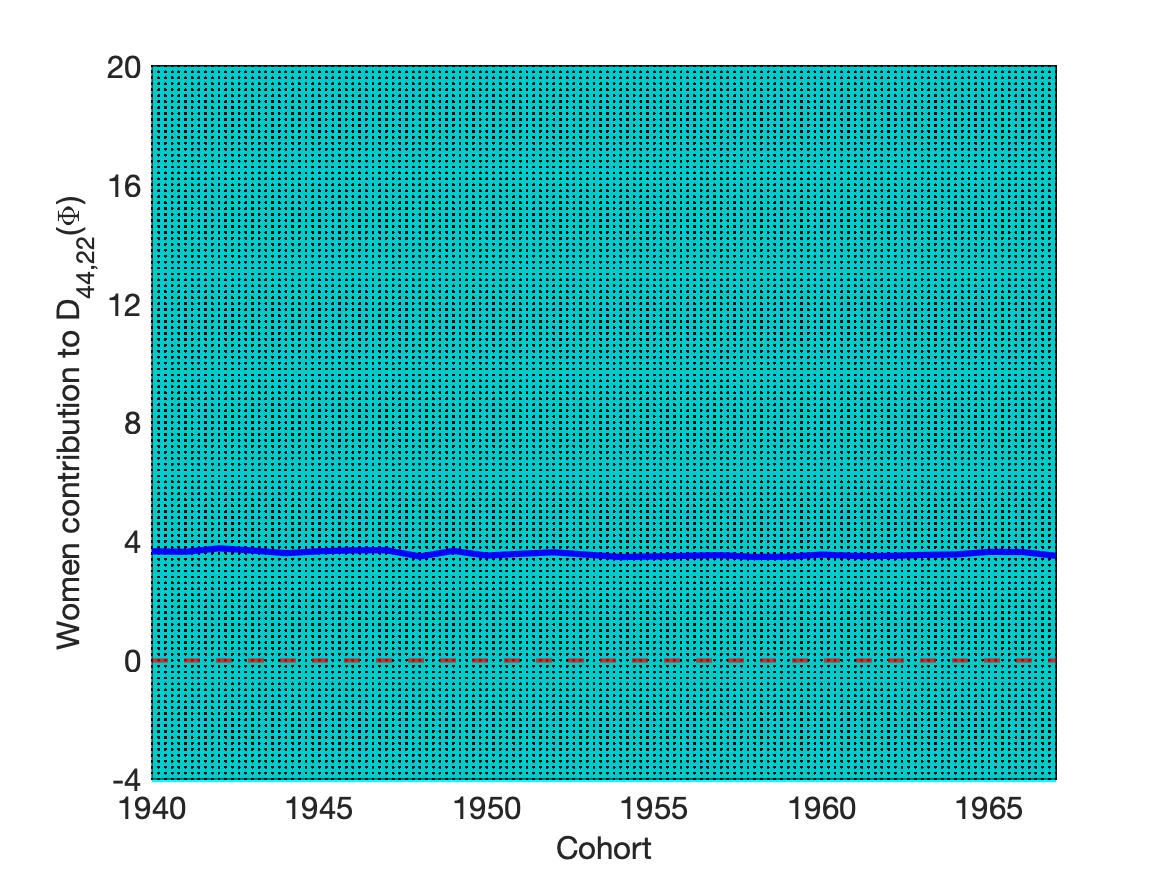}
            \subcaption{\hspace*{4.2em}}
    \end{subfigure}
    \end{adjustwidth}
\caption{The  blue and dotted regions are the estimated identified sets of $V_{yy}+V_{\tilde{y}\tilde{y}}-V_{y\tilde{y}}-V_{\tilde{y}y}$ for each $y,\tilde{y}\in \mathcal{Y}$ with $y>\tilde{y}$, under specifications [\text{A}] and [B], respectively.   The dark blue line represents the Logit estimates..}
\label{core_ours_W}
\end{figure}

\end{appendix}
\end{document}